\documentclass[runningheads]{llncs}
\usepackage[margin=1in]{geometry}
\usepackage[utf8]{inputenc} % allow utf-8 input
\usepackage[T1]{fontenc}    % use 8-bit T1 fonts
\usepackage{hyperref}       % hyperlinks
\usepackage{url}            % simple URL typesetting
\usepackage{booktabs}       % professional-quality tables
\usepackage{amsfonts}       % blackboard math symbols
\usepackage{nicefrac}       % compact symbols for 1/2, etc.
\usepackage{microtype}      % microtypography
\usepackage{xcolor}         % colors
\usepackage{verbatim} % block comment

\usepackage{wrapfig}
\RequirePackage{algorithm}%
\RequirePackage{algorithmicx}%
\RequirePackage{algpseudocode}%
\RequirePackage{listings}%
\usepackage{xr}
\usepackage{amsmath}
\usepackage{amssymb}
\usepackage{mathtools}
\usepackage{thmtools}
\usepackage{thm-restate}
\usepackage{enumitem}
\usepackage{bm}
\usepackage{soul}
\usepackage{subcaption}
\usepackage{graphicx}
\usepackage{epstopdf}
\usepackage{mathtools}
\usepackage{algorithm}
\usepackage{epstopdf}
\usepackage{dsfont}
\usepackage{multirow}
\usepackage{wrapfig}
\usepackage{adjustbox}
\usepackage{soul}
\usepackage[title,titletoc]{appendix}
\usepackage{hyperref}

\makeatletter

\renewcommand\section{\@startsection{section}{1}{\z@}%
  {-1.2ex plus -.2ex minus -.2ex}% space BEFORE
  {0.8ex plus .2ex}%             space AFTER
  {\normalfont\Large\bfseries}}

\renewcommand\subsection{\@startsection{subsection}{2}{\z@}%
  {-1.0ex plus -.2ex minus -.2ex}%
  {0.6ex plus .2ex}%
  {\normalfont\large\bfseries}}

\makeatother

\renewcommand{\sectionautorefname}{Appendix} 

\graphicspath{{Fig/}}

\begin{document}

\def\phc{{\sc pHapCompass}}

%\subtitle{Subject Section}
\title{pHapCompass: Probabilistic Assembly and Uncertainty Quantification of Polyploid Haplotype Phase}

\author{Marjan Hosseini\inst{1}$^{,*}$\and
Ella Veiner\inst{1}\and
Thomas Bergendahl\inst{1}\and
Tala Yasenpoor\inst{1}\and
Zane Smith\inst{2}\and
Margaret Staton\inst{2}\and
Derek Aguiar\inst{1,3}$^,$\thanks{To whom correspondence should be addressed. E-mail: \url{marjan.hosseini@uconn.edu} and \url{derek.aguiar@uconn.edu}}
}
\authorrunning{Hosseini et al.}
\institute{School of Computing, University of Connecticut\\
\and
Department of Entomology and Plant Pathology, University of Tennessee\\
\and
Institute for Systems Genomics, University of Connecticut}

\maketitle

\begin{abstract}
\thispagestyle{empty}

Computing haplotypes from sequencing data, i.e. haplotype assembly, is an important component of molecular and population genetics problems, including interpreting the effects of genetic variation on complex traits and reconstructing genealogical relationships. 
Assembling the haplotypes of polyploid genomes remains a significant challenge due to the exponential search space of haplotype phasings and read assignment ambiguity;
the latter challenge is particularly difficult for haplotype assemblers since the information contained within the observed sequence reads is often insufficient for unambiguous haplotype assignment in polyploid genomes.
We present \phc{}, probabilistic haplotype assembly algorithms for diploid and polyploid genomes that explicitly model and propagate read assignment ambiguity to compute a distribution over polyploid haplotype phasings.
We develop graph theoretic algorithms to enable statistical inference and uncertainty quantification despite an exponential space of possible phasings.
Since prior work evaluates polyploid haplotype assembly on synthetic genomes that do not reflect the realistic genomic complexity of polyploidy organisms, we develop a computational workflow for simulating genomes and DNA-seq for auto- and allopolyploids.
Additionally, we generalize the vector error rate and minimum error correction evaluation criteria for partially phased haplotypes.
Benchmarking of \phc{} and several existing polyploid haplotype assemblers 
shows that \phc{} yields competitive performance across varying genomic complexities and polyploid structures while retaining an accurate quantification of phase uncertainty.
The source code for \phc{}, simulation scripts, and datasets are freely available at \href{https://github.com/bayesomicslab/pHapCompass}{https://github.com/bayesomicslab/pHapCompass}.

% \textbf{Contact:} \{marjan.hosseini, thomas.palmer, derek.aguiar\}@uconn.edu\\
\end{abstract}


\newpage

\section{Introduction}
% \setcounter{page}{1}

% what is polyploid haplotype assembly and why is it important?
Genomic variability is completely characterized by the sequences of genetic variant alleles along a single chromosome, or \textit{haplotypes}~\cite{lancia2001snps}.
Haplotypes are typically inferred computationally, since their experimental determination requires specialized protocols that are costly and challenging to implement~\cite{browning2011haplotype}.
However, this inference problem is inherently combinatorial: there exist $2^{n-1}$ possible haplotype explanations for a single diploid genotype with $n$ heterozygous variants, and the number of  explanations grows even faster when the number of chromosome copies per cell exceeds $2$ (i.e., for \textit{polyploid} genomes)~\cite{aguiar2013haplotype}.

\begin{wrapfigure}[20]{r}{0.5\textwidth}
\vspace{-30pt}
\begin{center}
    \includegraphics[width=0.49\textwidth]{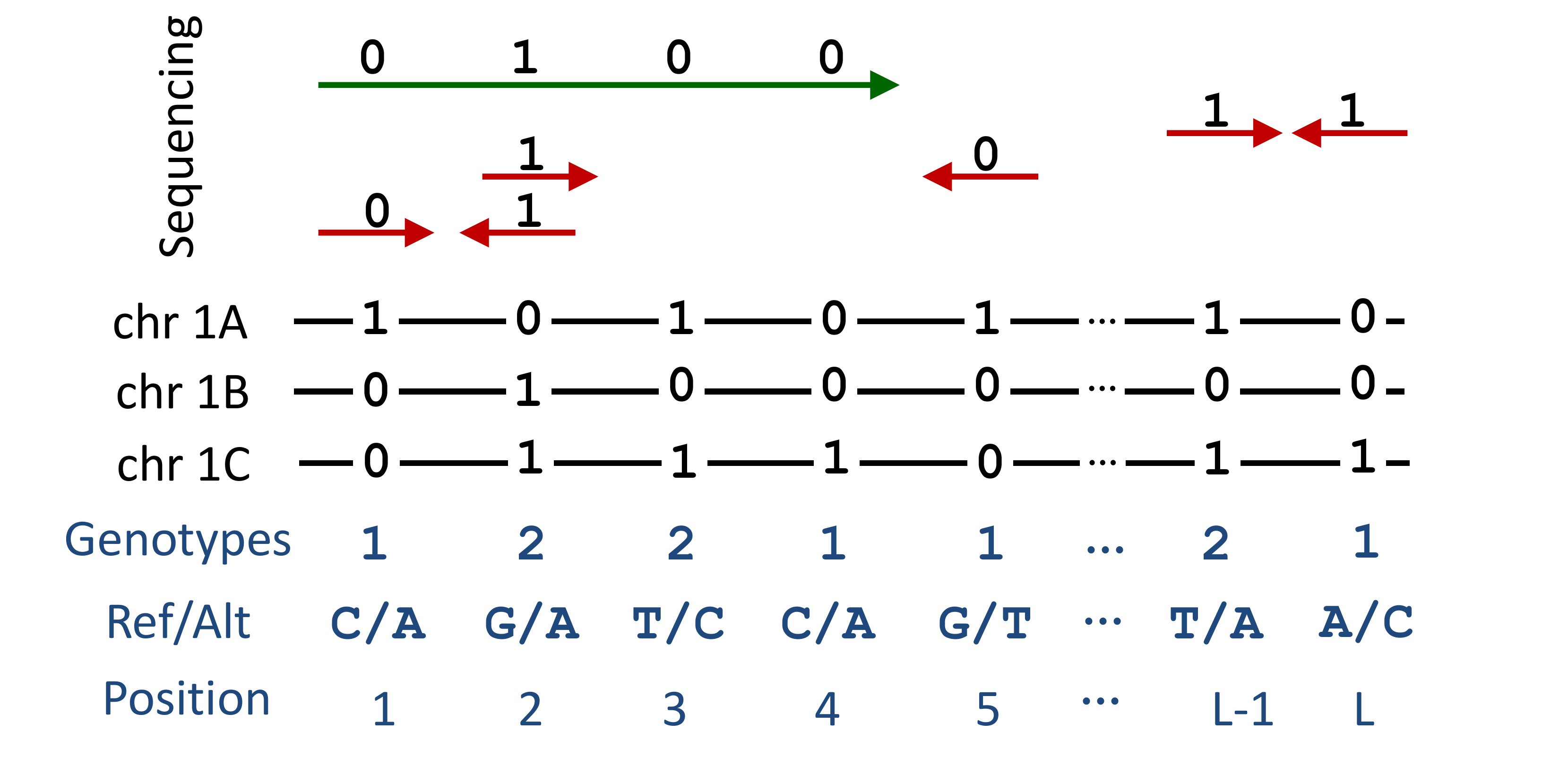}
\end{center}
\vspace{-10pt}
    \caption{\textbf{Haplotype assembly input for a triploid genome.}     Long (green) or paired-end short (red) reads that cover two or more heterozygous SNPs (encoded as 0 or 1 for major or minor alleles) are informative of haplotype phase.
    A collection of overlapping reads (e.g., positions 1-4) can be assembled into a single phasing (haplotype block) but their haplotype phase relative to other blocks is undetermined (e.g., positions $L-1$ and $L$).  }
    \label{fig:input}
\end{wrapfigure}

Haplotype inference algorithms fall into two classes: \textit{haplotype phasing} and \textit{haplotype assembly}.
Haplotype phasing infers haplotypes from genotype sequences derived from genotyping arrays~\cite{dunning2007beadarray} or DNA sequencing~\cite{evrony2021applications}.
These methods require population-level data, relying on haplotypes shared identical-by-descent to enable accurate phasing of common variants, but consequently perform poorly for rare or \textit{de novo} mutations~\cite{browning2011haplotype,hofmeister2023accurate}.
In contrast, haplotype assembly reconstructs haplotypes for a single genome using aligned and overlapping sequence reads~\cite{schwartz2010theory}. 
Because haplotype assembly operates on individual samples rather than a population, it is well defined without a reference panel and has been extended to polyploid genomes~\cite{aguiar2013haplotype}.
%%% short description of how read length influence model design
Polyploidy has been associated with adaptive potential under environmental stress and with evolutionary transitions leading to increased biological complexity~\cite{van2017autoallo}. 
Resolving polyploid haplotypes is therefore critical for understanding these processes and for applications in breeding and evolutionary studies in economically important species such as wheat, strawberry, and potato.
% Resolving polyploid haplotypes enables trait mapping, phylogenetic analysis, and crop improvement in economically important species such as wheat, strawberry, and potato~\cite{van2017autoallo}.
The structure of sequencing data strongly shapes the computational formulation of haplotype assembly. 
Short reads span only a small number of SNPs, motivating models that aggregate local pairwise phasings across overlapping reads, while long reads are typically lower coverage and require methods that integrate sparse, long-range constraints (Fig.~\ref{fig:input}).

Polyploid genomes add complexity to the haplotype assembly task in two significant ways. 
First, haplotypes of heterozygous single nucleotide polymorphisms (SNPs) are no longer complementary (Fig.~\ref{fig:input}). 
In diploid genomes with a known genotype, the resolution of one haplotype inherently produces the resolution of the complementary haplotype, such that the combination of the haplotypes is consistent with the genotype. 
In $K$-ploidy genomes, $(K-1)$ haplotypes must be resolved before leveraging genotype consistency to infer the final haplotype. % maybe this complement part isn't necessary, but I think it's incorrect to assert that polyploid genomes aren't complementary, instead more than one haplotype resolution is required before using its complement.
Second, haplotype segments are not unique.
Polyploid genomes can contain long segments (even chromosome length) of genetic material that is duplicated from the same (autopolyploidy) or similar (allopolyploidy) species.
As a result, a single sequence read cannot be deterministically mapped to its chromosome of origin (e.g., the read in positions 0 and 1 in Fig.~\ref{fig:input}) and runs of homozygosity on any subset of haplotypes can result in a haplotype assembly fragmented into phased haplotype blocks (Fig.~\ref{fig:input}). 
In other words, uncertainty exists at multiple levels: both in terms of haplotype phase but also haplotype mapping uncertainty in the presence of identical haplotype segments.
Statistical models provide a principled framework for quantifying polyploid phase uncertainty, but their formulation is challenging due to the exponentially large, discrete space of possible haplotype configurations.
HapTree and Poly-Harsh are the only two existing statistical polyploid haplotype assemblers, but neither quantify haplotype phase uncertainty and considered only short-read sequencing~\cite{berger2014haptree,he2018efficient}.

In this work, we develop \phc{}-short and \phc{}-long, two complementary statistical models for polyploid haplotype assembly tailored to short-read and long-read sequencing data, respectively. 
\phc{}-short targets cost-effective, high-coverage short-read technologies commonly used in population studies.
To scale with the large number of reads, \phc{}-short adopts a SNP-centric formulation, constructing a Markov random field whose vertices represent discrete distributions over haplotype phasings for SNP pairs, with edges connecting phasings that share a SNP.
Inference then leverages the pCompass graph, an exponentially sized factor graph, to adaptively select subsets of SNP pairs for incorporation into the current phasing.
\phc{}-long is designed for long-read sequencing data, which are lower throughput and higher cost than short reads, but provide long-range phasing information that enables more contiguous haplotype assemblies. 
To scale with the large number of SNP–SNP relationships induced by long reads, \phc{}-long adopts a read-centric formulation.
\phc{}-long defines a chain graph (hybrid graphical model combining directed and undirected dependencies) that captures both the causal (directed) and mutual (undirected) dependencies. 
Explicit random variables encode both the haplotype sequences and the mapping of reads to haplotypes.
Both algorithms allow for computing a single haplotype phasing using the Viterbi algorithm or quantifying phasing uncertainty using forward-filtering backward-sampling. 
Further, we develop the first pipeline to simulate realistic autopolyploid or allopolyploid genomes and benchmark \phc{} along with several leading polyploid haplotype assemblers with varied ploidy, coverage, and mutation rates on both short- and long-read data.
Lastly, we produce the first assembly of an allo-octoploid Strawberry chromosome.
Source code, simulation scripts, and data are freely available at \href{https://github.com/bayesomicslab/pHapCompass}{https://github.com/bayesomicslab/pHapCompass}.

\section{Related Work}
\label{sec:relatedwork}

Most haplotype assembly methods build a deterministic representation (typically a graph) whose complexity either scales with the number of sequence reads or genetic variants~\cite{schwartz2010theory}. 
The first diploid human genome assembly was constructed using an algorithm that considers the assignment of reads to haplotypes by alternating two steps: (1) assign sequence reads to haplotypes and (2) compute haplotypes based on a majority vote of its assigned reads~\cite{levy2007}.
Although the Levy \textit{et al.} algorithm is greedy, it was accurate and scaled well for the high quality and low throughput Sanger-based sequencing that was common in early genome assemblies.

The widespread adoption of short-read, high-throughput sequencing resulted in a transition from read-based to variant-based methods.
HapCut defines a graph where vertices are heterozygous SNPs, edges connect two SNPs connected by at least one sequence fragment, and edge weights are proportional to the number of fragments that are inconsistent with a current phasing~\cite{bansal2008hapcut}.
An iterative algorithm improves the haplotype assembly through maximum graph cuts; the successor algorithm HapCut2 extends HapCut by introducing a more sophisticated error model and support for different sequencing technologies~\cite{edge2017hapcut2}. 
In contrast, HapCompass optimizes over the cycles of a graph where sequence reads (edges) provide evidence for the two phasings that can exist between two heterozygous SNP positions (vertices)~\cite{Aguiar2012a}.

Extending haplotype assembly to polyploid genomes introduces additional uncertainty due to sequence similarity among haplotypes and the lack of haplotype complementarity.
The first polyploid haplotype assembly algorithm leveraged the HapCompass model, but the loss of the complement haplotype and haplotype uniqueness constraints made their cycle optimization approach scale poorly with the ploidy~\cite{aguiar2013haplotype}.
The first statistical model for polyploid haplotype assembly, HapTree, searches for a maximum-likelihood solution by iteratively extending polyploid assemblies one SNP at a time~\cite{berger2014haptree}; its probabilistic priors were later extended in HapTree-X~\cite{Berger2020}.
Another statistical assembler, Poly-Harsh, alternately samples haplotypes and read assignments~\cite{he2018efficient}, though neither method quantifies phasing uncertainty.

Motivated by increasing PacBio and nanopore read lengths, recent \textit{long-read assembly} methods have returned to read-centric formulations, modeling reads directly as graph nodes and improving performance in long-read settings.
Sharing some similarities with HapCompass, Ranbow resolves conflicting cycles and paths in a multipartite graph~\cite{moeinzadeh2020ranbow}.
Other methods such as SDhap~\cite{das2015sdhap}, H-PoP and H-PoPG~\cite{xie2016h}, nPhase~\cite{abou2021nphase}, and WhatsHap~\cite{schrinner2020haplotype} formalize polyploid assembly using generalized graph cuts or sequence read partitioning.
Most recently, deep learning-based approaches are able to produce single haplotype block assemblies by leveraging correlations between non-overlapping reads, although the interpretation of this phenomenon is unclear~\cite{consul2023xhap,10.1093/bib/bbae656}.
In contrast to prior work, \phc{} provides a unified framework for diploid and polyploid haplotype assembly that explicitly quantifies phasing uncertainty, with two complementary statistical models designed to scale to high-coverage short reads (\phc{}-short) and low-coverage long reads (\phc{}-long).

\section{Methods}
\label{sec: Problem Formulation}

% \subsection{Problem Formulation}
\label{sec:problemformulation}
The haplotype assembly problem aims to reconstruct the haplotypes $H = \{h_1, \ldots, h_K\}$ of an organism given a set of aligned sequencing reads $\mathcal{R} = \{r_j\}_{j=1}^{\lvert \mathcal{R} \rvert}$ spanning single nucleotide polymorphism (SNP) positions $\{1,\ldots,L\}$ indexed by $\ell$, a ploidy $K$, and genotype information $\mathcal{G} = \{g_1, \ldots, g_L\}$ where $g_{\ell}$ denotes the count of alternate alleles at position $\ell$ ($g_{\ell} \in \{1, \ldots, K-1\}$).
Sequence reads are informative of haplotype phase only if they cover at least two heterozygous SNP positions (Fig.~\ref{fig:input}).
For simplicity, we assume all SNPs are biallelic, encoded as $0$ (reference) and $1$ (alternate), though the proposed framework extends naturally to multiallelic variants.
The reconstructed haplotypes are defined as $H^* = \{h^*_1, \ldots, h^*_K\}$ where $h^*_k \in \{0, 1\}^L$. 
% By convention: 
% $$
% h^*_k(\ell) =
% \begin{cases} 
% 0 & \text{if the allele is reference,} \\
% 1 & \text{if the allele is alternate.}
% \end{cases}
% $$
The haplotypes must satisfy the genotype constraint: $\sum_{k=1}^{K} h^*_k(\ell) = g_{\ell}, \quad \forall \ell \in \{1, \ldots, L\}$.

\subsection{\phc{}-short}
\label{sec:method}

% \paragraph{\textbf{Graph formalization and introducing the pCompass graph.}}

The pCompass graph is the foundational data structure for the \phc{}-short algorithm and is constructed from two related graphs. 
Let $\{\ell_0,\ell_1, \ldots \}$ denote arbitrary SNP indices. \begin{wrapfigure}[22]{R}{0.4\textwidth}
    \vspace{-1cm}
    \centering
    \includegraphics[width=0.39\textwidth]{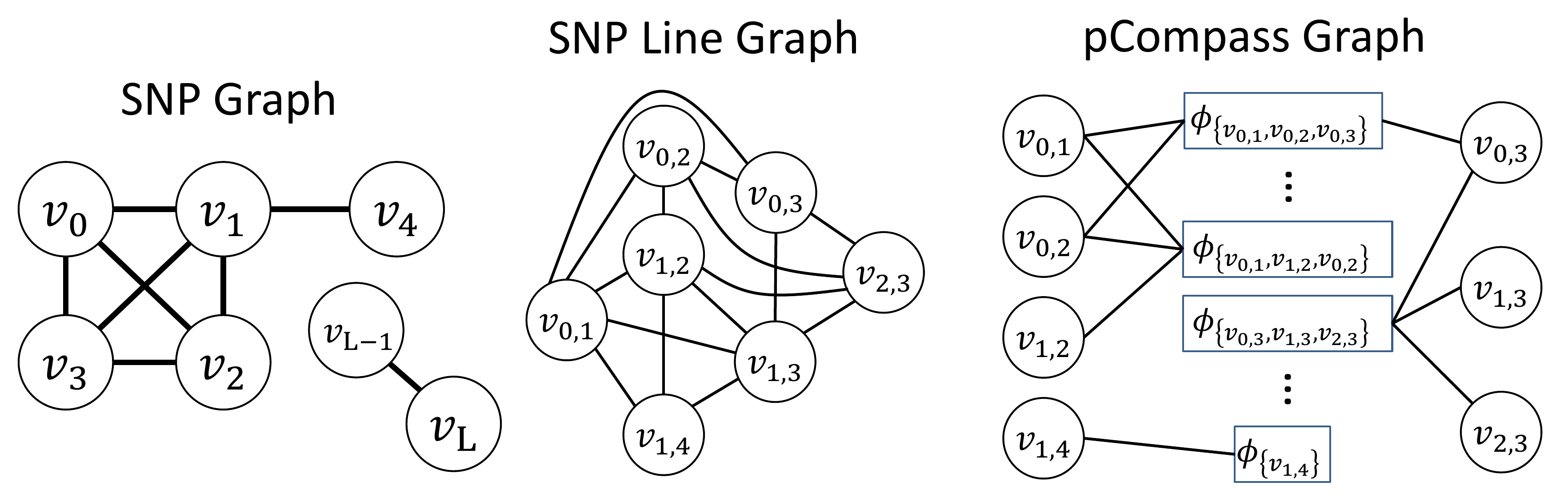}
    \caption{\textbf{Graph constructions for pHapCompass-short inference.}  
    \textit{Left}: The SNP graph $G = (V, E)$ where vertices $v_0, \ldots, v_L$ represent heterozygous SNP positions and edges connect positions covered by at least one sequencing read. \textit{Middle}: The SNP line graph $Q = (U, E_Q)$, where each node $v_{i,j}$ corresponds to an edge $(i, j)$ in $G$, and two nodes are adjacent if their corresponding edges share exactly one SNP position. \textit{Right}: The pCompass graph, a factor graph over $Q$ where node potentials $\phi_{v_{i,j}}$ encode the likelihood of phasings for SNP pairs given read evidence, and clique potentials (e.g., $\phi_{v_{0,1}, v_{0,2}, v_{1,2}}$) capture phasing evidence from reads spanning three or more positions.}
    \label{fig:model}
\end{wrapfigure}

\paragraph{SNP Graph.} 
The SNP graph $G = (V, E)$ represents the read overlaps where $V = \{1, \ldots, L\}$ and an edge $(\ell_0, \ell_1) \in E$ if at least one read covers both (not necessarily adjacent) SNP positions $\ell_0$ and $\ell_1$ (Fig.~\ref{fig:model}, Left). 
A read $r$ ``covers'' SNP $\ell$ if the read overlaps with the position of SNP $\ell$ and the called base is $0$ or $1$; an uncovered SNP is denoted by $\text{`}-\text{'}$~\cite{schwartz2010theory}.

% \vspace{-5pt}

\paragraph{SNP Line Graph.}
The SNP line graph $Q = (U, E_Q)$ is the line graph of $G$. 
Formally, each node $u_{\ell_0\ell_1} \in U$ corresponds to an edge $(\ell_0, \ell_1) \in E$ of the SNP graph and two nodes in $Q$ are adjacent if they share exactly one SNP position (Fig.~\ref{fig:model}, Middle). Formally,
\begin{align*}
    E_Q = \left\{\left((\ell_0, \ell_1), (\ell_2, \ell_3)\right) : \lvert \{\ell_0, \ell_1\} \cap \{\ell_2, \ell_3\} \rvert = 1 \right\} .
\end{align*}

\paragraph{pCompass Graph.} 
The pCompass graph is a factor graph constructed from $Q$ that assigns potentials to all (not necessarily maximal) cliques
%edges 
based on observed read evidence (Fig.~\ref{fig:model}, Right). 
These potentials encode the compatibility between different phasings and the sequencing data.
The number of factors in the pCompass graph is exponentially sized with respect to $\lvert U \rvert = O(L^2)$, thus we only enumerate the 1- and 2-cliques of the graph.
For each node $u_{\ell_0\ell_1} \in U$, we define a \textit{node potential} to be a function mapping elements in the space of valid phasings $\Phi_{\ell_0\ell_1} \subset \{0,1\}^{K \times 2}$ to $\mathbb{R}_+$, where each phasing $\phi_{\ell_0 \ell_1}$ satisfies the genotype constraints:
$$
\sum_{k=1}^{K} \phi_{\ell_0\ell_1}^{(k)}(\ell_0) = g_{\ell_0} \quad \text{and} \quad \sum_{k=1}^{K} \phi_{\ell_0\ell_1}^{(k)}(\ell_1) = g_{\ell_1} ,
$$
where $\phi_{\ell_0\ell_1}^{(k)} \in \{0,1\}^2$ is the phasing for the pair of positions for haplotype $k$ and $ \phi_{\ell_0\ell_1}^{(k)}(\ell_0)$ is the allele at position $\ell_0$.
% By parameterizing the space of valid phasings by the count of haplotypes containing $(1,1)$ allele pattern, we can enumerate all $|\Phi_{\ell_0\ell_1}| = O(K)$ distinct phasings efficiently (see Section~\ref{sec:suppEnumerating} for details).
By parameterizing the space of valid phasings by $t_{11} = |\{k : \phi_{\ell_0\ell_1}^{(k)} = (1,1)\}|$, the count of haplotypes in the phasing with allele pattern $(1,1)$, we can enumerate all $|\Phi_{\ell_0\ell_1}| = O(K)$ distinct phasings efficiently (see ~\autoref{sec:suppEnumerating}).
For each phasing $\phi_{\ell_0\ell_1} \in \Phi_{\ell_0\ell_1}$, we compute its node potential as:

\begin{equation*}
f(\phi_{\ell_0\ell_1})
\;=\;
\sum_{r \in \mathcal{R}_{\ell_0\ell_1}}
\mathcal{L}\!\left(r,\phi_{\ell_0\ell_1}\right),
\label{eq:f_phi}
\end{equation*}
\begin{equation*}
\mathcal{L}\!\left(r,\phi_{\ell_0\ell_1}\right)
\;=\;
\sum_{k=1}^{K}
\varepsilon^{\,d\!\left(\phi_{\ell_0\ell_1}^{(k)},\, r[\ell_0,\ell_1]\right)}
(1-\varepsilon)^{\,2-d\!\left(\phi_{\ell_0\ell_1}^{(k)},\, r[\ell_0,\ell_1]\right)},
\label{eq:L_r_phi}
\end{equation*}
where $\mathcal{R}_{\ell_0\ell_1} = \{r \in \mathcal{R} : r[\ell_0] \neq \text{`}-\text{'} \land r[\ell_1] \neq \text{`}-\text{'}\}$ is the set of reads covering both positions, 
$r[\ell_0, \ell_1]$ denotes the observed alleles in read $r$ at these positions, $d(\cdot, \cdot)$ is the Hamming distance, and $\varepsilon$ is the sequencing error rate.
For each edge $e \in E_Q$ whose endpoints together cover three SNPs $\ell_0, \ell_1, \ell_2$, we define the \textit{edge potential} over phasings of the three SNPs $\phi_{\ell_0\ell_1\ell_2} \in \Phi_{\ell_0\ell_1\ell_2} \subset \{0,1\}^{K \times 3}$:
\begin{align*}
    f(\phi_{\ell_0\ell_1\ell_2}) = \sum_{r \in \mathcal{R}_{\ell_0\ell_1\ell_2}} \mathcal{L}(r, \phi_{\ell_0\ell_1\ell_2}),
\end{align*}
where $\mathcal{R}_{\ell_0\ell_1\ell_2} = \{r \in \mathcal{R} : r[\ell_0] \neq \text{`}-\text{'} \land r[\ell_1] \neq \text{`}-\text{'} \land r[\ell_2] \neq \text{`}-\text{'}\}$ is the set of reads covering all three positions, and
\begin{align*}
    \mathcal{L}(r, \phi_{\ell_0\ell_1\ell_2}) = \sum_{k=1}^{K} \varepsilon^{d(\phi_{\ell_0\ell_1\ell_2}^{(k)}, r[\ell_0,\ell_1,\ell_2])}(1-\varepsilon)^{3-d(\phi_{\ell_0\ell_1\ell_2}^{(k)}, r[\ell_0,\ell_1,\ell_2])}.
\end{align*}
For  any edge in $E_Q$ representing three SNPs $\ell_0, \ell_1, \ell_2$ for whom $\mathcal{R}_{\ell_0\ell_1\ell_2} = \varnothing$, we define $f(\phi_{\ell_0 \ell_1 \ell_2}) \propto 1 ~ \forall \phi_{\ell_0 \ell_1 \ell_2} \in \Phi_{\ell_0 \ell_1 \ell_2}$. Informally, node and edge potentials represent unnormalized likelihoods of a read being generated conditioned on a particular haplotype phasing (see~\autoref{sec:suppNotation} for a summary of notation). 

\paragraph{Inference Algorithm Leveraging the pCompass Graph.}
\label{sec:inference}

The pCompass graph represents a probability distribution over haplotype assemblies. 
We perform inference using the Viterbi algorithm for maximum a posteriori (MAP) estimation or the forward-filtering backward-sampling (FFBS) for posterior sampling and uncertainty quantification.
%\paragraph{Topological Ordering.}
To make probabilistic inference on $Q$ tractable, we impose a topological ordering on the nodes of $U$ according to ascending genomic position, first by the lesser-indexed of the two SNPs represented by a node, breaking ties by considering the second SNP. 
We then direct the edges of $E_Q$ to respect this ordering.
% We process nodes such that if node $u_{\ell_0\ell_1}$ precedes $u_{\ell_2\ell_3}$, then $\max(\ell_0, \ell_1) \leq \min(\ell_2, \ell_3)$. 
When the SNP graph contains multiple connected components, each component is ordered independently (see ~\autoref{sec:suppViterbi}
and ~\autoref{sec:suppFFBS} for details).

\paragraph{Constructing the Full Assembly.}

\label{sec:matching}
After MAP estimation or FFBS, we obtain a local phasing $\phi_t \in \Phi_t$ for each node $u_t \in U$. 
However, these local phasings cannot be directly concatenated into a global haplotype solution $H^*$ due to \textit{haplotype label ambiguity}: 
since haplotypes are biologically unordered, any permutation of haplotype indices represents a valid solution. 
When two adjacent nodes $u_{\ell_0\ell_1}$ and $u_{\ell_1\ell_2}$ share position $\ell_1$, their phasings must be consistent at the shared position, but %achieving this consistency requires determining the correct permutation alignment between their respective haplotype labelings. 
this consistency can be achieved through multiple different permutations of haplotype labels. % I think I should bring an example.  
We resolve this combinatorial matching problem through a greedy connectivity-based assembly approach that incrementally selects the next variant to be phased and constructs the global haplotype configuration $H^*$ variant-by-variant (see \autoref{sec:matching_app}). 

The \textit{variant selection algorithm} maintains four sets: $U^{\text{phased}}$ and $U^{\text{unphased}}$ tracking which nodes in the SNP line graph $Q$ have been processed, and $S^{\text{phased}}$ and $S^{\text{unphased}}$ tracking which SNP positions have been assigned to global haplotype $H^*$. 
The algorithm starts by selecting the first node in $Q$ according to the topological order and assigning both of its positions using the node's inferred phasing.  % talk about the init in one sentence 
At each iteration, we identify unphased nodes in $Q$ that are adjacent to phased nodes $U^{\text{phased}}$, then we select the unphased position $\ell^*$ appearing in the most such nodes. 
Selecting positions with high connectivity ensures that the phasing decision is well-constrained by multiple node pairs in $Q$, progressively extending the phased region along paths of strong graph connectivity (see \S\ref{sec:alg1}, Fig.~\ref{fig:algorithmsvis}, and Algorithm \ref{alg:varselect}).

The \textit{phasing position algorithm} determines the haplotype assignment of a selected position $\ell^*$ through candidate generation and scoring. 
Candidates are constructed from nodes containing $\ell^*$ that are connected to phased positions, using node and edge potentials computed during graph construction. %  
% Each candidate must % satisfy genotype constraints and 
% maintain label consistency in already phased position under some permutation of rows. 
Each candidate must maintain consistency with the already-assigned alleles in $H^*[S^{\text{phased}}]$, i.e. the haplotype labels in $S^{\text{phased}}$ are fixed, but the candidates may permute haplotype rows to achieve consistency at shared positions.
Candidates are scored using a weighted combination of three criteria: likelihood (how well the candidate explains observed reads), minimum error correction (the number of sequence read allele corrections required to map reads to the phasing), and inference agreement (consistency with Viterbi or FFBS results). 
After normalizing each metric within the candidate set, we select the highest-scoring candidate and assign its alleles for position $\ell^*$ to the global haplotype matrix $H^*$. 
When no valid candidates exist due to sparse coverage, we use the inferred phasing from any node containing $\ell^*$ (see \S\ref{sec:alg2} and Algorithm~\ref{alg:phase_position}).

\subsection{\phc{}-long}
\label{sec:pCompass_mixture_model}

Since the number of vertices in the SNP line graph grows quadratically with the length of a read, \phc{}-short scales poorly with long sequence reads.
Here, we introduce \phc{}-long, a partially directed hierarchical haplotype mixture model whose complexity scales linearly with the number of reads. 

%For long read data, notably when considering distributions over all pairs of co-covered SNPs may be computationally expensive due to the number of vertices in the SNP line graph associated with a single read growing quadratically with the length of the read, we introduce the hierarchical haplotype mixture model \phc{}-long. 

\begin{wrapfigure}[16]{R}{0.31\textwidth}
\vspace{-30pt}
\begin{center}
    \includegraphics[width=0.31\textwidth]{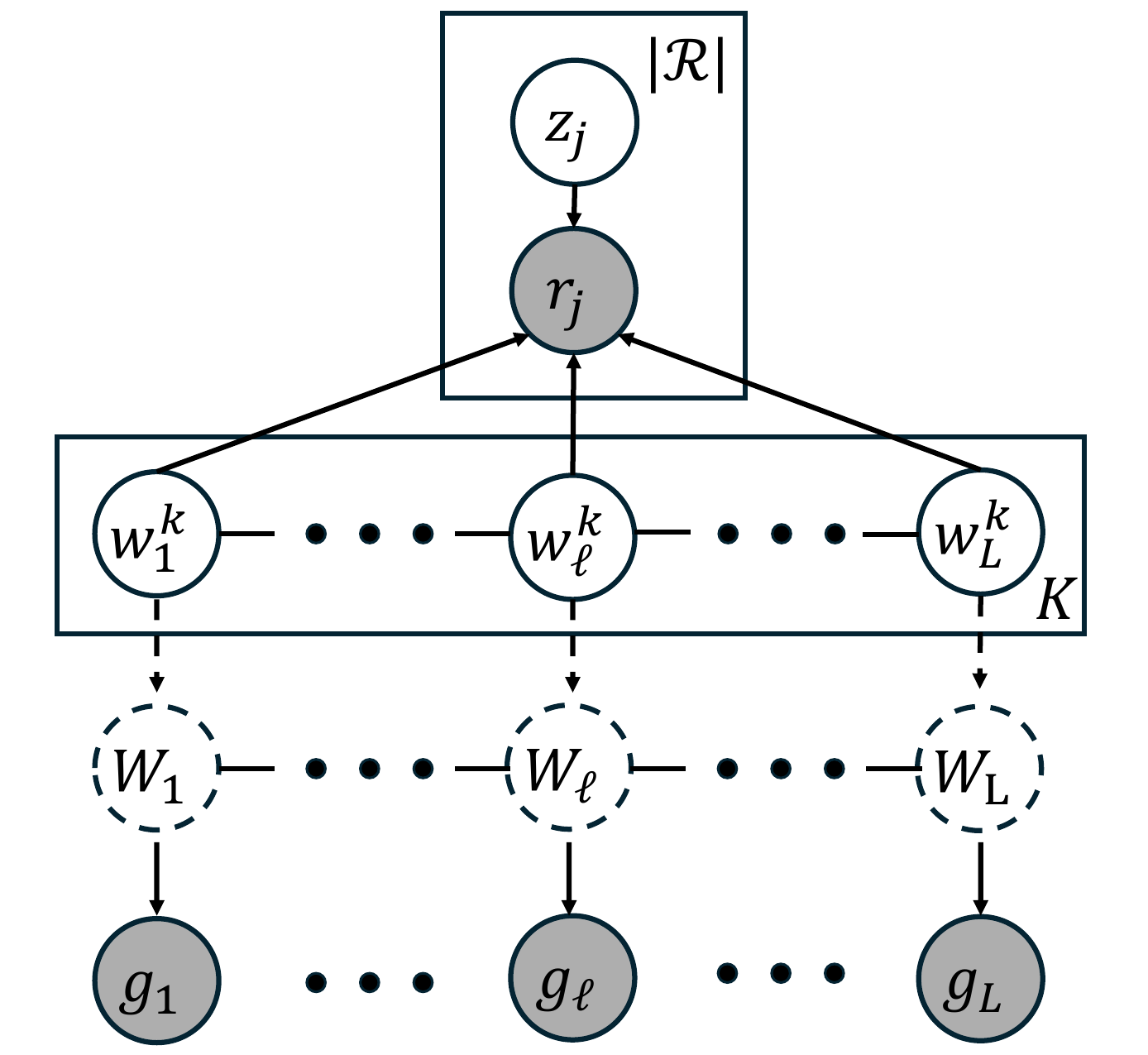}
\end{center}
\vspace{-15pt}
    \caption{\textbf{\phc{}-long graphical model.} 
     The dashed undirected edges from each $w_\ell^k$ to $W_\ell$ and around each $W_\ell$ represent a deterministic dependence between $w_\ell^k$ and $W_\ell$ (rather than statistical).  
    }
    \label{fig:mixture model graphical model}
\end{wrapfigure}

\paragraph{Defining the Mixture Model.} 
The \phc{}-long model defines how a read set $\mathcal{R}$ is generated from a mixture of $K$ conditional random fields (CRFs) $\{h^*_k\}_{k=1}^K$ each representing a single haplotype along the genome (Fig.~\ref{fig:mixture model graphical model}). % for the graphical model 
Each haplotype CRF $h^*_k$ is a chain over $L$ variables $w^k_1, \ldots, w^k_L$ that each emit a single allele in $\{0,1\}$ with 2 states per variable, one that emits a 0 allele with probability $1-\varepsilon$ and another that emits a 0 allele with probability $\varepsilon$, where $\varepsilon\in [0,1]$ is the sequencing error rate. 
The read assignments, $z_j$, and hidden states, $w^k_\ell$, are unobserved random variables. 
For convenience, define vector function $\epsilon(r[\ell]) = \left\langle \epsilon_0(r[\ell]), \epsilon_1(r[\ell]) \right\rangle$, where $\epsilon_a(r[\ell]) = 1 - \varepsilon$ if $r[\ell] = a$ and $\varepsilon$ if $r[\ell] = a' \neq a$. 

For each haplotype $h^*_k$ and each allele $a \in \{0,1\}$, we compute the marginal potentials as
\vspace{-2mm}
\begin{align*}
    \psi_\ell^k(a) \propto \prod_{\substack{r_j \in \mathcal{R}: \\ z_j = k ~ \land ~ r_j[\ell] \neq \text{`}-\text{'}}} \epsilon_a(r_j[\ell]) ,
\end{align*}
and the transitions as
\begin{align*}
    & \psi_{\ell, \ell+1}^k(a_1, a_2) \propto  \\ &
    \frac{\lvert \{r_j \in \mathcal{R} : z_j = k \land r_j[\ell] = a_1 \land r_j[\ell+1] = a_2 \} \rvert}{\lvert \{r_j \in \mathcal{R} : z_j = k \land r_j[\ell] \neq \text{`}-\text{'} \land r_j[\ell+1] \neq \text{`}-\text{'} \} \rvert} + \delta ,
\end{align*}
where $\delta \geq 0$ is a hyperparameter used to smooth transition potentials.

To enforce genotype constraints, we define another chain-shaped conditional random field $H^*$ over variables $W_1, \ldots, W_L$, where each $W_\ell$ takes states over the $2^K$ configurations of $w^1_\ell, \ldots, w^K_\ell$.
The marginal and transition potentials for each $W_\ell$ and each pair $(W_\ell,~W_{\ell+1})$ along $H^*$ are defined by first taking the $K$-ary products over potentials from $h^*_1, \ldots, h^*_K$, and then only assigning non-zero mass in marginals to the states of $W_\ell$ that agree with the genotype as called in $g_\ell$.
%We consider the called genotype at each SNP as an observed variable dependent on $W_\ell$, materialized as the sum of alternate alleles in the state that variable $W_\ell$ is in.

For a (not necessarily contiguous) read $r_j$ that is covering a subset of SNPs within $\ell_0, \ldots, \ell_1$, we can define the conditional posterior probability of $r_j$ given $h^*_k$, or colloquially the ``likelihood'' of $r_j$ being emitted from $h^*_k$, as 
\begin{equation}
\begin{aligned}
P(r_j | h^*_k) = ~ &\alpha_k(\ell_0) \cdot \left(\psi_{\ell_0}^k \cdot \epsilon\left(r_j[\ell_0]\right)\right) \cdot \\
&\left(\prod_{\ell = \ell_0 + 1}^{\ell_1} \psi_{\ell - 1, \ell}^k \cdot  \left(\psi_{\ell}^k \cdot \epsilon\left(r_j[\ell]\right)\right) \right) \cdot \beta_k(\ell_1),
\label{eq_read_likelihood}
\end{aligned}
\end{equation}
where
\begin{itemize}
    \item $\alpha_k(\ell_0)$ is the forward message into SNP $\ell_0$,
    \item $\beta_k(\ell_1)$ is the backward message into SNP $\ell_1$, 
    \item $\psi_\ell^k$ is the normalized marginal distribution over alleles for SNP $\ell$,  
    \item $\psi_{\ell, \ell + 1}^k$ is the transition probability between adjacent SNP variables $\ell$ and $\ell + 1$, and
    \item $\epsilon\left(r_j[\ell]\right)$ is the emission probability for the data along the read
\end{itemize}
all along haplotype $h^*_k$. 
To accommodate missing alleles in the read due to low base quality, we extend the sequencing error function as $\epsilon_a(r[\ell]) = 0.5$ if $r[\ell] = \text{`}-\text{'}$.
Intuitively, we first compute the probability of emitting the observed alleles from the read over all possible paths through the hidden states given the potentials, approximating missing data within the read as uniform over the two alleles. 
Then, we marginalize over unobserved alleles before the beginning and after the end of the read using the forward and backward messages, respectively.

In the generative process, read assignments to haplotypes $z_j$ are drawn uniformly from $\{1,\ldots,K\}$, reads over a fixed set of SNPs are emitted from haplotypes $h^*_{z_j}$ as in Equation~\ref{eq_read_likelihood}, and genotypes $g_\ell$ are emitted as the sum of alternate alleles represented in the phase of the state of variable $W_\ell$. 
Because adjacent SNPs have correlative relationships but are not causal in one direction or another, the edges between adjacent $w_\ell^k$ and adjacent $W_\ell$ are undirected. 
Notice that since $h_k^*$ and $H^*$ are parts of the undirected portion of the model, their nodes have conditional distributions as described above, but are not generated.

\paragraph{Gibbs Sampling Algorithm.}
To train \phc{}-long, we develop an iterative algorithm inspired by the Levy~\textit{et al.} algorithm and Gibbs sampling~\cite{levy2007}. 
We draw each $P(z_j = k | \cdot) \propto \frac{1}{K}\cdot P(r_j | h^*_k)$ and then update potentials for the two affected haplotype CRFs if the read assignment $z_j$ has changed. 
To extract a complete haplotype assembly, we take a Viterbi path through $H^*$ starting at the first SNP implicitly directing edges to the last SNP and concatenating the phasings per SNP as determined by the states for each $W_\ell$ along the path. 
During model training, we share information between separate $h^*_k$ by augmenting the first 20 iterations of the Gibbs sampler with a gradient descent update on the potentials; that is, we sample a single path using FFBS through the $H^*$ and take a linear combination of the existing potential distributions of $h^*_k$ with the Dirac distribution of marginals and transitions defined by the single FFBS sample. 
We initialize each $z_j$ randomly, and run the iterative algorithm until the log likelihood of the model, defined as $\log \left( \prod_{r_j \in \mathcal{R}}   P(r_j | h^*_{z_j}) \right)$, converges. 
We remark that in \phc{}-long, the likelihood of the genotype is omitted from the model likelihood because we strictly enforce its discrete consistency by nullifying atoms in $W_\ell$ inconsistent with $g_\ell$.
Uncertainty quantification is achieved by sampling multiple paths through the CRF $H^*$ using FFBS.

\section{Results}
\label{sec:results}

We evaluated \phc{}, H-PoPG, WhatsHap, and HapTree-X on synthetic and experimental data with respect to haplotype assembly quality and uncertainy quantification.

\subsection{Simulated Data}
\label{main:simdata}
We generated synthetic polyploid genomes from \textit{S. tuberosum} reference haplotypes~\cite{Sun2021tuberosum} using HaploSim~\cite{Motazedi2018Haplosim}. 
For autopolyploids, we simulated ploidies in $\{2,3,4,6\}$ with mutation rates $\mu \in \{0.001, 0.005, 0.01\}$ and 20 replicates per configuration. 
For allopolyploids, we generated structured subgenomes (AAB, AABB, AABBCC) with inter-subgenome divergence $\mu_{sub} \in \{0.0001, 0.0005\}$ and within-subgenome variation $\mu_{within} \in \{0.00005, 0.0001\}$ (10 replicates per configuration).
Paired-end 125bp short reads were simulated using ART with Illumina HiSeq 2500 profiles at coverages in $[3\times,40\times]$ per haplotype~\cite{huang2012art}. 
Long reads were simulated using PBSIM3 with ONT R10.4 chemistry profiles (mean length $\sim 9kb$, Q13-14 base quality) with coverages in $[5\times,40\times]$ per haplotype~\cite{ono2022pbsim3}. 
Reads were aligned using BWA-MEM (short) or Minimap2 (long), and fragments were extracted using Hap10 with base quality thresholds of Q13 and Q4 for short and long reads, respectively~\cite{majidian2020hap10}. 
In total, we generated 1,200 autopolyploid short-read samples, 960 autopolyploid long-read samples, and 480 samples each for allopolyploid short and long reads (see \autoref{sec:suppsim}, Table~\ref{tab:simdatasetconf}, and Figs.~\ref{fig:auto-snp-density}-\ref{fig:allo-longread} for details).

\subsection{Experimental Data}
Cultivated octoploid strawberry (\textit{Fragaria x ananassa}) is an allopolyploid consisting of four subgenomes (A,B,C,D) originating from four diploid progenitor species. 
To benchmark allopolyploid haplotype assembly, raw Fragaria x ananassa reads were obtained from the European Nucleotide Archive (Biosample: SAMN13059213) \cite{han2025chromosome}. 
Raw reads were subsequently adapter- and quality-trimmed using fastp v0.23.4 \cite{chen2018fastp}. Duplicates were marked using picard v3.4.0 and the GATK MarkDuplicates v4.6.2.0 tool \cite{mckenna2010}. 
Subgenome A of the telomere-to-telomere Fragaria x ananassa 'Seolhyang' assembly \cite{han2025chromosome} was extracted from the reference using samtools v1.20 \cite{danecek2021twelve}. 
Trimmed reads were then aligned to subgenome A using BWA-MEM v0.7.19~\cite{li2009}. % to evaluate allopolyploid subgenome phasing with pHapCompass. 
% Alignments were sorted using samtools, and biallelic variants were called using the freebayes v1.3.10 with ploidy (-p) set to eight \cite{garrison2012haplotype}. 
% Alignments were sorted using samtools, and biallelic variants were called using freebayes v1.3.10 with ploidy (-p) set to eight \cite{garrison2012haplotype}.
Alignments were sorted using samtools, and biallelic variants were called using freebayes v1.3.10 with ploidy (-p) set to eight \cite{garrison2012haplotype}. 
%; in general, other polyploid-aware variant callers such as GATK \cite{mckenna2010} or Octopus \cite{cooke2021unified} can also be used for genotype calling.
Variants were filtered for quality (minimum depth 10, mapping quality $\geq$20). 
The biallelic assumption maintains comparability with other assemblers and compatibility with upstream and downstream tools (e.g., GWASpoly~\cite{https://doi.org/10.3835/plantgenome2015.08.0073}, updog~\cite{gerard2018genotyping}, or rrBLUP~\cite{endelman2011ridge}). 
In our experimental data, multiallelic SNPs represent only 2.2\% of raw variants (154,259 out of 7,020,992 total variants) and are likely overrepresented due to homeologous alignments (Table~\ref{tab:variant_filtering}).
To reduce the effects of linkage disequilibrium, we thinned our VCF along static windows of 200bp based on the estimated rate of linkage disequilibrium decay (120bp) for cosmopolitan F. x ananassa \cite{hardigan2021unraveling}. We increased our window size to account for variability in linkage disequilibrium decay rates both along each chromosome and among homeologs, while maintaining sensitivity to attempt to detect local haplotype blocks (see \S\ref{res:expprocess} for additional data processing details).

\subsection{Evaluation Criteria}

\newcommand{\setK}{\{1, \ldots, K\}}
\newcommand{\VE}{\operatorname{VE}}
\newcommand{\VER}{\operatorname{VER}}
\newcommand{\MEC}{\operatorname{MEC}}

To enable fair comparison across assemblers that produce discontiguous assemblies and varying levels of phase uncertainty, we generalize the standard evaluation metrics of minimum error correction and vector error rate.
Specifically, we extend the metrics used in prior works to consider the following scenarios observed in the output of polyploid assemblers:
\begin{enumerate}
    \item the reconstructed phasing of a heterozygous SNP deviates from the ground truth genotype;
    \item less than $K$ haplotypes are assembled at a given SNP; and % because polyploid genomes do not have complementary haplotypes (that is, one haplotype's content does not ``imply'' the other), and 
    \item the assembled haplotypes are not contiguous, but rather reported as a series of one or more \textit{blocks} where  assemblers produce a different number of phased blocks.
\end{enumerate}

\paragraph{Generalized Vector Error Rate.}
Given ground truth haplotypes, the vector error rate ($\VER$) is the polyploid generalization of switch error for reconstructed phasings that obey genotype constraints.
First, we define vector error for a contiguous and complete phasing. 
Let $H = \{h_1, \ldots, h_K\}$ be the true phasing and $H^* = \{h^*_1, \ldots, h^*_K\}$ be the reconstructed phasing, where each $h_k$ and $h^*_k$ are $L$-length vectors representing one true or phased haplotype. 
Let $\phi_\ell : \setK \to \setK$ be a bijection from the predicted phasing at SNP $\ell$ to the true phasing at SNP $\ell$ such that $h_k[\ell] = h^*_{\phi_\ell(k)}[\ell]$; that is, $\phi_\ell$ maps 0 alleles to 0 alleles and 1 alleles to 1 alleles.
There are $g[\ell]!(K-g[\ell])! = O(K!)$ such permutations that exist when the genotype of the reconstruction matches the genotype of the ground truth; vector error is not well-defined for assemblies that do not respect the reported genotype. 
For the sake of comparison with assemblies that fail to align with genotype information, we perturb the computed assembly by introducing the minimal number of allele flips (chosen uniformly at random from the offending alleles) to make genotypes consistent and compute vector error rate on the modified assembly with a per flip penalty. 

Let $\Phi = (\phi_1, \ldots, \phi_L)$ be a sequence of valid matchings. 
We define the vector error associated with a matching of the entire reconstructed phasing as 
\begin{align}
    \VE(\Phi) = \sum_{\ell=2}^{L} \lvert \{k : \phi_{\ell}(k) \neq \phi_{\ell-1}(k)\} \rvert , \label{eqn:ve1}
\end{align}
or the number of times that the matching ``switches'' which predicted haplotype maps to which true haplotype over the length of the assembly. 
The overall vector error for an assembly is then
\begin{align}
    \VE(H, H^*) = \min_{\Phi}\{\VE(\Phi)\}.
    \label{eqn:ve2}
\end{align}
The vector error can be computed exactly using a dynamic program over transitions between valid matchings of adjacent SNPs in time $O(L(K!)^2)$.
The vector error rate is defined as $\VER(H, H^*) = \VE(H, H^*)/L$. 

% BLOCK TYPES

Special considerations must be taken to compute a single $\VER$ for an assembly with many blocks. 
A block is a sub-assembly over which the phasing between all SNPs in the block is computed, such that the union of SNPs over all blocks is the disjoint union of $\{1, \ldots, L\}$. 
Further, we define a block such that the number of haplotypes phased at each SNP within the block is constant. %, even if the phase between some of the haolotypes is specified in the model's output. 
There exist four types of blocks: (1) fully phased ($K$ haplotypes phased); (2) mostly phased ($K-1$ haplotypes phased), which can be augmented with the final haplotype needed to respect the called genotype and treated as fully phased; (3) partially phased (between 1 and $K-2$ haplotypes phased, inclusive); and (4) completely unphased.

We generalize vector error over multiple blocks by defining vector error for each block type and adding a penalty for each additional block reported beyond the first.
The vector error for blocks of type 1 or 2 can be computed as described in Equations~\ref{eqn:ve1} and \ref{eqn:ve2}.
The vector error for blocks of type 4 is computed as an expectation: we consider all of the valid bijections at each SNP position besides the first uniformly and compute the expected number of non-fixed points from an arbitrary but fixed permutation considered at the previous position, and sum over positions besides the first.  
The vector error for type 3 blocks is computed as a combination of the vector error (Equations~\ref{eqn:ve1} and \ref{eqn:ve2}) and expected vector error.
See $\S$\ref{sup:ec} for details on vector error computation for block types 3 and 4.
% PENALTY
Additionally, in order to report a single value, we penalize reconstructed phasings that phase into more than one block by computing the expected number of vectors errors ``into'' the first SNP of each additional block beyond the first using $K - \mathbb{I}\{g[\ell] > 1\} - \mathbb{I}\{K - g[\ell] > 1\}$. 
%We do not penalize the first block's first SNP because there is no previous permutation to conflict against. 
The vector error of an assembly is the sum of the within-block vector errors plus the additional block penalties.
Finally, we note that because all assemblers are provided with the same input, differences in the number of reported haplotype blocks reflect algorithmic design choices rather than differences in available data.
% MEC

\paragraph{Block-adjusted Minimum Error Correction.}
The minimum error correction ($\MEC$) criterion is often used when ground truth haplotypes are unknown.
Traditionally, $\MEC$ measures the proportion of called alleles over all reads that have to be flipped for each read to map perfectly onto the reconstructed phasing. 
Formally,
\begin{align*}
    \MEC(\mathcal{R}, H^*) = \frac{1}{\sum_{r \in \mathcal{R}} \lvert r \rvert} \sum_{r \in \mathcal{R}} \text{argmin}_{k \in \setK} \left\{d(r,h^*_k[r]) \right\} ,
\end{align*}
where $\lvert r \rvert$ is the number of alleles that read $r$ covers, $d(u,v)$ is the Hamming distance between vectors $u$ and $v$, and $h^*_k[r]$ is the $k^{\text{th}}$ reconstructed haplotype at the set of SNPs that read $r$ covers. 
The Hamming distance between an allele along a read and an unphased allele in the reconstruction is taken as 1, naturally giving the analog of $\MEC$ for partially phased blocks as long as the read is contained within one phased block.
% PENALTY
%For each read spanning more than one distinct block, we penalize the $\MEC$ for each block covered beyond the first. 
Because $\MEC$ considers the error correction over a \textit{single} row of the reconstruction, we penalize for each read that covers more than one block. % with the probability that the optimal haplotype chosen in each block is different for all covered blocks. 

Notice that a reconstruction over multiple blocks is less informative of a full phase than an unbroken phasing over the entire span of the read, necessitating a block-cutoff penalty for the former. To illustrate in the extreme, any read over heterozygous SNPs can be mapped back with 0 error correction on a within-block basis if every block is one SNP wide.
So, for a read covering $B$ blocks, we apply a penalty of $(B-1)\cdot(1 - \frac{1}{K})$, or the expected number of times that the optimal haplotype for each pair of adjacent blocks differs, assuming a single optimally matching haplotype for each block. 
Under this penalty, the interpretation of $\MEC$ as a rate must be relaxed.
A read covering $B$ blocks as seen in isolation will contribute $O(B)$ to the unnormalized block adjusted MEC, where in the single-block setting this is $O(1)$ as expected from the traditional definition of the metric. % NOTICE THIS IS THE NON GEOMETRIC MEC

\subsection{Synthetic Data Results}
We evaluated \phc{} against three state-of-the-art polyploid haplotype assembly methods, WhatsHap, H-PoPG, and HapTree-X, using simulated data across multiple ploidies (2, 3, 4, 6), coverage depths (3×-40×), and genomic structures (autopolyploid and allopolyploid). 
Performance was measured using the generalized VER and block-adjusted MEC. %, which extend standard diploid metrics to handle partial phasing, multiple blocks, and non-complementary haplotypes.
We attempted to evaluate the deep learning method XHap~\cite{consul2023xhap}, but were unable to obtain stable results due to persistent software errors across multiple configurations.
All results represent means across multiple replicates with shaded regions indicating standard error.
% talk about hyperparams and grid search here
%WhatsHap and HapTree-X received VCF and BAM files, while H-PoPG and \phc{} used fragment files extracted from the same alignments using Hap10's extractHAIRS. 
%Both input formats represent the same underlying sequencing data.
For \phc{}, we tuned hyperparameters using a held-out autopolyploid validation dataset: with fixed mutation rate of $0.001$, ploidy $\in \{2, 3, 4, 6\}$, 3 coverages (5×, 10×, 20×), and 5 samples per configuration (60 samples in total). 
The selected parameters for \phc{}-short were $w_F=1/12$, $w_M=10/12$, $w_L=1/12$ (FFBS weight, MEC weight, likelihood weight respectively), and for \phc{}-long were $\delta=5$ (transition smoothing), $\varepsilon=0.00001$ (error rate), and learning rate $=0.02$ (gradient descent step size for FFBS augmentation).
Competing methods were run with default parameters as specified in their respective publications. 
These parameters were held fixed across all test evaluations (see \autoref{sec:simressupp} for further details).

\begin{figure*}[h!]
    \centering
\includegraphics[width=0.99\linewidth]{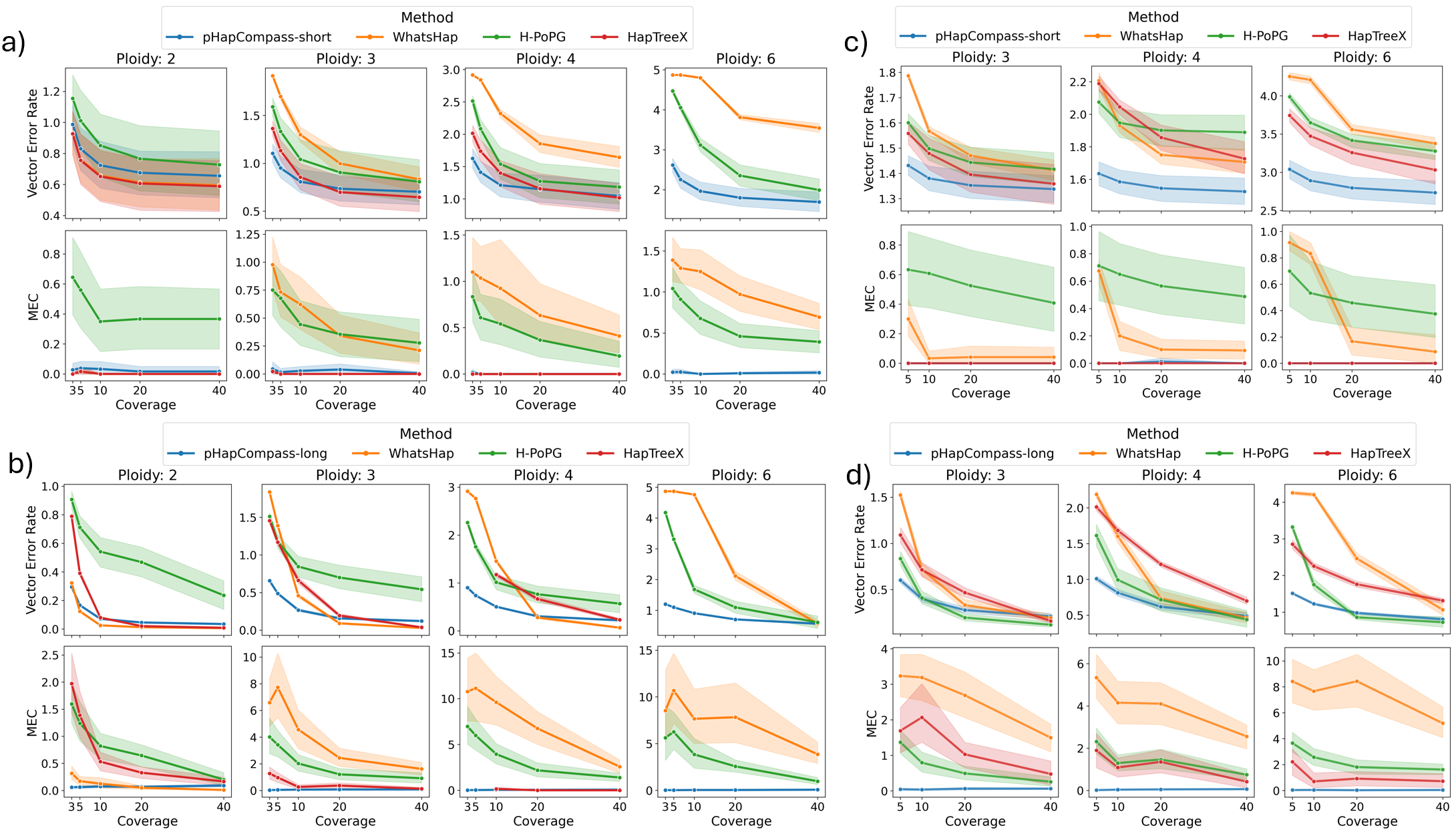}
    \caption{\textbf{Haplotype assembly performance on simulations.} (a-d) VER (top) and MEC (bottom) for \phc{} (blue), WhatsHap (orange), H-PoPG (green), and HapTree-X (red) on: (a) autopolyploid short reads, (b) autopolyploid long reads, (c) allopolyploid short reads, (d) allopolyploid long reads.}
    \label{fig:res_sim}
    \vspace{-20pt}
\end{figure*}

Autopolyploid genomes present the most challenging phasing scenario: all K haplotypes originate from the same ancestral species and accumulate mutations independently, creating scenarios where haplotypes can share long stretches of high similarity. 
This violates the fundamental assumption of most read clustering methods that similar reads originate from the same haplotype. 
Haplotype similarity heatmaps illustrate this challenge (Fig.~\ref{fig:auto-heatmap}), which is coupled with lower SNP densities at smaller mutation rates (Fig.~\ref{fig:auto-snp-density}). 
% This similarity increases with ploidy: at K=6, all pairwise distances fall within 0.32-0.42, creating fundamental ambiguity for methods relying on read similarity.
In short-read autopolyploid data, the two probabilistic assemblers, HapTree-X and \phc{}-short, demonstrated the lowest $\VER$ and $\MEC$, though the latter specifically performs well in lower coverage and higher ploidy scenarios (Fig.~\ref{fig:res_sim}a).
HapTree-X did not complete execution on K=6 due to segmentation faults.  %, suggesting algorithmic instability at high ploidy.
At K=2, \phc{}-short performed slightly worse than HapTree-X, likely because the greedy matching algorithm builds the global phasing incrementally and can commit to early suboptimal choices.

Long-read autopolyploid results show that \phc{}-long maintained consistently low $\VER$ and $\MEC$ across all ploidies and coverages (Fig.~\ref{fig:res_sim}b). 
The \phc{}-long model iteratively refines both haplotype sequences and read assignments which enables the model to handle the ambiguities in autopolyploid long-read data, particularly in lower coverages where other methods' performance drops.

Allopolyploid genomes differ from autopolyploids through subgenome divergence: haplotypes originating from different ancestral species accumulate systematic sequence differences that create structured heterozygosity.
Even at low inter-subgenome mutation rates, haplotypes from different subgenomes showed Hamming distances of 0.85-0.95, while haplotypes within the same subgenome remained similar (Fig.~\ref{fig:allo-heatmap}).
%The allopolyploid heatmaps  quantify this structure: 
On short-read allopolyploid data, all methods improved relative to autopolyploid performance, especially at higher ploidies, validating that subgenome divergence aids assembly (Fig.~\ref{fig:res_sim}c). 
% At $K=2$ and $K=3$, the performance difference between autopolyploid and allopolyploid was minimal, as expected since these lower ploidies have limited or no subgenome structure to exploit.
\phc{}-short maintained lowest $\VER$ across ploidies and coverages, with particularly stable performance at K=6, even in lower coverages. 
HapTree-X achieved competitive performance at K=6, likely because its $\MEC$ objective aligns well with the distinct error signatures between divergent subgenomes. % (See block level results in Supplementary 

Long-read allopolyploid results show convergence of top methods at high coverage, where abundant long-range information makes the problem well-constrained (Fig.~\ref{fig:res_sim}d). 
% At K=6 and 40× coverage, \phc{}-long (VER≈0.7), H-PoPG (≈0.8), and HapTree-X (≈0.8) achieve similar accuracy. 
However, performance diverged sharply at low coverage, where \phc{}-long achieved the lowest VER and MEC. 
By sharing information between haplotypes via $\{W_\ell\}_{\ell=1}^L$, read coverage for some but not all haplotypes can be propagated to inform the phase of unobserved haplotype segments.
% Additionally, block-level analysis reveals that \phc{} produces fewer and longer phased blocks than competitors (Figs.~\ref{fig:auto_short_block_info}-\ref{fig:allo_long_block_info}).
%\phc{} produces longer, more contiguous blocks with highest block N50 in most configurations (Fig.~\ref{fig:Block_N50_sim}).

Lastly, block-level analysis reveals that \phc{} produced fewer and longer phased blocks than competitors (Figs.~\ref{fig:auto_short_block_info}-\ref{fig:allo_long_block_info}), achieving the highest block N50 in most configurations (Fig.~\ref{fig:Block_N50_sim}).
Detailed block statistics and results of evaluation with standard metrics are provided in \S\ref{sec:detailed_sim_results} (Tables~\ref{tab:sim_auto_short_collapsed_ploidy2}-\ref{tab:sim_allo_long_collapsed}) and \S\ref{sec:standardmetrics} (Figs.~\ref{fig:auto_short_standard}-\ref{fig:allo_long_standard}), respectively.

\subsection{Uncertainty Quantification}
\label{sec:uncertanty_quant}

\begin{wrapfigure}[13]{R}{0.65\textwidth}
\vspace{-15pt}
\begin{center}
\vspace{6 pt}
\includegraphics[width=1\linewidth]{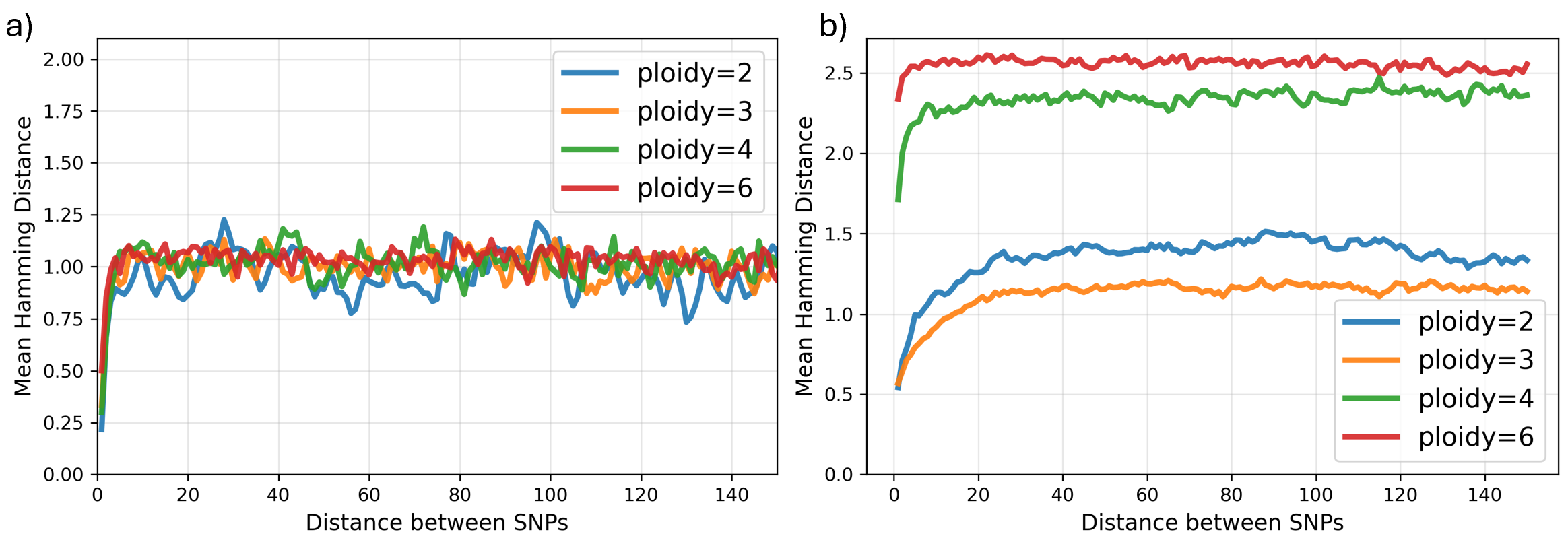}
\end{center}
\vspace{-15pt}
\caption{\textbf{Uncertainty quantification via FFBS sampling.} Mean Hamming distance versus SNP distance for (a) \phc{}-short and (b) \phc{}-long across ploidies.}
\label{fig:uncertainty}
\end{wrapfigure}
An important feature of the \phc{} probabilistic models is the ability to quantify uncertainty over phasings.
The potential functions for both models imply an exponentially sized distribution over phasings, which we approximate by collecting multiple FFBS samples. 
In \phc{}-short, an FFBS path through the connected components (blocks) of $Q$ corresponds to phasings over pairs of SNPs, joined using Algorithm \ref{alg:phase_position}. 
%Each connected component of $Q$ corresponds to a block. 
In \phc{}-long, a path through $H^* = (W_1, \ldots, W_L)$ yields a single block assembly. 
In both models, the probability of a phasing is proportional to the number of consistent FFBS samples.

We evaluated uncertainty quantification by considering the phase between SNP pairs at different distances.
After optimally matching the sampled phasings to $H$, we computed the phase accuracy as the mean Hamming distance between each sample's phase to the ground truth, with Hamming distance extended to pairs of matrices in $\{0,1\}^{K \times 2}$ as the number of elements in the matrices that differ. 
For example, a Hamming distance of 0 represents a perfect phase, and a Hamming distance of 2 represents two vector errors.
%To see how pair phase accuracy varies over the assembly, we plot average Hamming distance against SNP distance (that is, for SNPs $\ell_0 < \ell_1$, the value $\ell_1 - \ell_0$). 
The pair phase accuracy decreases (Hamming distance and uncertainty increases) as SNP distance increases before plateauing around a similar accuracy due to the evaluation of SNPs across blocks (Fig.~\ref{fig:uncertainty}).
%We can interpret this local window of decreasing accuracy as the distance where reads typically co-cover SNPs; the direct evidence over the pair tends the Hamming distance towards 0, and farther away SNPs have shared evidence less frequently. 
We remark that the difference in the asymptotes is due to Hamming distance being undefined for unassembled SNPs and \phc{}-short assembling fewer SNPs than \phc{}-long.

Additionally, we examined whether phasing uncertainty correlated with genotype balance by stratifying SNP pairs according to their alternate-allele counts $\{g_{\ell}, g_{\ell'}\}$. 
% Across both short-read and long-read settings, more balanced genotype pairs (those closer to $K/2$ alternate alleles) exhibited moderately higher uncertainty in most cases, consistent with the hypothesis that locally ambiguous allele balance reduces phasing certainty 
Generally, genotype pairs with more potential phasings exhibited higher uncertainty across both short-read and long-read settings
% consistent with the hypothesis that increased combinatorial phasing possibilities reduce phasing accuracy 
(\S\ref{sec:supp_Uncertaintyquant} and Fig.~\ref{fig:uncert_2}).

% \subsection{Experimental Data Results}
\vspace{15pt}
\subsection{\mbox{Experimental Data Results}}

To evaluate performance on experimental data, we applied \phc{}-short, WhatsHap, and H-PoPG to octoploid strawberry data at two coverage levels 8× and 16×. 
% HapTree-X could not be run due to computational resource limitations. 
We filtered variants along 200bp windows to reduce linkage disequilibrium effects and extracted chromosome regions for assembly (Table~\ref{tab:chromosome_info}).
% We evaluated the performance at 8× and 16× coverage to assess performance under low-coverage scenarios.
\phc{}-short achieved lower MEC than competing methods, indicating assemblies with minimal read-haplotype conflicts. 
Additionally, it produced %5-9× 
fewer phased blocks compared to WhatsHap and H-PoPG, demonstrating the continuity across longer genomic segments even with short-read data, where clustering-based methods make different blocks due to ambiguous read assignments
% (Tables~\ref{tab:experimental_results} and~\ref{tab:experimental_results_detailed}).
(see \autoref{sec:expressupp}, Tables~\ref{tab:experimental_results} and~\ref{tab:experimental_results_detailed}).
\begin{wraptable}[16]{r}{0.5\textwidth}
\vspace{-1 pt}
\centering
\small
\caption{Haplotype assembly performance on experimental octoploid strawberry data aggregated across seven chromosomes. MEC: block-adjusted Minimum Error Correction. Blocks: total number of phased segments.}
\begin{tabular}{lccc}
\toprule
\textbf{Method} & \textbf{Coverage} & \textbf{MEC} & \textbf{Blocks} \\
\midrule
\phc{}-short & 8×  & 0.50  & 126,461 \\
             & 16× & 0.00  & 84,522 \\
\midrule
WhatsHap     & 8×  & 6.12  & 691,508 \\
             & 16× & 6.12  & 754,252 \\
\midrule
H-PoPG       & 8×  & 4.75  & 485,542 \\
             & 16× & 5.50  & 527,045 \\
\bottomrule
\end{tabular}
\label{tab:experimental_results}
\end{wraptable}

\subsection{Runtime Analysis}
We measured the runtimes for all methods across synthetic datasets (see \autoref{sec:runtime_supp} for details). 
Although hardware heterogeneity makes direct comparisons difficult to interpret, the explicitly probabilistic methods of \phc{} and HapTree-X exhibited the longest runtimes (Fig.~\ref{fig:times}). 
Since SNP density varies across samples and confounds runtime, we normalized runtime by SNP count ($t/L$, seconds per SNP; Fig.~\ref{fig:times_norm}). 
The runtime characteristics differ between the two \phc{} models: 
\phc{}-short scales well for short reads but the SNP line graph becomes overly dense for long read data. 
Conversely, \phc{}-long performs Gibbs updates per read, scaling efficiently with long reads but poorly with high coverage.

\section{Discussion}
\label{sec:discussion}
The evaluation of \phc{} on realistic autopolyploid and allopolyploid genomes suggests several key insights.
First, realistic simulation %—capturing the true similarity structure of polyploid genomes—
exposes fundamental limitations of clustering-based methods that depend only on read similarity. 
Second, \phc{}'s probabilistic framework, which models uncertainty in both haplotype configurations and read assignments, provides robust performance across the full spectrum of polyploid structures, ploidies, and coverage regimes. 
The key algorithmic innovation is maintaining and propagating probabilistic evidence over all valid phasings at each SNP pair, %(via node and edge 
rather than committing to a single best phasing as in deterministic methods. 
Third, the distinction between autopolyploidy and allopolyploidy proves critical for understanding assembly performance. 
Allopolyploid subgenome divergence creates structured heterozygosity that aids all methods,
however, \phc{} maintains a consistent advantage across both genomic structures, suggesting that the probabilistic framework provides value beyond simply exploiting subgenome divergence.
Fourth, the convergence of methods at high coverage in allopolyploid long-read data
indicates that when evidence is abundant and subgenomes are divergent, the problem becomes well-constrained for multiple algorithmic approaches. 
The divergence at low coverage highlights the benefits of probabilistic inference: by maintaining distributions over candidate phasings, \phc{}-long can integrate sparse evidence, whereas deterministic approaches must commit to specific phasing decisions under limited information.
The reduced block counts for \phc{} relative to competitors demonstrate that this connectivity-based approach maintains phasing continuity.

Limitations of the current work suggest directions for future development. 
The hyperparameter tuning for \phc{} used a separate validation set with fixed mutation rate, which may not generalize optimally to all experimental conditions. 
The matching algorithm's greedy selection strategy, while effective, does not guarantee global optimality. 
Additionally, \phc{} is more computationally demanding than deterministic competitors. 
The SNP line graph density limits \phc{}-short on long reads, while iterative Gibbs sampling limits \phc{}-long on high-coverage data; we evaluated each method on its intended data type due to these computational differences.
Although both \phc{} models can, in principle, be applied to short- and long-read data, their differing computational characteristics currently limit their practical applicability across sequencing regimes.
Algorithmic improvements like approximating exponentially sized factors or variational inference methods could be used to improve scalability for \phc{}-short and \phc{}-long, respectively.

\section{Conclusions}
\label{sec:conclusion}
In this work, we made contributions to the simulation, probabilistic computation, and evaluation of polyploid genome assemblies.
First, we extended the VER and block-adjusted MEC evaluation criteria to handle the complexities of polyploid assemblies: partial phasing, multiple blocks, and non-complementary haplotypes. 
Second, we developed a pipeline for realistic polyploid genome simulation, which captures the true similarity structure that makes polyploid phasing fundamentally challenging; both the software and simulated datasets for both autopolyploid and allopolyploid genomes are freely available to the community to facilitate benchmarking of future polyploid assembly methods~\cite{pHapCompass2025}. % ADD GITHUB CITATION HERE MARJAN
Lastly, we presented pHapCompass, a probabilistic framework for polyploid haplotype assembly that explicitly models and quantifies uncertainty in  both haplotype phase and read-to-haplotype assignments. 
Unlike existing polyploid assemblers that output single point estimates, \phc{} uses FFBS to generate distributions over haplotype configurations. 
Each sample represents a plausible phasing weighted by its posterior probability given the observed reads, which we report in the VCF formatted output and include probability scores for each generated phasing. 
\phc{} demonstrated reliable performance across diverse polyploid configurations, with particular advantages in low-coverage scenarios where deterministic methods are more limited.

\bibliographystyle{plain}
\bibliography{ref}

\clearpage
\newpage
\onecolumn

\setcounter{section}{0}
\setcounter{figure}{0}
\setcounter{table}{0}
\setcounter{algorithm}{0}

\renewcommand{\thesection}{\Alph{section}}
\renewcommand{\thesubsection}{\Alph{section}.\arabic{subsection}}
\renewcommand{\thefigure}{\Alph{section}.\arabic{figure}}
\renewcommand{\thetable}{\Alph{section}.\arabic{table}}
\renewcommand{\thealgorithm}{\Alph{section}.\arabic{algorithm}}

% --- Force appendix-style section headings: "Appendix A: Title" ---
\renewcommand{\sectionautorefname}{Appendix} % for \autoref

\let\oldsection\section
\renewcommand{\section}[1]{%
  \refstepcounter{section}%
  \setcounter{subsection}{0}%
  \oldsection*{Appendix~\thesection: #1}%
}
% ---------------------------------------------------------------

% \appendix

% \section{Notation}
% \section{Appendix \thesection: Notation}
\section{Notation}
\label{sec:suppNotation}

\subsection*{Indices}
\begin{itemize}
\item $j$: Read index, $j \in \{1, \ldots, |\mathcal{R}|\}$
\item $\ell$: SNP position index, $\ell \in \{1, \ldots, L\}$
\item $k$: Haplotype index, $k \in \{1, \ldots, K\}$
% \item $\ell_0, \ell_1, \ell_2$: Specific SNP positions in ordered pairs/triples
\end{itemize}

\subsection*{Problem Setup}
\begin{itemize}
\item $K$: Ploidy level (number of haplotypes)
\item $L$: Total number of heterozygous SNP positions
\item $\mathcal{R} = \{r^{(1)}, r^{(2)}, \ldots, r^{(|\mathcal{R}|)}\}$: Set of sequencing reads
% \item $r^{(j)}[\ell] \in \{0, 1, \text{`-'}\}$: Observed allele of read $j$ at SNP position $\ell$, where `\texttt{-}' indicates the position is not covered
\item $\mathcal{G} = \{g_1, \ldots, g_L\}$: Reference genotype information, where $g_{\ell} \in \{0, \ldots, K\}$ is the count of alternate alleles at position $\ell$.
\item $H = \{h^{(1)}, \ldots, h^{(K)}\}$: Set of $K$ haplotypes to be reconstructed (ground truth).
\item $H^* = \{h^{(1)}, \ldots, h^{(K)}\}$: Set of $K$ predicted haplotypes.
\item $h^{(k)}[\ell] \in \{0, 1\}$: Allele of haplotype $k$ at position $\ell$ (0=reference, 1=alternate)
\item $\varepsilon$: Sequencing error rate.
\end{itemize}

\subsection*{Graphs}
\begin{itemize}
\item $G = (V, E)$: \textbf{SNP graph}, where $V = \{1, \ldots, L\}$ are SNP positions and $(\ell_0, \ell_1) \in E$ if at least one read covers both positions
\item $Q = (U, E_Q)$: \textbf{Line graph of $G$}, where each node $u_{\ell_0\ell_1} \in U$ represents an edge $(\ell_0, \ell_1) \in E$, and two nodes in $U$ are adjacent in $E_Q$ if their corresponding edges in $G$ share exactly one SNP
\item $Q' = (U, E_Q')$: \textbf{Directed acyclic graph (DAG)} derived from $Q$ by directing edges according to SNP order
\end{itemize}

\subsection*{Phasings}
\begin{itemize}
\item $\phi_{\ell_0\ell_1} \in \{0,1\}^{K \times 2}$: A \textbf{phasing configuration} for two SNPs $\ell_0$ and $\ell_1$, represented as a $K \times 2$ binary matrix where row $k$ contains the alleles of haplotype $k$ at positions $\ell_0$ and $\ell_1$
\item $\phi_{\ell_0\ell_1}^{(k)} \in \{0,1\}^2$: The $k$-th row of phasing $\phi_{\ell_0\ell_1}$, representing the allele pair from haplotype $k$
\item $\phi_{\ell_0\ell_1}^{(k)}(\ell)$: The allele at position $\ell$ in haplotype $k$ within phasing $\phi_{\ell_0\ell_1}$
\item $\Phi_{\ell_0\ell_1} \subset \{0,1\}^{K \times 2}$: The \textbf{space of valid phasings} for SNPs $\ell_0$ and $\ell_1$ that satisfy genotype constraints
\item $\phi_{\ell_0\ell_1\ell_2} \in \{0,1\}^{K \times 3}$: A phasing configuration for three SNPs $\ell_0$, $\ell_1$, and $\ell_2$
\item $\Phi_{\ell_0\ell_1\ell_2} \subset \{0,1\}^{K \times 3}$: The space of valid phasings for three SNPs
\end{itemize}

% \subsection*{Phasing Enumeration (Section~\ref{sec:suppEnumerating})}
% \begin{itemize}
% \item $\mathbf{t} = [t_{00}, t_{01}, t_{10}, t_{11}]$: \textbf{Count vector} for two-SNP phasings, where $t_{ab}$ is the number of haplotypes with allele pattern $(a,b)$
% \item $t_{11}$: Count of haplotypes carrying the $(1,1)$ allele pattern; used as the free parameter to enumerate all valid phasings
% \end{itemize}

\subsection*{Read Sets}
\begin{itemize}
\item $\mathcal{R}_{\ell_0\ell_1} = \{r \in \mathcal{R} : r[\ell_0] \neq \text{`}-\text{'} \land r[\ell_1] \neq \text{`}-\text{'}\}$: Set of reads covering both SNP positions $\ell_0$ and $\ell_1$
\item $\mathcal{R}_{\ell_0\ell_1\ell_2} = \{r \in \mathcal{R} : r[\ell_0] \neq \text{`}-\text{'} \land r[\ell_1] \neq \text{`}-\text{'} \land r[\ell_2] \neq \text{`}-\text{'}\}$: Set of reads covering all three SNP positions
\end{itemize}

\subsection*{Likelihood and Potentials}
\begin{itemize}
\item $d(\cdot, \cdot)$: Hamming distance between two binary vectors
\item $\mathcal{L}(r, \phi_{\ell_0\ell_1})$: Likelihood of observing read $r$ given phasing $\phi_{\ell_0\ell_1}$
\item $f(\phi_{\ell_0\ell_1})$: \textbf{Node potential} (emission probability) for phasing $\phi_{\ell_0\ell_1}$, computed as the sum of likelihoods over all reads in $\mathcal{R}_{\ell_0\ell_1}$
\item $f(\phi_{\ell_0\ell_1\ell_2})$: \textbf{Edge potential} (transition probability) for three-SNP phasing $\phi_{\ell_0\ell_1\ell_2}$
\end{itemize}

\newpage

% \section{Enumerating Valid Phasings for Two SNPs}
% \section{Appendix \thesection: Enumerating Valid Phasings for Two SNPs}
\section{Enumerating Valid Phasings for Two SNPs}
\label{sec:suppEnumerating}

Consider a tetraploid organism ($K=4$) with genotypes $g_{\ell} = 2$ and $g_{\ell'} = 2$ at two adjacent SNP positions $\ell$ and $\ell'$. 

Each haplotype can have one of four possible allele patterns at these two positions:
\begin{itemize}
    \item $(0,0)$: reference allele at both positions
    \item $(0,1)$: reference at $\ell$, alternate at $\ell'$
    \item $(1,0)$: alternate at $\ell$, reference at $\ell'$
    \item $(1,1)$: alternate allele at both positions
\end{itemize}

A phasing configuration is characterized by the count vector $\mathbf{t} = [t_{00}, t_{01}, t_{10}, t_{11}]$, where $t_{ab}$ denotes the number of haplotypes with pattern $(a,b)$.
\begin{itemize}

\item Genotype constraints:

These counts must satisfy:
\begin{align}
t_{10} + t_{11} &= g_{\ell} = 2 \tag{total 1's at position $\ell$}\\
t_{01} + t_{11} &= g_{\ell'} = 2 \tag{total 1's at position $\ell'$}\\
t_{00} + t_{01} + t_{10} + t_{11} &= K = 4 \tag{total haplotypes}\\
t_{ab} &\geq 0 \quad \forall a,b \tag{non-negativity}
\end{align}

\item Parameterization:

We parameterize by $t_{11}$, the count of $(1,1)$ patterns. Given $t_{11}$, the other counts are determined:
\begin{align}
t_{10} &= g_{\ell} - t_{11} = 2 - t_{11}\\
t_{01} &= g_{\ell'} - t_{11} = 2 - t_{11}\\
t_{00} &= K - g_{\ell} - g_{\ell'} + t_{11} = 4 - 2 - 2 + t_{11} = t_{11}
\end{align}

\item Feasible range:

For all counts to be non-negative:
\begin{align}
t_{11} &\geq 0\\
t_{10} = 2 - t_{11} &\geq 0 \implies t_{11} \leq 2\\
t_{01} = 2 - t_{11} &\geq 0 \implies t_{11} \leq 2\\
t_{00} = t_{11} &\geq 0 \implies t_{11} \geq 0
\end{align}

Thus $t_{11} \in \{0, 1, 2\}$, yielding exactly \textbf{three valid phasings}:

\begin{center}
\begin{tabular}{cccccl}
\toprule
$t_{11}$ & $t_{00}$ & $t_{01}$ & $t_{10}$ & $\mathbf{t}$ & Phasing Configuration \\
\midrule
0 & 0 & 2 & 2 & $[0,2,2,0]$ & $\begin{bmatrix}0&1\\0&1\\1&0\\1&0\end{bmatrix}$ \\[0.5em]
1 & 1 & 1 & 1 & $[1,1,1,1]$ & $\begin{bmatrix}0&0\\0&1\\1&0\\1&1\end{bmatrix}$ \\[0.5em]
2 & 2 & 0 & 0 & $[2,0,0,2]$ & $\begin{bmatrix}0&0\\0&0\\1&1\\1&1\end{bmatrix}$ \\
\bottomrule
\end{tabular}
\end{center}

Each row in the phasing configuration matrix corresponds to one haplotype, with columns representing the two SNP positions.

\item Verification:

For $t_{11} = 1$ (middle row):
\begin{itemize}
    \item Count of 1's at position $\ell$ (first column): $0+0+1+1 = 2 = g_{\ell}$ 
    \item Count of 1's at position $\ell'$ (second column): $0+1+0+1 = 2 = g_{\ell'}$ 
    \item Total haplotypes: $1+1+1+1 = 4 = K$ 
\end{itemize}

\item General case

For arbitrary genotypes $(g_{\ell}, g_{\ell'})$ and ploidy $K$, the feasible range is:
$$
t_{11} \in \left[\max(0, g_{\ell} + g_{\ell'} - K), \min(g_{\ell}, g_{\ell'})\right]
$$
The number of valid phasings is therefore:
$$
|\Phi_{\ell\ell'}| = \min(g_{\ell}, g_{\ell'}) - \max(0, g_{\ell} + g_{\ell'} - K) + 1 = O(K)
$$

This represents a reduction from the $2^{2K} = 2^8 = 256$ possible binary matrices without genotype constraints.

\end{itemize}
\newpage

\section{Viterbi Details}
\label{sec:suppViterbi}

The Viterbi algorithm computes the maximum a posteriori (MAP) sequence of states in the directed pHapCompass graph $Q' = (U, E'_Q)$:
\begin{equation*}
  (\phi^*_1, \ldots, \phi^*_{|U|}) = \arg\max_{\phi_1,\ldots,\phi_{|U|}} P(\phi_1, \ldots, \phi_{|U|} \mid \mathcal{R}).  
\end{equation*}

This is solved via dynamic programming. For each node $u_t \in U$ and state $\phi_t \in \Phi_t$, we compute:
$$
V_t(\phi_t) = \max_{\phi_1,\ldots,\phi_{t-1}} P(\phi_1, \ldots, \phi_t, \mathcal{R}_{1:t}),
$$
the maximum probability of any path ending in state $\phi_t$ given all observations up to node $t$, where $\mathcal{R}_{1:t}$ denotes all reads covering SNP pairs in nodes $u_1, \ldots, u_t$.

\subsection{Base Case ($t = 1$)} 

\begin{equation*}
    V_1(\phi_1) = P(\phi_1) \prod_{r \in \mathcal{R}_{\ell_0\ell_1}} L(r, \phi_1),
\end{equation*}
where $P(\phi_1)$ is the prior probability of phasing $\phi_1$ (uniform), and $L(r, \phi_1)$ is the likelihood from Equation (2).

\subsection{Recursion ($t = 2, \ldots, |U|$)} 
\begin{align*}
V_t(\phi_t) &= \left( \prod_{r \in \mathcal{R}_t} L(r, \phi_t) \right) \times \\
&\quad \max_{\phi_{Pa(u_t)}} 
\left[ \prod_{u_p \in Pa(u_t)} V_{p}(\phi_p) \times P(\phi_t \mid \phi_{Pa(u_t)}) \right],
\end{align*}
where:
\begin{itemize}
    \item $\mathcal{R}_t$ is the set of reads covering the SNP pair at node $u_t$
    \item $Pa(u_t)$ denotes the parent nodes of $u_t$ in the directed graph $Q'$
    \item $\phi_{Pa(u_t)}$ denotes the collection of parent phasings
    \item $P(\phi_t \mid \phi_{Pa(u_t)})$ is the transition probability derived from edge potentials (Equation 3), normalized over all three-SNP phasings consistent with parent states
\end{itemize}

The recursion proceeds in topological order. 
At each step, the maximizing parent configuration is recorded via a backpointer to reconstruct the optimal path after completion. 
Each connected component of $Q$ is processed independently.

% \section{Appendix \thesection: FFBS Details}

\clearpage
\section{FFBS Details}
\label{sec:suppFFBS}

We direct the edges of the pHapCompass graph $Q = (U, E_Q)$ to obtain $Q' = (U, E'_Q)$ by imposing a topological ordering on nodes according to ascending genomic position. 
Each vertex $u_{\ell_0\ell_1} \in U$ has state space $\Phi_{\ell_0\ell_1}$.

The forward-filtering backward-sampling algorithm computes forward messages in topological order, then samples states in reverse order.

\subsection*{Forward Pass}

The forward message for each vertex $u_t$ represents the joint probability of state $\phi_t$ and all observations up to node $t$:
$$
\alpha_t(\phi_t) = P(\phi_t, \mathcal{R}_{1:t}),
$$
where $\mathcal{R}_{1:t}$ denotes all reads covering SNP pairs in nodes $u_1, \ldots, u_t$.

\textbf{Base Case ($t=1$):}
$$
\alpha_1(\phi_1) = P(\phi_1) \prod_{r \in \mathcal{R}_{\ell_0\ell_1}} L(r, \phi_1),
$$
where $P(\phi_1)$ is the prior (uniform) and $L(r, \phi_1)$ is the likelihood defined in Equation (2).

\textbf{Recursion ($t = 2, \ldots, |U|)$:}
\begin{align}
\alpha_t(\phi_t) &= P(\phi_t, \mathcal{R}_{1:t}) \notag \\
&= \prod_{r \in \mathcal{R}_t} L(r, \phi_t) \sum_{\phi_{Pa(u_t)}} P(\phi_t \mid \phi_{Pa(u_t)}) \prod_{u_p \in Pa(u_t)} \alpha_{p}(\phi_p),
\label{eq:ffbs_forward}
\end{align}
where:
\begin{itemize}
    \item $\mathcal{R}_t$ is the set of reads covering the SNP pair at node $u_t$
    \item $L(r, \phi_t)$ is the emission probability from Equation (2)
    \item $P(\phi_t \mid \phi_{Pa(u_t)})$ is the transition probability derived from edge potentials in Equation (3), normalized over all three-SNP phasings consistent with parent states
    \item $Pa(u_t)$ denotes the parent nodes of $u_t$ in the directed graph $Q'$
\end{itemize}

Forward messages are computed only after all parent messages are available, following the topological order.

\subsection*{Backward Sampling}

We sample states sequentially in reverse topological order from the posterior distribution $P(\phi_{1:|U|} \mid \mathcal{R})$.

\textbf{Base Case ($t=|U|$):}

Sample the final node proportional to its normalized forward message:
$$
\phi_{|U|} \sim \frac{\alpha_{|U|}(\phi_{|U|})}{\sum_{\phi'_{|U|}} \alpha_{|U|}(\phi'_{|U|})}.
$$

\textbf{Recursion ($t = |U|-1, \ldots, 1$):}

For each node $u_t$, sample from the conditional posterior given all sampled children:
$$
P(\phi_t \mid \{\phi_c : u_c \in Ch(u_t)\}, \mathcal{R}) \propto \alpha_t(\phi_t) \prod_{u_c \in Ch(u_t)} P(\phi_c \mid \phi_t),
\label{eq:ffbs_backward}
$$
where $Ch(u_t)$ denotes the children of $u_t$ in $Q'$. 
Each node is sampled only after all its children have been sampled, proceeding in reverse topological order. Each connected component of $Q$ is processed independently.

\newpage
% \section{Appendix \thesection: Matching Problem}
\section{Matching Problem}
\label{sec:matching_app}
% \section{Matching Problem}

\subsection{Variant Selection Algorithm}
\label{sec:alg1}

The variant selection algorithm (Algorithm~\ref{alg:varselect}) constructs a final haplotype solution by iteratively selecting one variant to be phased at a time, using a greedy connectivity-based strategy. 
The algorithm maintains four sets throughout execution:
\begin{itemize}
\item $U^{\text{phased}}$: nodes in $Q$ whose phasings have been incorporated into the global solution
\item $U^{\text{unphased}}$: nodes in $Q$ not yet processed
\item $S^{\text{phased}}$: individual SNP positions that have been assigned to global haplotype rows
\item $S^{\text{unphased}}$: SNP positions not yet assigned
\end{itemize}

Note that a node $u_{\ell_0\ell_1} \in U$ represents a pair of SNP positions, while elements of $S$ are individual positions. 
A node is considered phased or processed once both of its positions have been assigned to the final haplotype matrix, or once it has been used to constrain the phasing of its positions.

\begin{figure}[!ht]
    \centering
    \includegraphics[width=0.8\linewidth]{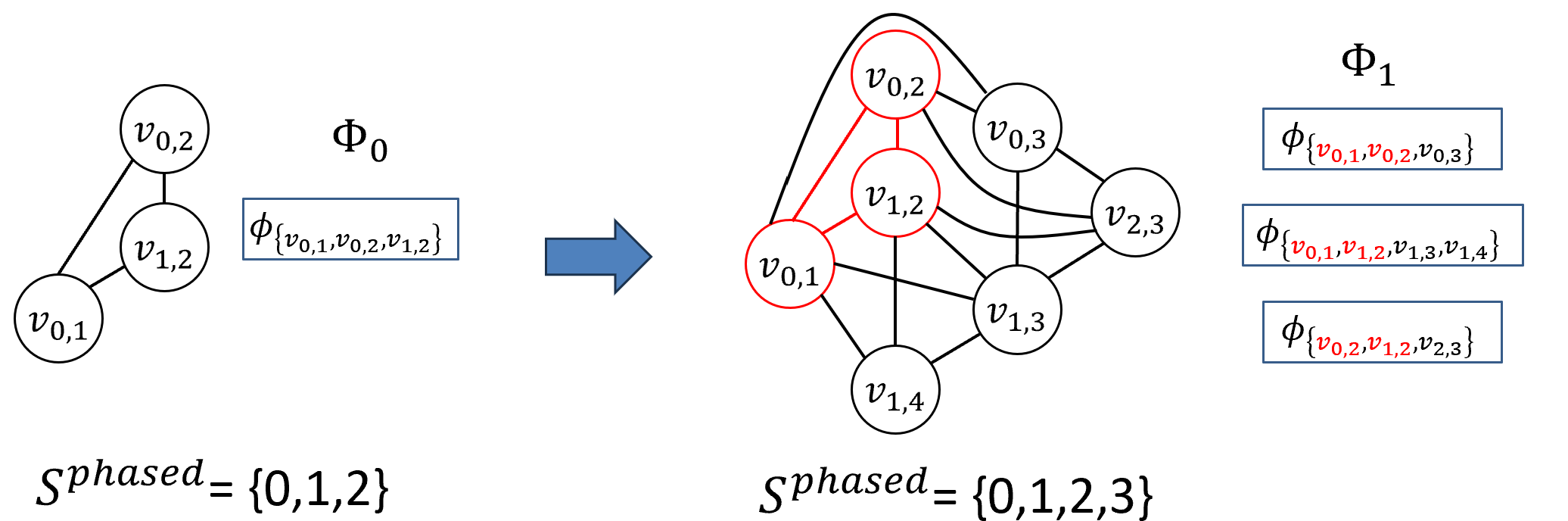}

    \caption{\textbf{Variant selection algorithm example.} Initially $U^{\text{phased}} = {v_{0,1}}$. Then the variant selection algorithm selects position $\ell^* = 2$ with maximum connectivity (2 edges) to already-phased nodes $U^{\text{phased}}$.
    The phasing is done in the phasing space of $\Phi_0$ covering positions $\{0,1, 2\}$ (left). After phasing position $2$, the algorithm updates $U^{\text{phased}} = \{v_{0,1}, v_{0,2}, v_{1,2}\}$ and $S^{\text{phased}} = \{0,1,2\}$ (middle). Then we the next highest-connectivity position $\ell^* = 3$ with 6 edges compared to $4$ with $2$ edges. Please note that the phasing space now has changed to $\Phi_1$, over cliques partially covering $U^\text{phased}$ (right).}
    \label{fig:algorithmsvis}
\end{figure}

\subsubsection{Initialization (Lines 2-7).}
The algorithm begins with all four sets initialized appropriately: the phased sets are empty, and the unphased sets contain all nodes and all positions, respectively. Block identifiers track which positions belong to the same connected component, starting with block\_id $= 1$.

\subsubsection{Starting Node Selection (Lines 8-13).}
To begin phasing, we select the first node $u_{\ell_0\ell_1}$ according to the topological ordering of $Q$ (typically the leftmost pair in genomic coordinates). 
We assign both positions $\ell_0$ and $\ell_1$ to the global haplotype matrix $H^*$ using the inferred phasing $\phi_{u_{\ell_0\ell_1}}^*$ obtained from either Viterbi or FFBS. 
The notation $\phi_u^*$ denotes the best or sampled phasing for node $u$: for Viterbi, this is the most probable state; for FFBS, this is a sample from the posterior distribution. 
Both positions are marked as phased, the node is moved from $U^{\text{unphased}}$ to $U^{\text{phased}}$, and both positions are assigned to the current block.

\subsubsection{Main Loop (Lines 14-34).}
The algorithm continues until all positions have been phased ($S^{\text{unphased}} = \emptyset$).

\subsubsection{Finding Neighbor Nodes (Line 16).}
At each iteration, we identify the set $\mathcal{N}$ of unphased nodes that are adjacent (in the graph $Q$) to at least one of the nodes in the phased node set ($U^{\text{phased}}$). 
These are the nodes that can be used to extend the current phased set, as they connect already-phased positions to new positions.

\subsubsection{Handling Disconnected Components (Lines 18-25).}
If $\mathcal{N} = \emptyset$, no unphased nodes connect to the current phased set, indicating that we have exhausted the current connected component of $Q$. 
In this case, we increment the block identifier to start a new phasing block, select the next unphased node $u_{\ell_i\ell_j}$ in topological order, initialize its phasing using $\phi_{u_{\ell_i\ell_j}}^*$, update all four sets accordingly, and continue to the next iteration. 
Positions in different blocks are phased independently and may have inconsistent haplotype labelings across blocks.

\subsubsection{Selecting the Next Position (Lines 27-31).}
If neighbor nodes exist, we select the next position to phase based on connectivity. 
For each unphased position $\ell$ that appears in at least one node in $\mathcal{N}$, we compute its connectivity as the number of nodes in $\mathcal{N}$ containing $\ell$. 
The position $\ell^*$ with maximum connectivity is selected, as it has the most connection to the already-phased set.

\subsubsection{Phasing the Selected Position (Line 33).}
The \textsc{PhasePosition} subroutine (Algorithm~\ref{alg:phase_position}) is called to determine the haplotype assignment $\mathbf{h}_{\ell^*} \in \{0,1\}^K$ for position $\ell^*$. 
This subroutine generates and scores candidate phasings based on read evidence, genotype constraints, and agreement with the inferred phasings $\{\phi_u^*\}$. 
The inputs include: the selected position $\ell^*$, the set of phased positions $S^{\text{phased}}$, the current global haplotype matrix $\mathcal{H}$, the set of neighbor nodes $\mathcal{N}$, the inferred phasings $\{\phi_u^*\}$, and the set of reads $\mathcal{R}$.

\subsubsection{Updating the Global Solution (Lines 35-40).}
The selected haplotype column $\mathbf{h}_{\ell^*}$ is assigned to position $\ell^*$ in the global matrix. P
Position $\ell^*$ is moved from $S^{\text{unphased}}$ to $S^{\text{phased}}$ and assigned to the current block. 
Finally, all nodes in $\mathcal{N}$ that contain position $\ell^*$ are moved from $U^{\text{unphased}}$ to $U^{\text{phased}}$, as these nodes have now been fully utilized in the phasing process.

\subsubsection{Termination and Output (Line 42).}
The algorithm terminates when $S^{\text{unphased}} = \emptyset$, meaning all positions have been phased. 
The output consists of the global haplotype matrix $H^* \in \{0,1, \text{`}-\text{'}\}^{K \times L}$ and the block assignment vector $\mathbf{b} \in (\mathbb{N} \cup \{\text{NaN}\})^L$, where $\mathbf{b}[\ell]$ indicates which phasing block contains position $\ell$. 
Positions that were not covered by any reads remain unphased and can be assigned NaN in the block vector, though in practice we assign them to the nearest block or use genotype-based imputation.

\subsubsection{Note on Unphased Positions.}
In practice, positions not covered by any informative reads, remain unphased by the algorithm. 
These positions can be assigned $\mathbf{b}[\ell] = \text{NaN}$ to indicate missing phase information, similarly for such variants, $H^*[\cdot, \ell] = \text{`}-\text{'}$.

\begin{algorithm}[!h]
\caption{Variant Selection Algorithm}
\label{alg:varselect}
\begin{algorithmic}[1]
\Require SNP Line Graph $Q = (U, E_Q)$, inferred phasings $\{\phi_u^* : u \in U\}$, $\mathcal{R}$
\Ensure Global haplotype matrix $H^* \in \{0,1, \text{`}-\text{'}\}^{K \times L}$, block assignments $\mathbf{b} \in (\mathbb{N} \cup \{\text{NaN}\})^L$

\State \textbf{Initialize:}
\State $U^{\text{phased}} \gets \emptyset$ \Comment{Set of phased vertices in $Q$ (initially empty).}
\State $U^{\text{unphased}} \gets U$ \Comment{Set of phased vertices in $Q$. Note: Vertices in $Q$ include pairs of positions.}
\State $S^{\text{phased}} \gets \emptyset$\Comment{Set of phased variants in L. Note, this algorithm phases variant by variant}
\State $S^{\text{unphased}} \gets \{1, \ldots, L\}$
\State block\_id $\gets 1$

\State
\State Select first node $u_{\ell_0\ell_1} \in U^{\text{unphased}}$ in topological order
\State $H^*[\cdot, \{\ell_0, \ell_1\}] \gets \phi_{u_{\ell_0\ell_1}}^*$
\State $U^{\text{phased}} \gets \{u_{\ell_0\ell_1}\}$, $U^{\text{unphased}} \gets U^{\text{unphased}} \setminus \{u_{\ell_0\ell_1}\}$
\State $S^{\text{phased}} \gets \{\ell_0, \ell_1\}$, $S^{\text{unphased}} \gets S^{\text{unphased}} \setminus \{\ell_0, \ell_1\}$
\State $\mathbf{b}[\{\ell_0,\ell_1\}] \gets$ block\_id

\State
\While{$S^{\text{unphased}} \neq \emptyset$}
    \State $\mathcal{N} \gets \{u \in U^{\text{unphased}} : \exists v \in U^{\text{phased}}, (v, u) \in E_Q \text{ or } (u, v) \in E_Q\}$
    \Comment{Neighbor nodes to the phased node set.}
    
    \If{$\mathcal{N} = \emptyset$} \Comment{There is a disconnected component}
        \State block\_id $\gets$ block\_id $+ 1$
        \State Select the next node $u_{\ell_i\ell_j} \in U^{\text{unphased}}$ in topological order
        \State $H^*[\cdot, \{\ell_i, \ell_j\}] \gets \phi_{u_{\ell_i\ell_j}}^*$
        \State $U^{\text{phased}} \gets U^{\text{phased}} \cup \{u_{\ell_i\ell_j}\}$, $U^{\text{unphased}} \gets U^{\text{unphased}} \setminus \{u_{\ell_i\ell_j}\}$
        \State $S^{\text{phased}} \gets S^{\text{phased}} \cup \{\ell_i, \ell_j\}$, $S^{\text{unphased}} \gets S^{\text{unphased}} \setminus \{\ell_i, \ell_j\}$
        \State $\mathbf{b}[\{\ell_i,\ell_j\}] \gets$ block\_id
        \State \textbf{continue}
    \EndIf
    
    \State
    \For{each position $\ell \in S^{\text{unphased}}$ appearing in $\mathcal{N}$}
        \State connectivity$(\ell) \gets |\{u \in \mathcal{N} : \ell \in u\}|$
    \EndFor
    \State $\ell^* \gets \arg\max_{\ell}$ connectivity$(\ell)$
    
    \State
    \State $\mathbf{h}_{\ell^*} \gets$ \textsc{PhasePosition}$(\ell^*, S^{\text{phased}}, H^*, \mathcal{N}, \{\phi_u^*\}, \mathcal{R})$
    
    \State
    \State $H^*[\cdot, \ell^*] \gets \mathbf{h}_{\ell^*}$
    \State $S^{\text{phased}} \gets S^{\text{phased}} \cup \{\ell^*\}$, $S^{\text{unphased}} \gets S^{\text{unphased}} \setminus \{\ell^*\}$
    \State $\mathbf{b}[\ell^*] \gets$ block\_id
    \State
    \For{each $u \in \mathcal{N}$ where $\ell^* \in u$}
        \State $U^{\text{phased}} \gets U^{\text{phased}} \cup \{u\}$, $U^{\text{unphased}} \gets U^{\text{unphased}} \setminus \{u\}$
    \EndFor
\EndWhile

\State
\Return $H^*, \mathbf{b}$
\end{algorithmic}
\end{algorithm}

\clearpage

\subsection{Phasing Selected Variant}
\label{sec:alg2}

This algorithm determines the haplotype assignment for a selected position $\ell^*$ by evaluating candidate phasings based on read evidence and consistency with already-phased positions.

\subsubsection{Candidate Generation.}
A candidate phasing is a haplotype matrix $\mathbf{C} \in \{0,1\}^{K \times |\mathcal{V}|}$ where $\mathcal{V}$ contains position $\ell^*$ and a subset of positions from $S^{\text{phased}}$. 
The size of $\mathcal{V}$ varies by candidate depending on which nodes are used for generation. 
Candidates are constructed from the potentials on the edges that connect the nodes in the neighbor set $\mathcal{N}$ (unphased nodes adjacent to phased nodes in $Q$) that contain position $\ell^*$ to the phased node set and the nodes potentials.

For each node $u_{\ell_i\ell_j} \in \mathcal{N}$ where $\ell^* \in \{\ell_i, \ell_j\}$, we extract phasings $\phi_{\ell_i\ell_j}$ from the node's state space $\Phi_{\ell_i\ell_j}$ that satisfy two conditions: (1) the phasing has non-zero node potential $f(\phi_{\ell_i\ell_j}) > 0$, indicating support from at least one read, and (2) some permutations of the phasing are consistent with the already-assigned haplotypes at any shared positions in $S^{\text{phased}} \cap \{\ell_i, \ell_j\}$. %, accounting for haplotype permutations. 
When multiple nodes in $\mathcal{N}$ contain $\ell^*$, we combine their phasings to form candidates that maintain consistency across all shared positions under appropriate permutation alignments.
Each resulting candidate assigns values to position $\ell^*$ along with positions from the nodes used in its construction. % all while respecting genotype constraints at every position. We dont have to say it because any phasing already respected the genotype in graph construction.

\subsubsection{Candidate Scoring.}
Each candidate $\mathbf{C}$ is evaluated using three metrics, which are normalized to $[0,1]$ within each iteration and combined with weights summing to 1.

\textbf{Likelihood ($L(\mathbf{C})$):} Measures how well candidate $\mathbf{C}$ explains observed reads covering positions in $\mathcal{V}$. For each read $r$ covering at least one position in $\mathcal{V}$, we compute the likelihood as defined in Section~\ref{sec:method}, generalized to the positions in $\mathcal{V}$:
$$
L(\mathbf{C}) = \sum_{r \in \mathcal{R}(\mathcal{V})} \log \sum_{k=1}^{K} \varepsilon^{d(\mathbf{C}^{(k)}, r[\mathcal{V}])}(1-\varepsilon)^{|\text{cov}(r)|-d(\mathbf{C}^{(k)}, r[\mathcal{V}])},
$$
where $\mathbf{C}^{(k)}$ is haplotype $k$ in the candidate, $r[\mathcal{V}]$ are the observed alleles in read $r$ at positions $\mathcal{V}$, $d(\cdot,\cdot)$ is Hamming distance over covered positions, $|\text{cov}(r)|$ is the number of positions in $\mathcal{V}$ covered by $r$, and $\varepsilon$ is the sequencing error rate.

\textbf{Minimum Error Correction ($M(\mathbf{C})$):} Counts the minimum number of read alleles that must be corrected to achieve consistency:
$$
M(\mathbf{C}) = \sum_{r \in \mathcal{R}(\mathcal{V})} \min_{k=1,\ldots,K} d(\mathbf{C}^{(k)}, r[\mathcal{V}]).
$$
% Lower MEC indicates greater parsimony with the observed reads.

\textbf{Inference Agreement ($F(\mathbf{C})$):} Counts how many nodes in $\mathcal{N}$ have their inferred phasings $\phi_u^*$ (from Viterbi or FFBS) matched by the candidate under haplotype permutation. %:
%$$
%F(\mathbf{C}) = \left|\left\{u \in \mathcal{N} : \exists \text{ permutation } \pi, \mathbf{C}[\pi, u] = \phi_u^*\right\}\right|.
%$$

After normalizing each metric to the range $[0,1]$ by min-max scaling within the current candidate set, the final score is:
$$
S(\mathbf{C}) = w_L \cdot \tilde{L}(\mathbf{C}) - w_M \cdot \tilde{M}(\mathbf{C}) + w_F \cdot \tilde{F}(\mathbf{C}),
$$
where $\tilde{L}$, $\tilde{M}$, $\tilde{F}$ denote normalized scores, weights satisfy $w_L + w_M + w_F = 1$, and default values are $w_L = 1/12$, $w_M = 10/12$, $w_F = 1/12$. 
The candidate maximizing $S(\mathbf{C})$ is selected, and its column for position $\ell^*$ is assigned to the global haplotype matrix. % according to the .

\paragraph{Fallback Strategy.}
If no valid candidates can be generated due to sparse coverage or lack of connecting nodes, the algorithm uses the inferred phasing $\phi_u^*[\ell^*]$ from any node $u$ containing $\ell^*$. 
This ensures all positions receive assignments while maintaining consistency with the probabilistic inference in regions where read evidence is uninformative.

\begin{algorithm}[H]
\caption{\textsc{PhasePosition}: Phase Selected Position}
\label{alg:phase_position}
\begin{algorithmic}[1]
\Require Selected position $\ell^*$, phased positions $S^{\text{phased}}$, global haplotypes $H^*$, neighbor nodes $\mathcal{N}$, inferred phasings $\{\phi_u^*\}$, reads $\mathcal{R}$, weights $(w_L, w_M, w_F)$
\Ensure Haplotype column $\mathbf{h}_{\ell^*} \in \{0,1\}^K$

\State \textbf{Generate Candidates:}
\State candidates $\gets \emptyset$
\For{each node $u \in \mathcal{N}$ where $\ell^* \in u$}
    \For{each phasing $\phi_u$ with $f(\phi_u) > 0$}
        \If{$\exists$ permutation making $\phi_u$ consistent with $H^*[S^{\text{phased}}]$}
            \State Construct candidate $\mathbf{C}$ from $\phi_u$ and edge potentials
            \State candidates $\gets$ candidates $\cup \{\mathbf{C}\}$
        \EndIf
    \EndFor
\EndFor

\State
\If{candidates $= \emptyset$}
    \State \Return $\phi_u^*[\ell^*]$ for any $u \in \mathcal{N}$ containing $\ell^*$
\EndIf

\State
\State \textbf{Score Candidates:}
\For{each $\mathbf{C}$ in candidates}
    \State Compute $L(\mathbf{C})$, $M(\mathbf{C})$, $F(\mathbf{C})$
    \State Normalize to $\tilde{L}$, $\tilde{M}$, $\tilde{F} \in [0,1]$
    \State $S(\mathbf{C}) \gets w_L \cdot \tilde{L} - w_M \cdot \tilde{M} + w_F \cdot \tilde{F}$
\EndFor

\State
\State \Return column for $\ell^*$ from $\arg\max_{\mathbf{C}} S(\mathbf{C})$
\end{algorithmic}
\end{algorithm}

\newpage

% \section{Appendix \thesection: Simulated Data}
\section{Simulated Data}
\label{sec:suppsim}
\subsection{Polyploid Genomes Simulated from Potato Reference}
Using a reference haplotype from the \textit{S. tuberosum} dataset \cite{Sun2021tuberosum}, we simulate full genomes to emulate autopolyploid and allopolyploid configurations. We adapt a version of Haplogenerator from the HaploSim v1.8 software package \cite{Motazedi2018Haplosim}. Haplogenerator generates random mutations on a reference haplotype using a certain probabilistic model (we choose Poisson) with specific mutation rate parameters for the model. We specify mutation rates (\texttt{-s [$\mu$, 0, 0]}) such that $\mu$ is the rate of a single nucleotide mutation, and there are no insertions or deletions.

\subsubsection{Autopolyploid}
\label{sec:simauto}
In autopolyploid genomes, haplotypes are derived from the same species \cite{van2017autoallo}. To generate data for the autopolyploid case, we consider configurations of ploidies 2 (simple diploid, not polyploid), 3, 4, and 6. For each ploidy, $k$, we generate $k$ haplotypes from the original \textit{S. tuberosum} haplotype, each with mutation rate, $\mu$, where $\mu \in \{0.001, 0.005, 0.01\}$ (each $\mu$ is a separate configuration). We do this 20 times per ploidy-mutation rate configuration to generate 20 samples.

\subsubsection{Allopolyploid}
\label{sec:simallo}
Allopolyploid genomes occur when divergent species merge and their genomes are hybridized \cite{van2017autoallo}. In the allopolyploid case, first we generate two configurations of subgenomes, each set with divergent haplotypes $A$, $B$, and $C$ representing the ancestors of each subgenome. These subgenomes are simulated from the original haplotype with mutation rates $\mu_{\text{sub}} \in \{0.0001, 0.0005\}$. Given these subgenome configurations, we simulate further configurations of haplotypes with mutation rates $\mu_{\text{within}} \in \{0.00005, 0.0001\}$ from these diverged subgenome haplotypes. We examine three different allopolyploid structures: a triploid organism with unbalanced subgenome structure (AAB), a tetraploid organism with a balanced structure of two subgenomes (AABB), and a hexaploid organism with a balanced structure of three subgenomes (AABBCC)(Figure~\ref{fig:allo-snp-density}).

\subsection{Read Simulation}
\label{sec:simread}
For each simulated polyploid genome configuration, we generated both short-read and long-read sequencing data at multiple coverage depths to evaluate phasing performance across different sequencing technologies and data availability scenarios.

\subsubsection{Short-read Simulation (Illumina)}
Short paired-end reads were simulated using ART~\cite{huang2012art} (v2.5.8) with the HiSeq 2500 error profile (\texttt{-ss HS25}), which models Illumina sequencing characteristics including position-specific quality scores and systematic errors. 
We configured ART with the following parameters: read length of 125 bp (\texttt{-l 125}), mean fragment size of 350 bp (\texttt{-m 350}), standard deviation of 50 bp (\texttt{-s 50}), and coverage depths of 3×, 5×, 10×, 20×, and 40× per haplotype for autopolyploid short-read data, and 5×, 10×, 20×, and 40× for other configurations. % allopolyploid data. 
The target coverage was distributed equally across all haplotypes (e.g., for a tetraploid genome at 20× total coverage, each haplotype received 5× coverage). 
These parameters represent typical Illumina paired-end sequencing protocols with insert sizes suitable for standard library preparation. 
The simulated reads maintain consistent quality metrics across all configurations (Figure ~\ref{fig:auto-shortread}).
In total, we generated 1200 short-read samples for autopolyploid data, and 480 for short-read allopolyploid data (Table~\ref{tab:simdatasetconf}).

\subsubsection{Long-read Simulation (Oxford Nanopore)}
Long-read data were simulated using PBSIM3~\cite{ono2022pbsim3} with the QSHMM-ONT-HQ error model (\texttt{--qshmm QSHMM-ONT-HQ.model}), which accurately replicates Oxford Nanopore Technologies (ONT) high-quality read characteristics including length distribution, error profiles, and base quality scores typical of modern R10.4 chemistry. 
The simulation used whole-genome strategy (\texttt{--strategy wgs}) at coverage depths of 5×, 10×, 20×, and 40× per haplotype. 
The QSHMM-ONT-HQ model produces reads with mean base quality of Q13-14, consistent with contemporary ONT sequencing, and includes realistic error modes dominated by insertion and deletion errors in homopolymer regions. 
We generated 960 and 480 long-read samples for autopolyploid and allopolyploid data (Table~\ref{tab:simdatasetconf}, Figures~\ref{fig:auto-longread},~\ref{fig:read_qual}).

\subsubsection{Read Alignment and Fragment Extraction}

\begin{figure}[!ht]
    \centering
    \includegraphics[width=0.8\linewidth]{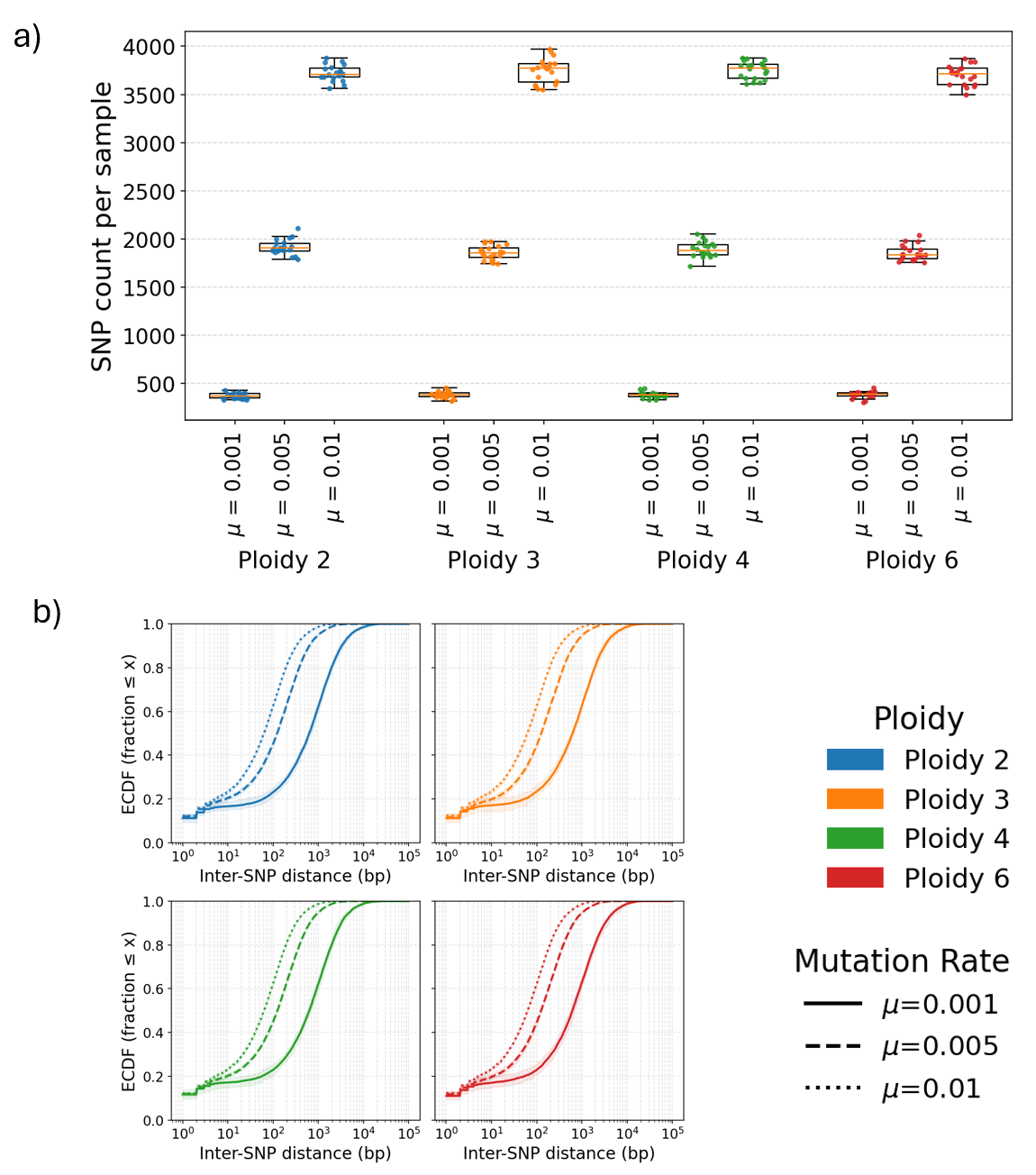}

    \caption{
    \textbf{Autopolyploid SNP density and spacing characteristics.}
    (a) SNP counts per simulated sample across ploidy levels (2, 3, 4, 6) and mutation rates 
    $\mu \in \{0.001, 0.005, 0.01\}$. 
    Boxplots summarize variation across replicates, and points show individual samples,
    colored by ploidy. 
    Higher mutation rates lead to increased SNP counts, while differences due to ploidy are more modest.
    (b) Empirical cumulative distribution functions (ECDFs) of inter-SNP distances on a $\log_{10}$ genomic scale. 
    Each curve reflects the mean ECDF across samples for a given ploidy and mutation rate, with shaded bands showing the 
    $10-90\%$ range across replicates. 
    }
    \label{fig:auto-snp-density}
\end{figure}

\begin{figure}[!ht]
    \centering
    \includegraphics[width=0.8\linewidth]{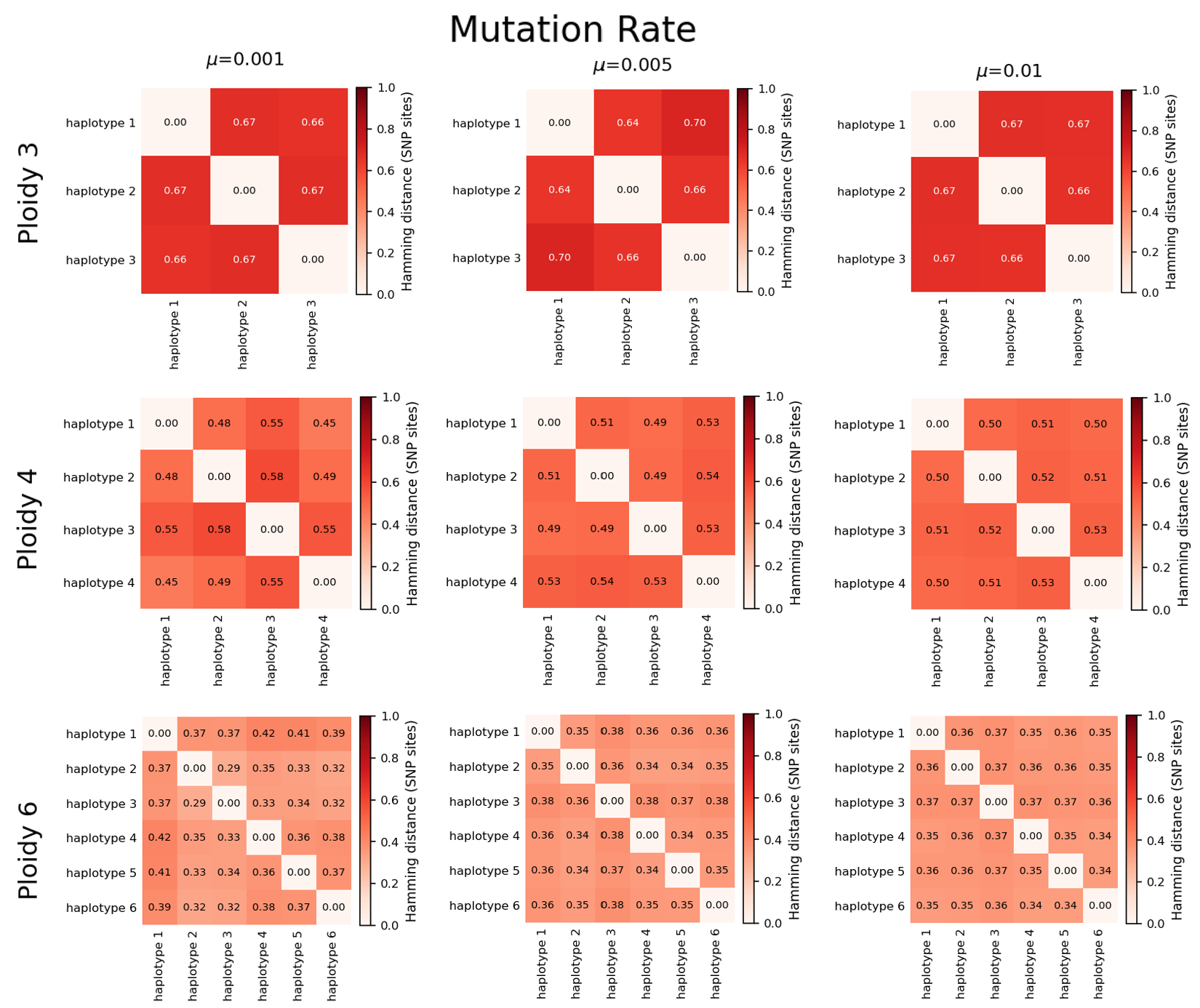}

    \caption{\textbf{Haplotype pairwise sequence similarity for autopolyploid simulations.} Hamming distance heatmaps showing pairwise distances between simulated haplotypes for autopolyploid configurations (Ploidy 3, 4, 6) across three mutation rates ($\mu \in \{0.001, 0.005, 0.01\}$) for one sample. Values closer to 0 indicate high sequence similarity between haplotypes, while higher values indicate greater divergence. }
    \label{fig:auto-heatmap}
\end{figure}

\begin{figure}[!ht]
    \centering
    \includegraphics[width=0.8\linewidth]{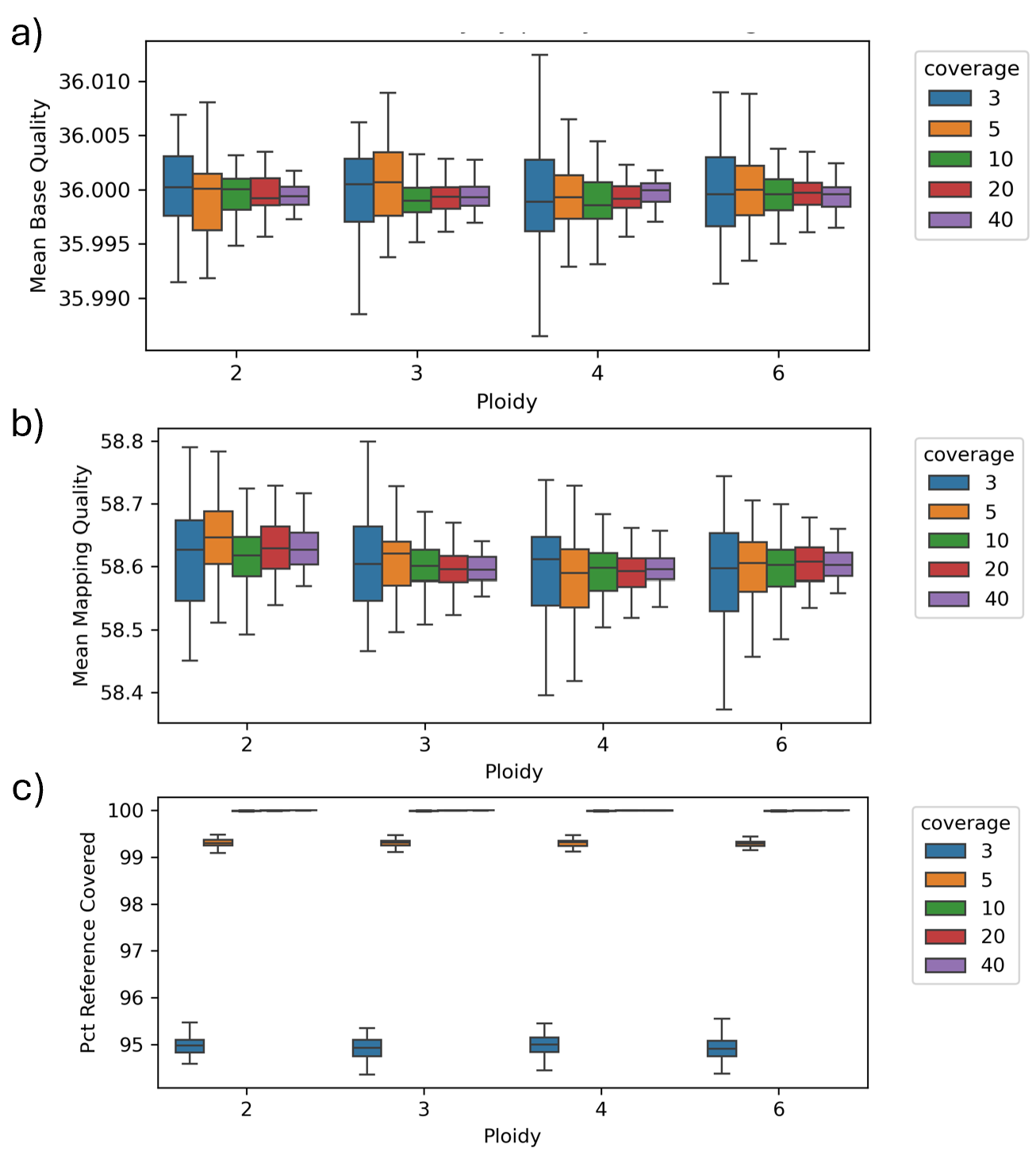}

    \caption{\textbf{Read and mapping quality characteristics for autopolyploid short-read simulations.} (a) Mean base quality scores, (b) mean mapping quality scores, and (c) percentage of reference genome covered, stratified by ploidy (2, 3, 4, 6) and sequencing coverage (3×–40×). Quality metrics remain stable across ploidies and coverage levels, confirming consistent read simulation across all configurations.}
    \label{fig:auto-shortread}
\end{figure}

\begin{figure}[!ht]
    \centering
    \includegraphics[width=0.8\linewidth]{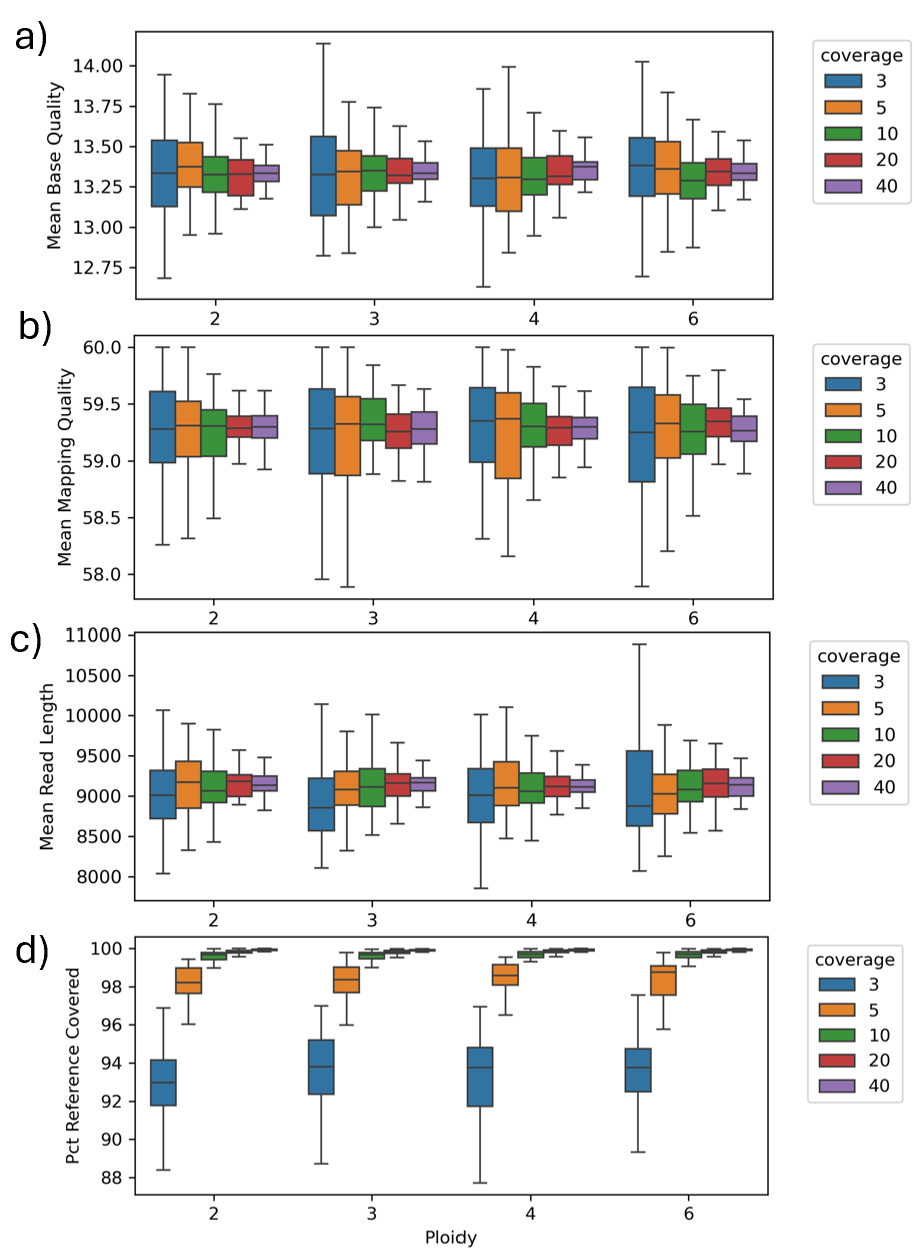}

    \caption{\textbf{Read metrics for autopolyploid long-read simulations.} (a) Mean base quality, (b) mean mapping quality, (c) mean read length, and (d) percentage of reference covered for autopolyploid samples across ploidies and coverages. }
    \label{fig:auto-longread}
\end{figure}

\begin{figure}[!ht]
    \centering
    \includegraphics[width=0.8\linewidth]{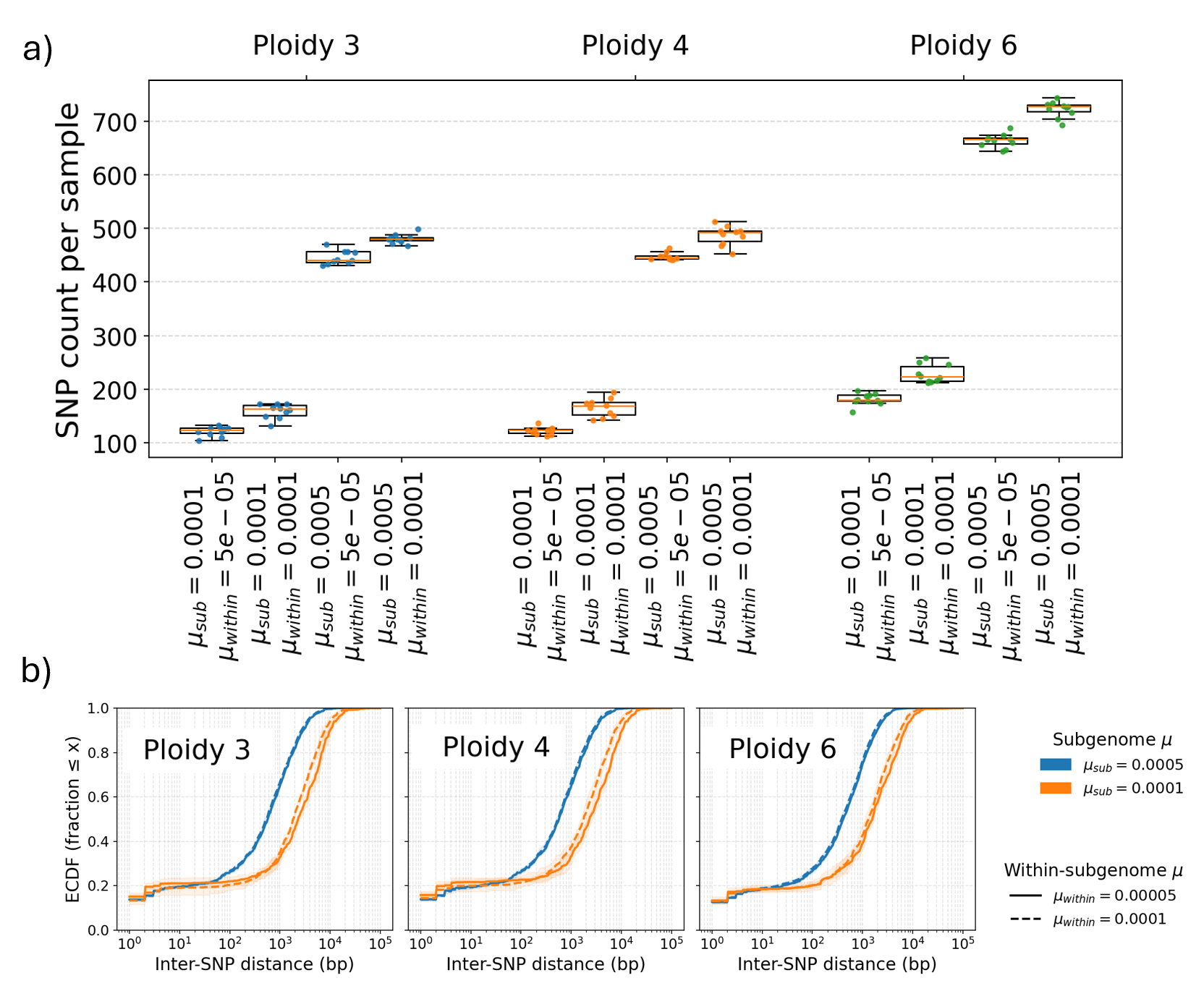}
    \caption{\textbf{Allopolyploid SNP density and spacing characteristics.} (a) SNP counts per simulated sample across ploidy levels (3: AAB, 4: AABB, 6: AABBCC) and inter-subgenome/within-subgenome mutation rate combinations. (b) Empirical cumulative distribution functions of inter-SNP distances on a $\log_{10}$ genomic scale, showing distinct spacing patterns for subgenome (higher divergence) versus within-subgenome (lower divergence) SNPs.}
    \label{fig:allo-snp-density}
\end{figure}

\begin{figure}[!ht]
    \centering
    \includegraphics[width=0.8\linewidth]{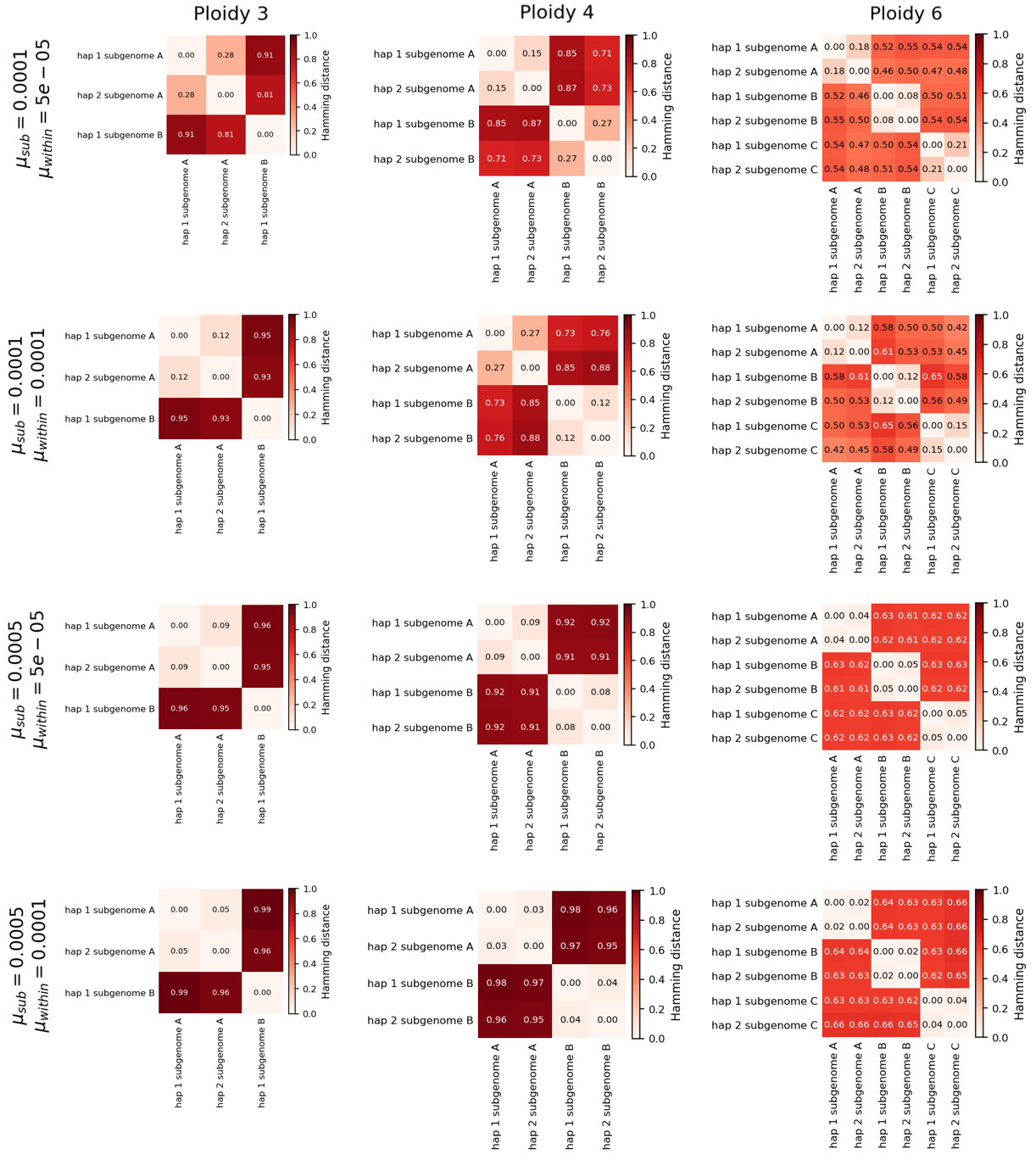}

    \caption{\textbf{Pairwise haplotype Hamming distances for allopolyploid simulations.} Columns: ploidies 3-6. Rows: inter-subgenome ($\mu_{sub}$) and within-subgenome ($\mu_{within}$) mutation rates. Block structure reflects subgenome organization.}
    \label{fig:allo-heatmap}
\end{figure}

\begin{figure}[!ht]
    \centering
    \includegraphics[width=0.8\linewidth]{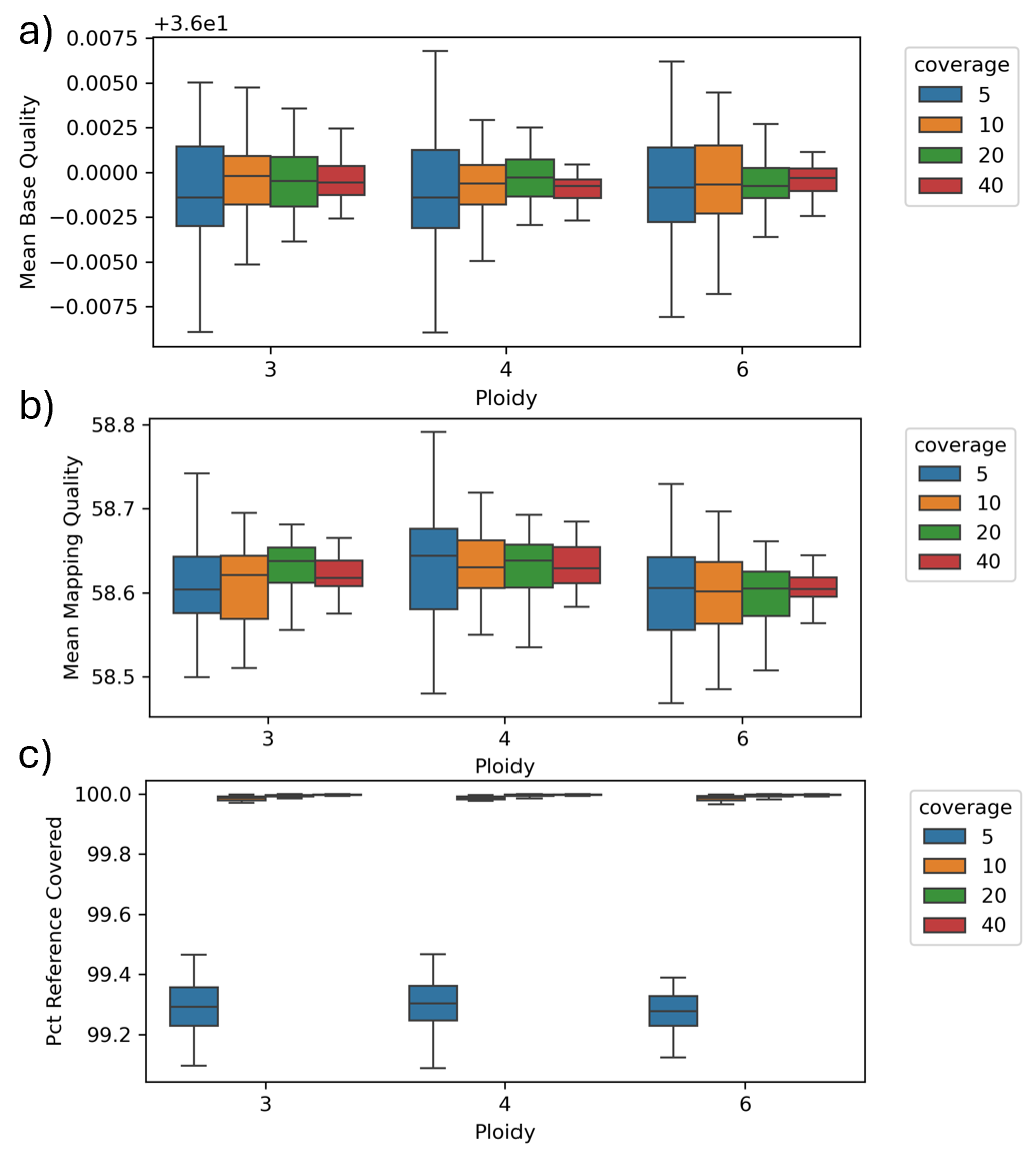}
    \caption{\textbf{Read and mapping quality metrics for allopolyploid short-read simulations.} (a) Mean base quality scores, (b) mean mapping quality scores, and (c) percentage of reference genome covered by ploidy (3, 4, 6) and sequencing coverage (5×-40×). }
    \label{fig:read_qual}
\end{figure}

\begin{figure}[!ht]
    \centering
    \includegraphics[width=0.8\linewidth]{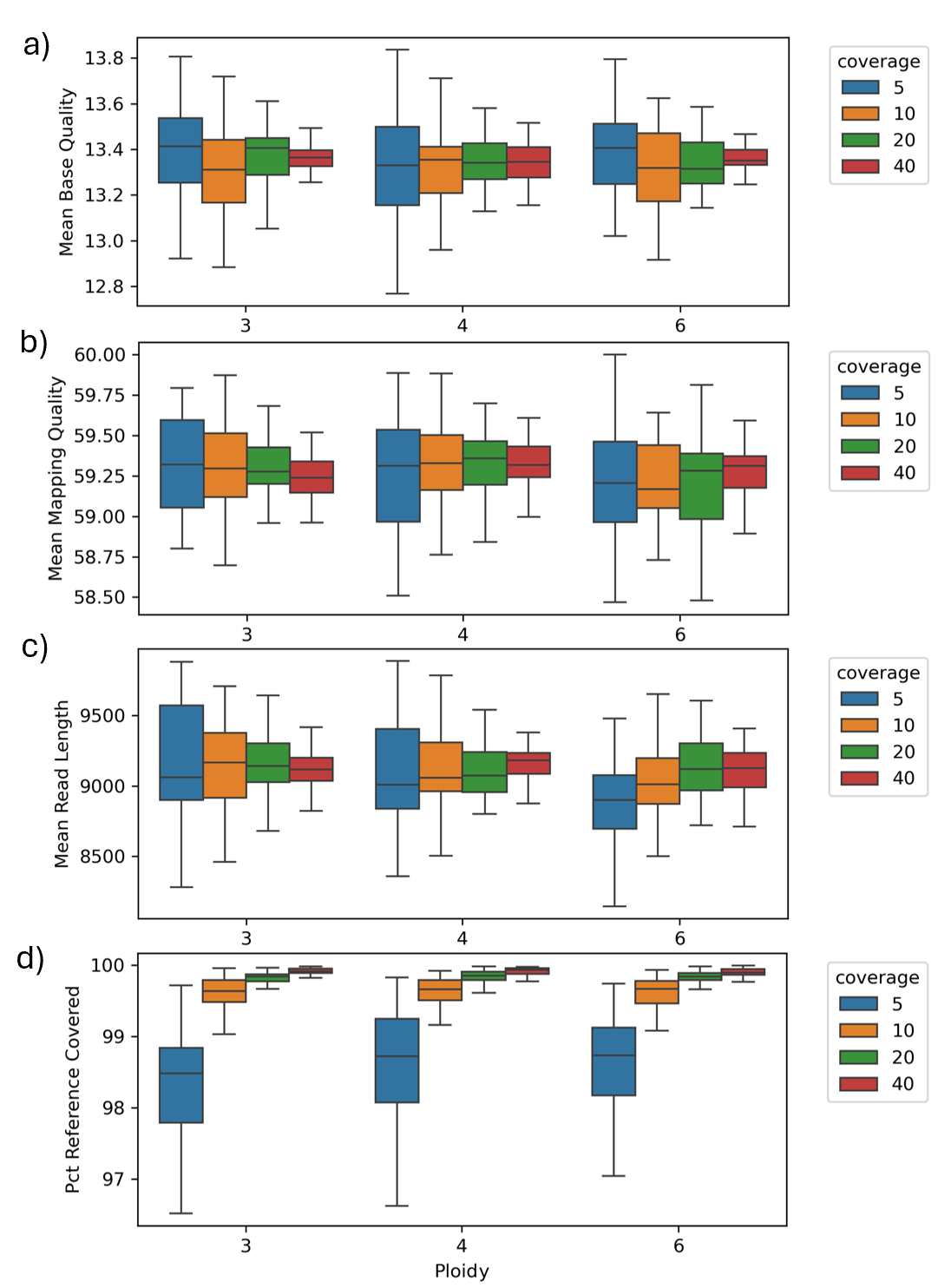}

    \caption{\textbf{Quality metrics for allopolyploid long-read simulations.} (a) base quality, (b) mapping quality, (c) read length, and (d) reference coverage across ploidies and coverages. Stable metrics confirm realistic ONT simulation.}
    \label{fig:allo-longread}
\end{figure}

Short reads were aligned to the reference genome using BWA-MEM (v0.7.17) with default parameters, and long reads were aligned using Minimap2 (v2.24) with the ONT preset (\texttt{-x map-ont}). Alignments were sorted and indexed using SAMtools (v1.15).
Then haplotype-informative fragments were extracted from aligned reads using extractHAIRS from the Hap10~\cite{majidian2020hap10} suite.
For short-read data, we used the default minimum base quality threshold of Q13 (\texttt{--mbq 13}), which is standard for high-quality Illumina data and filters bases with >95\% accuracy. 
For long-read ONT data, we set a more permissive threshold of Q4 (\texttt{--mbq 4}) to account for the characteristically lower per-base quality of ONT sequencing (mean Q13-14). 
% While Q4 retains bases with ~60\% accuracy, this threshold is justified for long-read haplotyping because: (1) read length and connectivity compensate for lower per-base accuracy through multiple overlapping observations, (2) overly stringent filtering (e.g., Q13) would discard ~50\% of bases and severely fragment the haplotype signal, and (3) the threshold allows maximum utilization of long-range information while excluding only the lowest-quality calls. 
% This approach aligns with ONT-specific tools like Longshot (default Q7) and is more conservative than the minimum allowed by extractHAIRS (Q4). 
(Tables \ref{tab:simdatasetconf} and Figures \ref{fig:auto-snp-density} to \ref{fig:allo-longread}.)

\begin{table}[h!]
\caption{Datasets Configurations}
\begin{tabular}{cccccc}
\hline
Dataset                                                                & Ploidy                                                                      & Mutation Rate                                                                                    & Coverage             & \# Samples & Total Samples \\ \hline
\begin{tabular}[c]{@{}c@{}}Autopolyploidy \\ (Short-read)\end{tabular} & 4 (2, 3, 4, 6)                                                              & 3 (0.001, 0.005, 0.01)                                                                            & 5 (3, 5, 10, 20, 40) & 20      & 1200          \\ \hline
\begin{tabular}[c]{@{}c@{}}Autopolyploidy \\ (Long-read)\end{tabular}  & 4 (2, 3, 4, 6)                                                              & 3 (0.001, 0.005, 0.01)                                                                            & 4 (5, 10, 20, 40)    & 20      & 960           \\ \hline
\begin{tabular}[c]{@{}c@{}}Allopolyploidy\\ (Short-read)\end{tabular} & \begin{tabular}[c]{@{}c@{}}3 (3: AAB, \\ 4: AABB, \\ 6:AABBCC)\end{tabular} & \begin{tabular}[c]{@{}c@{}}4: (0.0001, 0.0005) \\ combine with\\ (0.00005, 0.0001)\end{tabular} & 4 (5, 10, 20, 40)    & 10      & 480           \\ \hline
\begin{tabular}[c]{@{}c@{}}Allopolyploidy\\ (Long-read)\end{tabular}  & \begin{tabular}[c]{@{}c@{}}3 (3: AAB, \\ 4: AABB, \\ 6:AABBCC)\end{tabular} & \begin{tabular}[c]{@{}c@{}}4: (0.0001, 0.0005) \\ combine with\\ (0.00005, 0.0001)\end{tabular} & 4 (5, 10, 20, 40)    & 10      & 480           \\ \hline
\end{tabular}
\label{tab:simdatasetconf}
\end{table}

\subsection{Additional Experimental Data Preprocessing Details}
\label{res:expprocess}
Read quality was assessed using FASTQC v0.12.1 and multiqc v1.24.1~\cite{simon2010fastqc,ewels2016multiqc}.
Raw reads were subsequently adapter- and quality-trimmed using fastp v0.23.4 with the following quality filtering parameters: \texttt{--trim\_poly\_x} \texttt{--length\_required 30}.
The coordinate-sorted BAM files were filtered for a minimum mapping quality (MAPQ) of 20 using samtools v1.20. 
Using bcftools v1.21, we retained only heterozygous biallelic variants with a minimum read depth (FORMAT/DP) of 10 and a minimum primary and alternate allele mapping quality of 20 (INFO/MQM and INFO/MQMR, respectively)~\cite{garrison2012haplotype}. 
Multiallelic variants, multi-nucleotide polymorphisms (MNPs), and indels were excluded.
%~\cite{schrider2011pervasive}.

\begin{table}[h]
\centering
\caption{Variant statistics for octoploid strawberry sample 11C151P008 before and after filtering. Multiallelic SNPs and multi-nucleotide polymorphism (MNPs) constitute only 2.2\% of raw variants (154,259 out of 7,020,992) and are excluded to maintain biallelic assumptions. 
Multiallelic SNP and MNP counts are likely inflated due to homeologous alignments from the four Fragaria subgenomes to the representative subgenome assembly.
All multiallelic variants, MNP, and indels are removed during quality filtering. }
\label{tab:variant_filtering}
\begin{tabular}{lc}
\toprule
\textbf{Variant Type} & \textbf{11C151P008} \\
\midrule
\multicolumn{2}{l}{\textit{Raw variants}} \\
\quad Biallelic SNPs & 6,865,733 \\
\quad Multiallelic SNPs & 154,259 \\
\quad MNPs & 1,654,847 \\
\quad Indels & 1,001,831 \\
\midrule
\multicolumn{2}{l}{\textit{Filtered variants}} \\
\quad Biallelic SNPs & 5,483,977 \\
\quad Multiallelic SNPs & 0 \\
\quad MNPs & 0 \\
\quad Indels & 0 \\
\bottomrule
\end{tabular}
\end{table}

\subsection{Evaluation Criteria}
\label{sup:ec}
\textbf{\textit{Generalized vector error rate block types 3 and 4.}}
For blocks of type 4, we approximate the ``expected'' vector error rate by considering all of the valid bijections at each SNP position besides the first uniformly and compute the expected number of non-fixed points from an arbitrary but fixed permutation considered at the previous position, and sum over positions besides the first. We remark that a valid matching can be viewed as the combination of any permutation of the 0 alleles with any permutation of the 1 alleles; a specific consequence of this is that if there are one or fewer of a certain allele then the permutation of that allele type is over one or fewer elements so we cannot have any moved points. Using a well-known result from algebra, we have that the expected number of fixed points of a permutation of two or more elements is 1, so we get that the expected vector error over an empty block is
\begin{align}
    \sum_{\ell=2}^{L} \left( K - \mathbb{I}\{g[\ell] > 1\} - \mathbb{I}\{K - g[\ell] > 1\} \right) ,
\end{align}
where $\mathbb{I}\{g[\ell] > 1\}$ and $\mathbb{I}\{K - g[\ell] > 1\}$ are the number of haplotypes that we expect \textit{not} to contribute vector errors from the 1 alleles and 0 alleles respectively.

For blocks of type 3, let $\tilde{K},~1 \leq \tilde{K} < K$ be the number of haplotypes phased within the block. We first find the actual vector error by taking the best matching over what has been phased with the ground truth, where each $\phi_\ell : \{1, \ldots, \tilde{K}\} \to \setK$, obviously not surjective. Then, we take the remaining $K-\tilde{K}$ haplotypes left unphased, note the number of each allele type left over (reference and alternate), and compute the expected number of vector errors as we did for type 4 blocks, and report the total block vector error as the sum of the actual and expected contributions.

\newpage

\subsection{Autopolyploidy Data}

\subsubsection{SNP Counts and Inter-SNP Distances}
Here we plot the SNP count information among different ploidies and mutation rates in autopolyploidy data, ranging from almost 500 SNPs to less than 4000 SNPs per sample.
We also plot the SNP density in the simulated data.
Increasing mutation rate decreases inter-SNP gap sizes, while ploidy influences the spacing more weakly. 
Together, panels (a) and (b) characterize how mutation rate and ploidy shape SNP abundance and genomic distribution in simulated autopolyploid genomes.

\clearpage
\newpage
\subsubsection{Heatmaps}
Hamming distance heatmaps showing pairwise distances between simulated haplotypes for autopolyploid configurations (Ploidy 3, 4, 6) across three mutation rates ($\mu \in \{0.001, 0.005, 0.01\}$) for one sample. Values closer to 0 indicate high sequence similarity between haplotypes, while higher values indicate greater divergence. 
% High within-ploidy similarity at low mutation rates demonstrates the challenge of polyploid haplotype assembly when haplotypes share extensive genomic segments.

\clearpage
\subsubsection{Short-read Data Statistics}
The plots show the statistics of the simulated short-read sequencing data for autopolyploidy.

\clearpage
\subsubsection{Long-read Data Statistics}
The plots show the statistics of the simulated long-read sequencing data for autopolyploidy.

\clearpage
\subsection{Allopolyploidy Data}
Here we provid detailed information about the allopolyploidy data:
\subsubsection{SNP Counts and Inter-SNP Distances}
We first plot the SNP counts and density in the generated dataset.

\clearpage
\subsubsection{Heatmaps}
We plot pairwise Hamming distances between simulated allopolyploid haplotypes across ploidies 3-6 (columns) and mutation rate combinations of subgenome ($\mu_{sub}$) and within-subgenome ($\mu_{within}$) divergence (rows), where darker red indicates greater sequence divergence and block structure reflects subgenome organization.

\clearpage
\subsubsection{Short-read Data Statistics}
We validated that simulated allopolyploid short-read datasets maintain consistent quality characteristics across all configurations.
The plot shows quality metrics remain stable across ploidies and coverage levels, confirming consistent read simulation across all allopolyploid configurations.

\clearpage
\subsubsection{Long-read Data Statistics}
Allopolyploid long-read simulations show stable quality metrics across all ploidies and coverage depths, consistent with modern ONT sequencing characteristics.

\newpage

\clearpage

% \section{Appendix \thesection: Synthetic Results}
\section{Synthetic Results}
\label{sec:simressupp}
\subsection{Blocks Statistics}
\label{sec:blockstat}

\phc{}-short produces longer average blocks (a) with fewer total blocks (b) than WhatsHap and H-PoPG, particularly at higher ploidies, while maintaining comparable phasing coverage (c). 
This demonstrates that probabilistic inference maintains phasing continuity where clustering methods fragment due to read assignment ambiguity.

\begin{figure}[!ht]
    \centering
\includegraphics[width=1\linewidth]{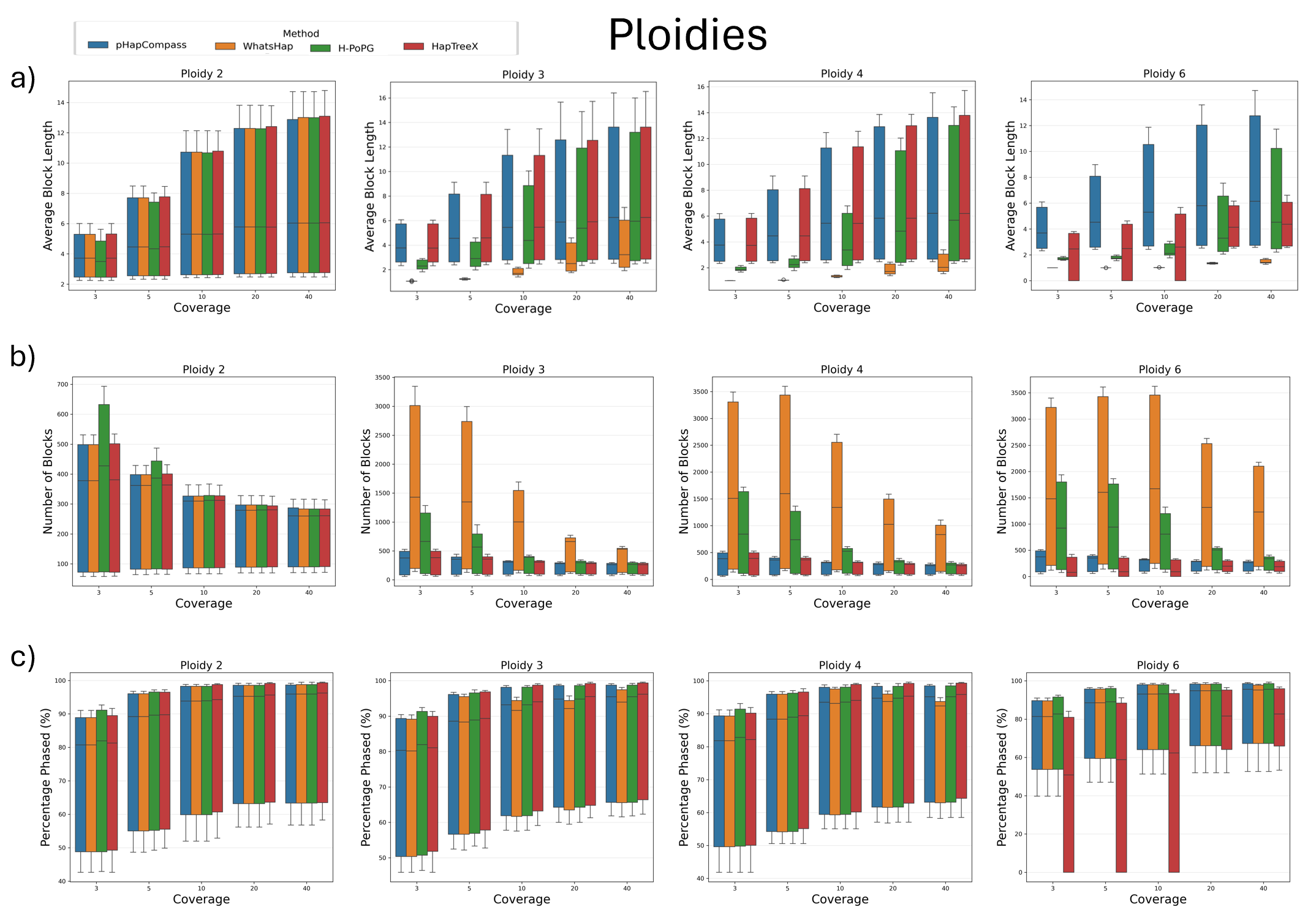}
\caption{\textbf{Autopolyploid short-read assembly block statistics.} (a) Average block length in SNPs, (b) total number of phased blocks, and (c) percentage of SNPs phased, by ploidy and coverage for \phc{}-short (blue), WhatsHap (orange), H-PoPG (green), and HapTree-X (red).}
\label{fig:auto_short_block_info}
\end{figure}

\clearpage

\phc{}-long produces single-block assemblies at high coverage across all ploidies.

\begin{figure}[!ht]
    \centering
    \includegraphics[width=1\linewidth]{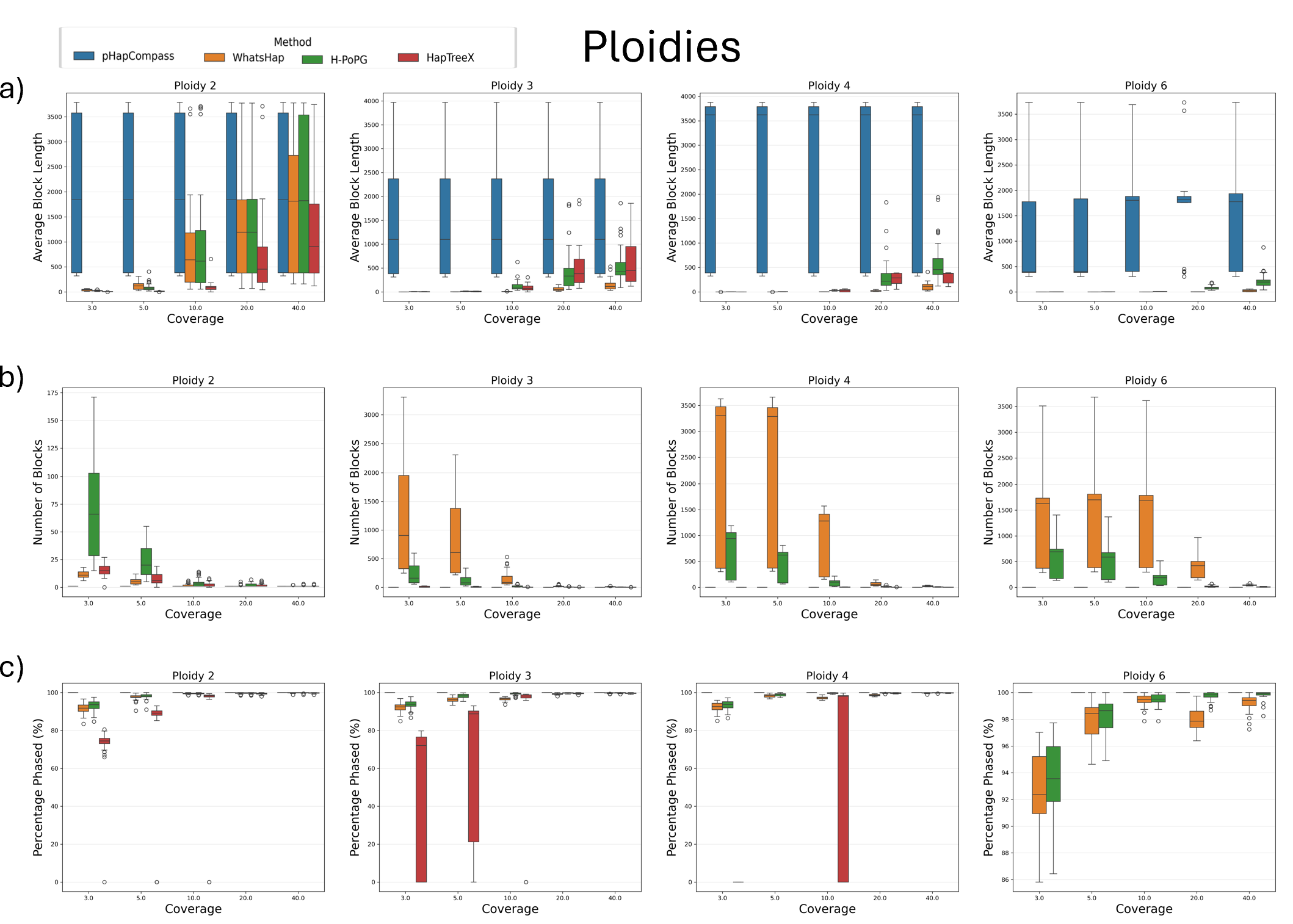}

    \caption{\textbf{Autopolyploid long-read assembly block statistics.} (a) average block length, (b) number of blocks, (c) percentage phased.}

    \label{fig:auto_long_block_info}
\end{figure}

\clearpage
\begin{figure}[!ht]
    \centering
    \includegraphics[width=1\linewidth]{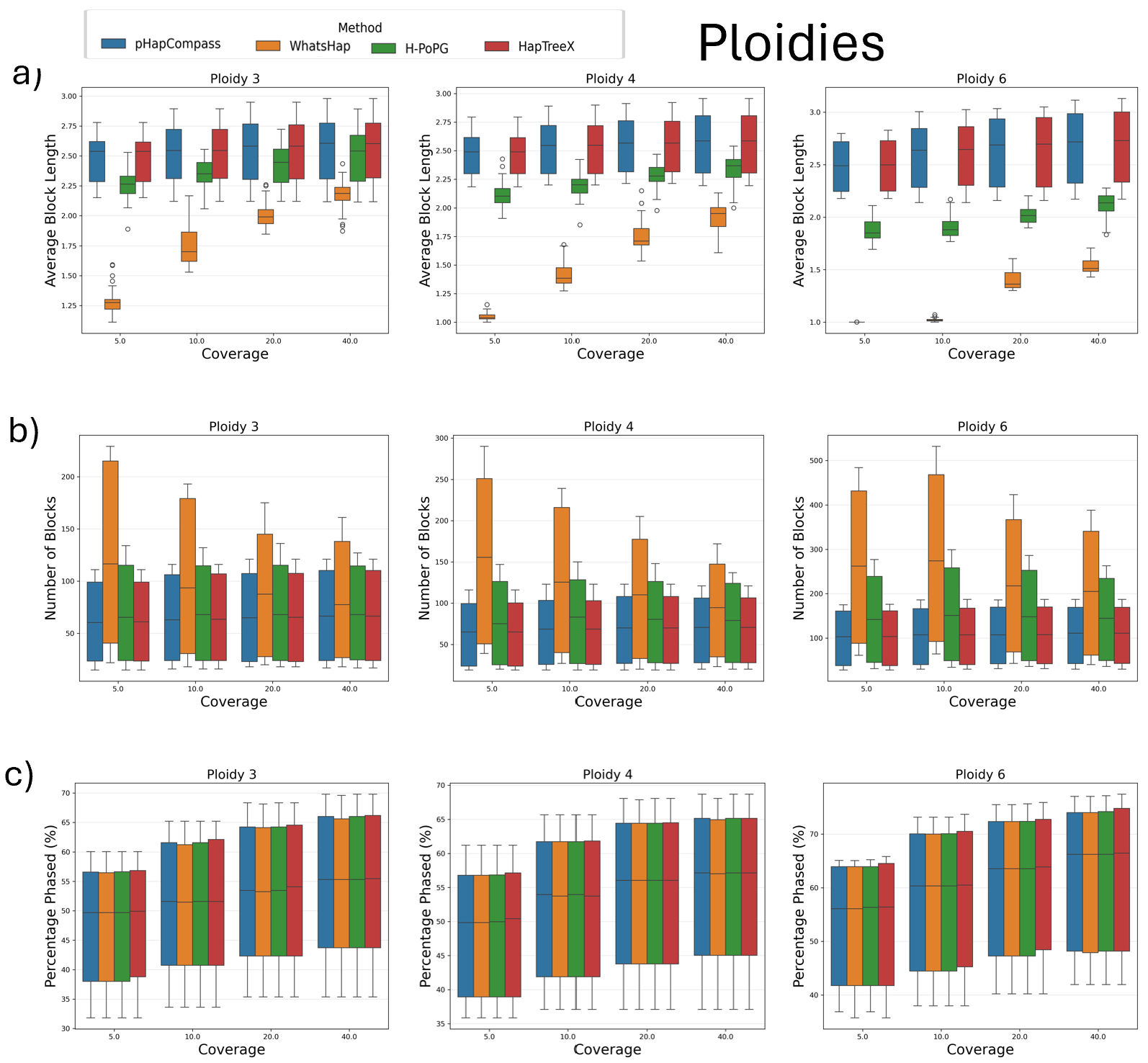}
\caption{\textbf{Allopolyploid short-read assembly block statistics.} (a) average block length, (b) number of blocks, (c) percentage phased.}
    \label{fig:allo_short_block_info}
\end{figure}

\clearpage
\begin{figure}[!ht]
    \centering
    \includegraphics[width=1\linewidth]{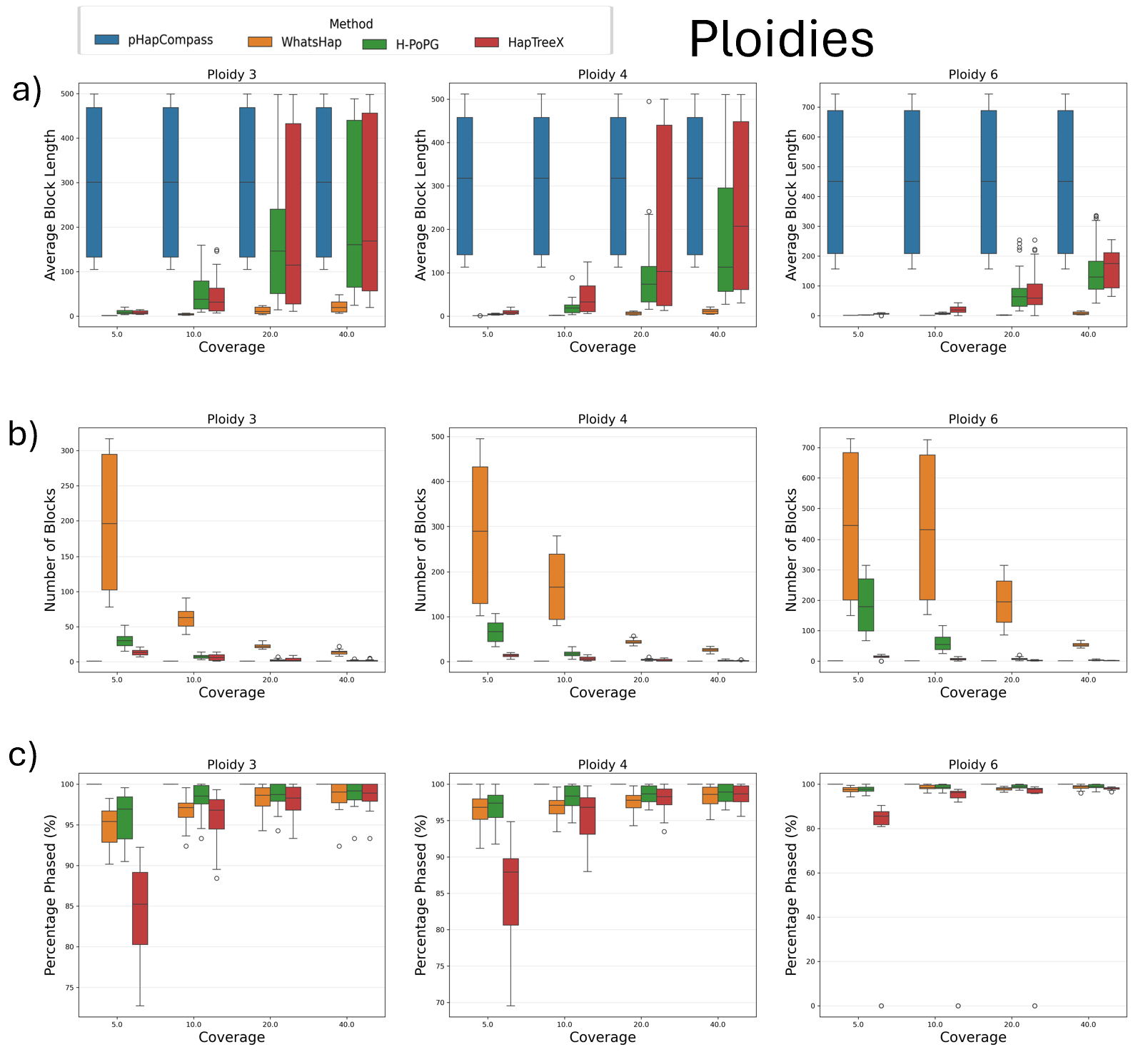}
\caption{\textbf{Allopolyploid long-read assembly block statistics.} (a) average block length, (b) number of blocks, (c) percentage phased.}
    \label{fig:allo_long_block_info}
\end{figure}

\clearpage

We additionally computed block N50 for all simulated datasets. 
Block N50 is the block length $L$ such that, the cumulative sum of sorted haplotype blocks (in descending order) is 50\% of the total phased length at $L$. 
%if you sort haplotype blocks from longest to shortest and accumulate their lengths, the cumulative sum reaches 
Larger block N50 indicates more contiguous (less fragmented) haplotype assemblies.

\begin{figure}[!ht]
    \centering
    \includegraphics[width=1\linewidth]{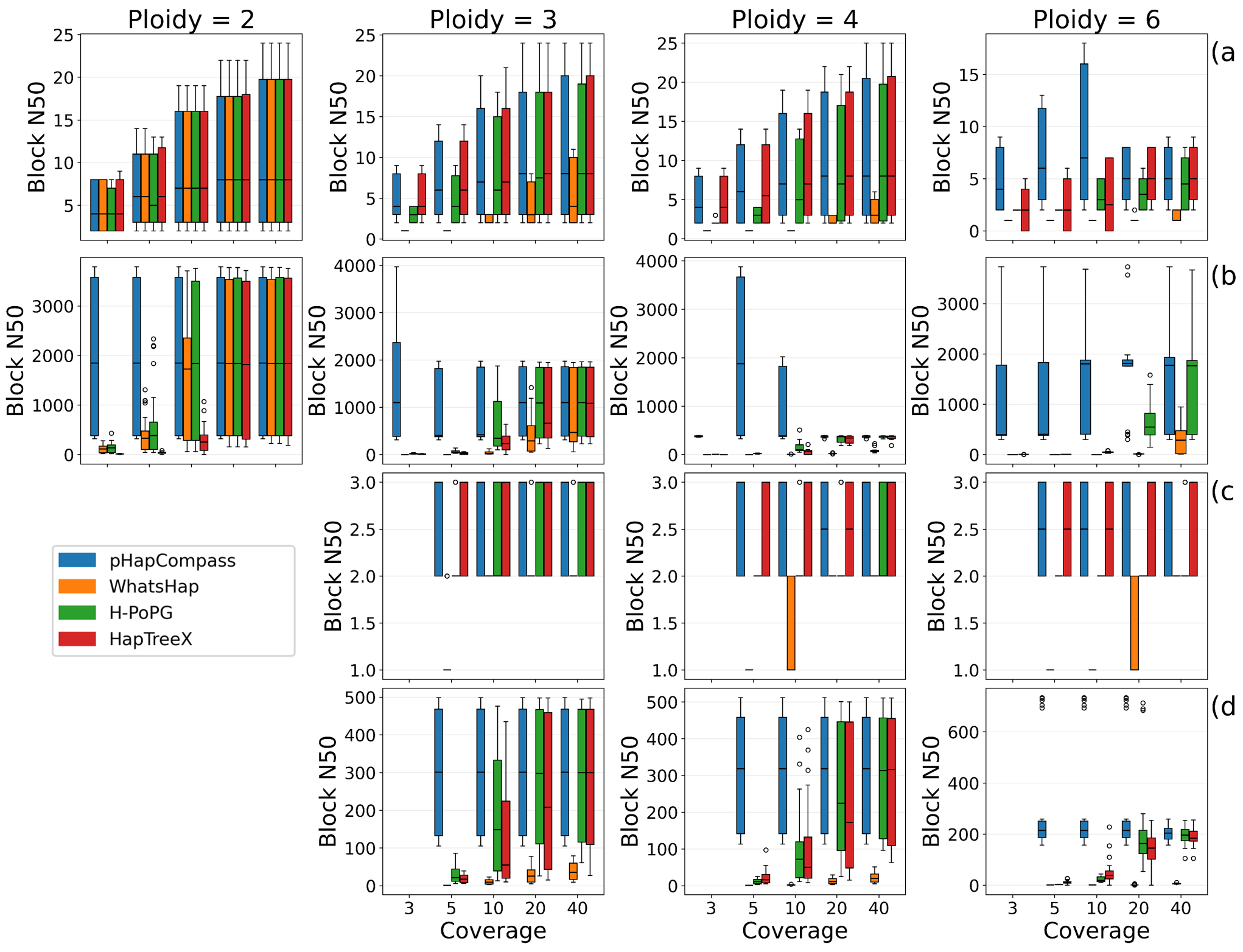}
\caption{\textbf{Block N50 for \phc{} (blue), WhatsHap (orange), H-PoPG (green), and HapTree-X (red) across ploidies and coverages.} 
Datasets include (a) autopolyploid short-reads, (b) autopolyploid long-reads, (c) allopolyploid short-reads, and (d) allopolyploid long-reads. }
    \label{fig:Block_N50_sim}
\end{figure}

\clearpage

\subsection{Detailed Synthetic Results}
\label{sec:detailed_sim_results}

We computed block-level phasing contiguity statistics for the simulated datasets (Table~\ref{tab:sim_auto_short_collapsed_ploidy2}. 
We derived phased blocks from the inferred block assignments along the ordered SNP list and computed summary measures which are then averaged over simulated samples, mutation rate, and sub-genomes (when applicable).

\begin{table*}[h]
\centering
\scriptsize
\setlength{\tabcolsep}{4pt}
\renewcommand{\arraystretch}{1.05}
\caption{Detailed haplotype assembly results for diploid autopolyploidy short-read simulated dataset. 
The values are averaged over mutation rates and samples.
The \textbf{\# Phased} column denotes the number of SNPs assigned to any block and \textbf{\% Phased} is the corresponding fraction of all SNPs ($100\times \#\text{Phased}/\#\text{SNPs}$).
The \textbf{\# Blocks} column gives the number of distinct phased blocks.
Block lengths are computed as the number of consecutive SNPs that are phased together.
\textbf{Block N50} is the block length $L$ such that, the cumulative sum of sorted haplotype blocks (in descending order) is 50\% of the total phased length at $L$. 
\textbf{Max Len.} is the length (in SNP count) of the largest phased block.}
\label{tab:sim_auto_short_collapsed_ploidy2}
\begin{tabular}{l c c r r r r r}
\toprule
\textbf{Method} & \textbf{Ploidy} & \textbf{Cov.} & \textbf{\# Phased} & \textbf{\% Phased} & \textbf{\# Blocks} & \textbf{Block N50} & \textbf{Max Len.} \\
\midrule
\phc{}-short & 2 & 3 & 1637.533 & 72.378 & 318.833 & 4.767 & 19.433 \\
 &  & 5 & 1776.433 & 79.613 & 282.267 & 6.700 & 25.133 \\
 &  & 10 & 1836.200 & 83.402 & 242.067 & 8.767 & 29.300 \\
 &  & 20 & 1853.167 & 85.050 & 222.200 & 9.933 & 31.867 \\
 &  & 40 & 1859.167 & 85.640 & 213.100 & 10.533 & 35.033 \\
\midrule
WhatsHap &  & 3 & 1637.500 & 72.377 & 318.833 & 4.767 & 19.433 \\
 &  & 5 & 1776.433 & 79.613 & 282.267 & 6.700 & 25.133 \\
 &  & 10 & 1836.200 & 83.402 & 242.067 & 8.767 & 29.300 \\
 &  & 20 & 1853.167 & 85.050 & 222.200 & 9.933 & 31.867 \\
 &  & 40 & 1860.933 & 85.689 & 212.233 & 10.667 & 35.100 \\
\midrule
H-PoPG &  & 3 & 1669.300 & 73.484 & 384.867 & 4.567 & 19.333 \\
 &  & 5 & 1784.333 & 79.931 & 304.733 & 6.567 & 25.133 \\
 &  & 10 & 1836.600 & 83.420 & 243.233 & 8.767 & 29.300 \\
 &  & 20 & 1853.267 & 85.052 & 222.333 & 9.933 & 31.867 \\
 &  & 40 & 1860.967 & 85.690 & 212.300 & 10.667 & 35.100 \\
\midrule
HapTreeX &  & 3 & 1648.600 & 72.894 & 320.533 & 4.800 & 19.433 \\
 &  & 5 & 1786.567 & 80.098 & 283.333 & 6.700 & 25.133 \\
 &  & 10 & 1845.133 & 83.844 & 242.833 & 8.767 & 29.500 \\
 &  & 20 & 1861.633 & 85.484 & 222.633 & 9.933 & 31.867 \\
 &  & 40 & 1869.133 & 86.132 & 212.467 & 10.733 & 35.367 \\
\midrule
\bottomrule
\end{tabular}
\end{table*}

\begin{table*}[!h]
\centering
\scriptsize
\setlength{\tabcolsep}{4pt}
\renewcommand{\arraystretch}{1.05}
\caption{Detailed haplotype assembly results for diploid autopolyploidy long-read simulated dataset. 
The values are averaged over mutation rates and samples.
The \textbf{\# Phased} column denotes the number of SNPs assigned to any block and \textbf{\% Phased} is the corresponding fraction of all SNPs ($100\times \#\text{Phased}/\#\text{SNPs}$).
The \textbf{\# Blocks} column gives the number of distinct phased blocks.
Block lengths are computed as the number of consecutive SNPs that are phased together.
\textbf{Block N50} is the block length $L$ such that, the cumulative sum of sorted haplotype blocks (in descending order) is 50\% of the total phased length at $L$. 
\textbf{Max Len.} is the length (in SNP count) of the largest phased block.}
\label{tab:sim_auto_long_collapsed_ploidy2}
\begin{tabular}{l c c r r r r r}
\toprule
\textbf{Method} & \textbf{Ploidy} & \textbf{Cov.} & \textbf{\# Phased} & \textbf{\% Phased} & \textbf{\# Blocks} & \textbf{Block N50} & \textbf{Max Len.} \\
\midrule
\phc{}-long & 2 & 3 & 1923.322 & 100 & 1 & 1923.322 & 1923.322 \\
 &  & 5 & 1923.322 & 100 & 1 & 1923.322 & 1923.322 \\
 &  & 10 & 1923.322 & 100 & 1 & 1923.322 & 1923.322 \\
 &  & 20 & 1923.322 & 100 & 1 & 1923.322 & 1923.322 \\
 &  & 40 & 1923.322 & 100 & 1 & 1923.322 & 1923.322 \\
\midrule
WhatsHap &  & 3 & 1768.814 & 91.735 & 11.542 & 115.153 & 267.237 \\
 &  & 5 & 1877.915 & 97.604 & 5.136 & 357.254 & 568.271 \\
 &  & 10 & 1913.678 & 99.508 & 1.746 & 1587.203 & 1591.017 \\
 &  & 20 & 1918.356 & 99.686 & 1.322 & 1,847 & 1847.136 \\
 &  & 40 & 1920.712 & 99.864 & 1.051 & 1897.254 & 1897.254 \\
\midrule
H-PoPG &  & 3 & 1800.390 & 93.274 & 70.864 & 133.169 & 321.898 \\
 &  & 5 & 1888.424 & 98.100 & 23.271 & 516.458 & 713.593 \\
 &  & 10 & 1914.864 & 99.548 & 3.593 & 1853.932 & 1856.339 \\
 &  & 20 & 1918.729 & 99.700 & 1.932 & 1900.136 & 1900.271 \\
 &  & 40 & 1920.847 & 99.869 & 1.186 & 1917.254 & 1917.254 \\
\midrule
HapTreeX &  & 3 & 1398.763 & 73.151 & 15.203 & 13.559 & 72.542 \\
 &  & 5 & 1448.780 & 81.787 & 7.559 & 36.119 & 136.339 \\
 &  & 10 & 1590.475 & 89.937 & 2.305 & 277.949 & 433.729 \\
 &  & 20 & 1913.746 & 99.436 & 1.593 & 1794.102 & 1794.271 \\
 &  & 40 & 1918.881 & 99.744 & 1.136 & 1907.508 & 1907.508 \\
\midrule
\bottomrule
\end{tabular}
\end{table*}

\clearpage

\begin{table*}[t]
\centering
\scriptsize
\setlength{\tabcolsep}{4pt}
\renewcommand{\arraystretch}{1.05}
\caption{Detailed haplotype assembly results for autopolyploidy short-read simulated dataset. The values are averaged over mutation rates and samples.
The \textbf{\# Phased} column denotes the number of SNPs assigned to any block and \textbf{\% Phased} is the corresponding fraction of all SNPs ($100\times \#\text{Phased}/\#\text{SNPs}$).
The \textbf{\# Blocks} column gives the number of distinct phased blocks.
Block lengths are computed as the number of consecutive SNPs that are phased together.
\textbf{Block N50} is the block length $L$ such that, the cumulative sum of sorted haplotype blocks (in descending order) is 50\% of the total phased length at $L$. 
\textbf{Max Len.} is the length (in SNP count) of the largest phased block.}
\label{tab:sim_auto_short_collapsed}
\begin{tabular}{l c c r r r r r}
\toprule
\textbf{Method} & \textbf{Ploidy} & \textbf{Cov.} & \textbf{\# Phased} & \textbf{\% Phased} & \textbf{\# Blocks} & \textbf{Block N50} & \textbf{Max Len.} \\
\midrule
\phc{}-short & 3 & 3 & 1675.033 & 72.969 & 318.300 & 5 & 19.800 \\
 &  & 5 & 1813.800 & 79.925 & 282.267 & 6.833 & 25.833 \\
 &  & 10 & 1876.567 & 84.062 & 239.500 & 9.100 & 32.467 \\
 &  & 20 & 1893.933 & 85.384 & 222.800 & 10.167 & 35.767 \\
 &  & 40 & 1901.100 & 86.063 & 211.933 & 10.900 & 38.700 \\
 & 4 & 3 & 1699.567 & 72.649 & 321.300 & 4.967 & 19 \\
 &  & 5 & 1831.700 & 79.343 & 282.967 & 6.767 & 26.733 \\
 &  & 10 & 1893.567 & 83.179 & 240.533 & 8.867 & 32.600 \\
 &  & 20 & 1908.600 & 84.376 & 222.233 & 10.133 & 36.800 \\
 &  & 40 & 1915.233 & 85.132 & 209.700 & 10.833 & 41.433 \\
 & 6 & 3 & 1656.467 & 73.820 & 317.867 & 5 & 19.133 \\
 &  & 5 & 1781.633 & 80.433 & 281.100 & 6.633 & 24.467 \\
 &  & 10 & 1843.867 & 84.195 & 245.133 & 8.667 & 31.800 \\
 &  & 20 & 998.500 & 78.966 & 197.200 & 5.200 & 17.500 \\
 &  & 40 & 1006.400 & 80.014 & 189.750 & 5.550 & 20.200 \\
\midrule
WhatsHap & 3 & 3 & 1673.567 & 72.922 & 1574.600 & 1 & 4.467 \\
 &  & 5 & 1804.833 & 79.636 & 1443.100 & 1 & 6.867 \\
 &  & 10 & 1819.300 & 82.192 & 909.233 & 2.333 & 11.067 \\
 &  & 20 & 1827.200 & 82.993 & 511.767 & 4.033 & 18.567 \\
 &  & 40 & 1877.300 & 85.120 & 398.500 & 5.633 & 21.367 \\
 & 4 & 3 & 1698.933 & 72.631 & 1693.067 & 1 & 2.600 \\
 &  & 5 & 1831.267 & 79.296 & 1765.600 & 1 & 4.133 \\
 &  & 10 & 1885.033 & 82.895 & 1375.233 & 1 & 7.300 \\
 &  & 20 & 1874.700 & 83.305 & 900.367 & 2.333 & 11.167 \\
 &  & 40 & 1839.400 & 82.597 & 669.400 & 3.267 & 14.567 \\
 & 6 & 3 & 1656.033 & 73.805 & 1656.033 & 1 & 1 \\
 &  & 5 & 1781.433 & 80.418 & 1781.200 & 1 & 1.200 \\
 &  & 10 & 1843.800 & 84.193 & 1808.800 & 1 & 4 \\
 &  & 20 & 998.300 & 78.956 & 762.900 & 1.050 & 4.850 \\
 &  & 40 & 1003.550 & 79.841 & 715 & 1.400 & 5.850 \\
\midrule
H-PoPG & 3 & 3 & 1706.500 & 74.075 & 655.467 & 3 & 14.300 \\
 &  & 5 & 1822.333 & 80.281 & 498.333 & 4.633 & 21 \\
 &  & 10 & 1876.833 & 84.074 & 299 & 8.300 & 31.167 \\
 &  & 20 & 1893.933 & 85.384 & 239.300 & 10.033 & 35.400 \\
 &  & 40 & 1901.667 & 86.078 & 219.133 & 10.700 & 38.633 \\
 & 4 & 3 & 1731.067 & 73.680 & 869 & 2.233 & 10.300 \\
 &  & 5 & 1839.300 & 79.608 & 720.433 & 3 & 15.467 \\
 &  & 10 & 1,894 & 83.196 & 404.500 & 6.667 & 28.833 \\
 &  & 20 & 1908.733 & 84.379 & 264.133 & 9.300 & 36.033 \\
 &  & 40 & 1915.733 & 85.147 & 225.633 & 10.600 & 41.467 \\
 & 6 & 3 & 1684.967 & 74.841 & 962.567 & 2 & 8.833 \\
 &  & 5 & 1789.733 & 80.745 & 967.300 & 2 & 10.567 \\
 &  & 10 & 1844.200 & 84.209 & 721.733 & 3.300 & 18.767 \\
 &  & 20 & 998.550 & 78.969 & 324.200 & 3.550 & 15.900 \\
 &  & 40 & 1006.400 & 80.014 & 246.100 & 4.650 & 18.950 \\
\midrule
HapTreeX & 3 & 3 & 1688.533 & 73.669 & 320.800 & 5.033 & 19.533 \\
 &  & 5 & 1827.033 & 80.621 & 284.033 & 6.833 & 26.133 \\
 &  & 10 & 1889.800 & 84.777 & 241 & 9.100 & 33.033 \\
 &  & 20 & 1906.900 & 86.079 & 224.100 & 10.233 & 36.133 \\
 &  & 40 & 1913.967 & 86.723 & 212.867 & 10.933 & 39.067 \\
 & 4 & 3 & 1713.867 & 73.231 & 323.467 & 5 & 19 \\
 &  & 5 & 1847.100 & 79.968 & 284.333 & 6.833 & 26.733 \\
 &  & 10 & 1907.900 & 83.783 & 241.500 & 8.867 & 32.600 \\
 &  & 20 & 1922.767 & 85.003 & 223.233 & 10.200 & 36.800 \\
 &  & 40 & 1929.267 & 85.767 & 210.133 & 10.933 & 41.533 \\
 & 6 & 3 & 568.600 & 44.035 & 154.633 & 2.167 & 7.233 \\
 &  & 5 & 622.933 & 48.736 & 148.633 & 2.667 & 8.867 \\
 &  & 10 & 657.533 & 51.734 & 138.400 & 3.167 & 10.433 \\
 &  & 20 & 1004.400 & 79.404 & 198.050 & 5.150 & 17.500 \\
 &  & 40 & 1012.300 & 80.500 & 190.650 & 5.550 & 20.200 \\
\midrule
\bottomrule
\end{tabular}
\end{table*}

\clearpage

\begin{table*}[t]
\centering
\scriptsize
\setlength{\tabcolsep}{4pt}
\renewcommand{\arraystretch}{1.05}
\caption{Detailed haplotype assembly results for autopolyploidy long-read simulated dataset. The values are averaged over mutation rates and samples.
The \textbf{\# Phased} column denotes the number of SNPs assigned to any block and \textbf{\% Phased} is the corresponding fraction of all SNPs ($100\times \#\text{Phased}/\#\text{SNPs}$).
The \textbf{\# Blocks} column gives the number of distinct phased blocks.
Block lengths are computed as the number of consecutive SNPs that are phased together.
\textbf{Block N50} is the block length $L$ such that, the cumulative sum of sorted haplotype blocks (in descending order) is 50\% of the total phased length at $L$. 
\textbf{Max Len.} is the length (in SNP count) of the largest phased block.}
\label{tab:sim_auto_long_collapsed}
\begin{tabular}{l c c r r r r r}
\toprule
\textbf{Method} & \textbf{Ploidy} & \textbf{Cov.} & \textbf{\# Phased} & \textbf{\% Phased} & \textbf{\# Blocks} & \textbf{Block N50} & \textbf{Max Len.} \\
\midrule
\phc{}-long & 3 & 3 & 1585.925 & 100 & 1 & 1585.925 & 1585.925 \\
 &  & 5 & 875.667 & 100 & 1 & 875.667 & 875.667 \\
 &  & 10 & 1025.565 & 100 & 1 & 1025.565 & 1025.565 \\
 &  & 20 & 1127.350 & 100 & 1 & 1127.350 & 1127.350 \\
 &  & 40 & 1127.350 & 100 & 1 & 1127.350 & 1127.350 \\
 & 4 & 3 & 377.600 & 100 & 1 & 377.600 & 377.600 \\
 &  & 5 & 2006.900 & 100 & 1 & 2006.900 & 2006.900 \\
 &  & 10 & 881 & 100 & 1 & 881 & 881 \\
 &  & 20 & 370.500 & 100 & 1 & 370.500 & 370.500 \\
 &  & 40 & 370.500 & 100 & 1 & 370.500 & 370.500 \\
 & 6 & 3 & 1023.172 & 100 & 1 & 1023.172 & 1023.172 \\
 &  & 5 & 1211.212 & 100 & 1 & 1211.212 & 1211.212 \\
 &  & 10 & 1505.781 & 100 & 1 & 1505.781 & 1505.781 \\
 &  & 20 & 1653.556 & 100 & 1 & 1653.556 & 1653.556 \\
 &  & 40 & 1625.600 & 100 & 1 & 1625.600 & 1625.600 \\
\midrule
WhatsHap & 3 & 3 & 1465.600 & 92.204 & 1307.925 & 1 & 18.900 \\
 &  & 5 & 840.800 & 96.299 & 522.533 & 1 & 30.367 \\
 &  & 10 & 992.609 & 96.736 & 94.609 & 40.304 & 103.478 \\
 &  & 20 & 1119.250 & 99.193 & 13.300 & 437.600 & 511.150 \\
 &  & 40 & 1,125 & 99.761 & 8.500 & 982.600 & 996.650 \\
 & 4 & 3 & 349.200 & 92.469 & 347.800 & 1 & 2 \\
 &  & 5 & 1966.883 & 97.918 & 1836.217 & 1 & 18.983 \\
 &  & 10 & 854.333 & 97.416 & 359.733 & 4.333 & 37.600 \\
 &  & 20 & 365 & 98.535 & 32.600 & 21.400 & 62.900 \\
 &  & 40 & 369 & 99.599 & 12.900 & 94 & 147.600 \\
 & 6 & 3 & 937.448 & 92.326 & 937.448 & 1 & 1 \\
 &  & 5 & 1183.212 & 97.825 & 1183.212 & 1 & 1 \\
 &  & 10 & 1497.688 & 99.481 & 1450.906 & 1 & 11.438 \\
 &  & 20 & 1612.852 & 97.721 & 450.333 & 13.185 & 92.667 \\
 &  & 40 & 1616.733 & 99.417 & 46.600 & 317.933 & 467.733 \\
\midrule
H-PoPG & 3 & 3 & 1489.700 & 93.703 & 245.225 & 19.700 & 123.425 \\
 &  & 5 & 858.467 & 98.027 & 71.533 & 57.100 & 152.567 \\
 &  & 10 & 1019.913 & 99.394 & 9.217 & 665.391 & 679.783 \\
 &  & 20 & 1123.650 & 99.730 & 3.450 & 1095.150 & 1095.150 \\
 &  & 40 & 1125.350 & 99.780 & 2.550 & 1108.100 & 1108.100 \\
 & 4 & 3 & 351.400 & 93.061 & 117.400 & 4.400 & 26.200 \\
 &  & 5 & 1977.450 & 98.417 & 367.200 & 20.950 & 125.083 \\
 &  & 10 & 878.533 & 99.755 & 34.800 & 153.400 & 266.800 \\
 &  & 20 & 370 & 99.872 & 3.200 & 321.400 & 321.400 \\
 &  & 40 & 370 & 99.870 & 1.600 & 369.100 & 369.100 \\
 & 6 & 3 & 946.345 & 93.128 & 398.379 & 3.207 & 21.448 \\
 &  & 5 & 1187.091 & 98.074 & 433.455 & 4.091 & 43.091 \\
 &  & 10 & 1499.344 & 99.591 & 186.250 & 46.531 & 144.250 \\
 &  & 20 & 1649.852 & 99.748 & 22.852 & 648.704 & 764.704 \\
 &  & 40 & 1,624 & 99.859 & 8.133 & 1489.333 & 1489.333 \\
\midrule
HapTreeX & 3 & 3 & 489.375 & 53.808 & 12.850 & 9.175 & 38.600 \\
 &  & 5 & 694.600 & 65.447 & 7.233 & 25.767 & 75.233 \\
 &  & 10 & 964.739 & 85.316 & 2.478 & 248.739 & 351.696 \\
 &  & 20 & 1121.350 & 99.547 & 1.600 & 1041.500 & 1,042 \\
 &  & 40 & 1123.450 & 99.626 & 1.300 & 1100.350 & 1100.350 \\
 % & 4 & 3 & 0 & 0 & 0 & 0 & 0 \\
 & 4 & 10 & 243.200 & 65.638 & 3.667 & 67 & 95.733 \\
 &  & 20 & 369.300 & 99.682 & 1.900 & 309.300 & 309.300 \\
 &  & 40 & 369.700 & 99.785 & 1.400 & 346.600 & 346.600 \\
\midrule
\bottomrule
\end{tabular}
\end{table*}

\clearpage

\begin{table*}[t]
\centering
\scriptsize
\setlength{\tabcolsep}{4pt}
\renewcommand{\arraystretch}{1.05}
\caption{Detailed haplotype assembly results for allopolyploidy short-read simulated dataset. The values are averaged over subgenomes, mutation rates and samples.
The \textbf{\# Phased} column denotes the number of SNPs assigned to any block and \textbf{\% Phased} is the corresponding fraction of all SNPs ($100\times \#\text{Phased}/\#\text{SNPs}$).
The \textbf{\# Blocks} column gives the number of distinct phased blocks.
Block lengths are computed as the number of consecutive SNPs that are phased together.
\textbf{Block N50} is the block length $L$ such that, the cumulative sum of sorted haplotype blocks (in descending order) is 50\% of the total phased length at $L$. 
\textbf{Max Len.} is the length (in SNP count) of the largest phased block.}
\label{tab:sim_allo_short_collapsed}
\begin{tabular}{l c c r r r r r}
\toprule
\textbf{Method} & \textbf{Ploidy} & \textbf{Cov.} & \textbf{\# Phased} & \textbf{\% Phased} & \textbf{\# Blocks} & \textbf{Block N50} & \textbf{Max Len.} \\
\midrule
\phc{}-short & 3 & 5 & 159.300 & 47.866 & 61.925 & 2.525 & 4.850 \\
 &  & 10 & 171.625 & 51.077 & 64.925 & 2.525 & 5.025 \\
 &  & 20 & 178.125 & 52.823 & 66.225 & 2.525 & 5.125 \\
 &  & 40 & 182.100 & 53.999 & 67.350 & 2.525 & 5.325 \\
 & 4 & 5 & 162.275 & 48.436 & 62.900 & 2.525 & 4.950 \\
 &  & 10 & 174.500 & 51.802 & 66.075 & 2.525 & 5.100 \\
 &  & 20 & 181.850 & 53.850 & 67.825 & 2.500 & 5.325 \\
 &  & 40 & 185.325 & 54.950 & 68.275 & 2.525 & 5.425 \\
 & 6 & 5 & 263.250 & 52.821 & 99.600 & 2.500 & 5.750 \\
 &  & 10 & 288.025 & 57.505 & 104.675 & 2.500 & 6.075 \\
 &  & 20 & 299.800 & 59.911 & 106.400 & 2.525 & 6.450 \\
 &  & 40 & 306.800 & 61.192 & 106.975 & 2.525 & 6.550 \\
\midrule
WhatsHap & 3 & 5 & 159.200 & 47.845 & 127.175 & 1.150 & 3.525 \\
 &  & 10 & 171.250 & 50.997 & 102.875 & 2 & 4.050 \\
 &  & 20 & 177.275 & 52.622 & 89.600 & 2.025 & 4.400 \\
 &  & 40 & 181.350 & 53.814 & 82.150 & 2.125 & 4.875 \\
 & 4 & 5 & 162.175 & 48.416 & 155.725 & 1 & 2.625 \\
 &  & 10 & 174.375 & 51.776 & 128.050 & 1.475 & 3.825 \\
 &  & 20 & 181.650 & 53.806 & 106.750 & 2 & 4.250 \\
 &  & 40 & 184.500 & 54.776 & 93.750 & 2 & 4.650 \\
 & 6 & 5 & 263.025 & 52.766 & 262.975 & 1 & 1.050 \\
 &  & 10 & 288 & 57.501 & 283.225 & 1 & 2.575 \\
 &  & 20 & 299.775 & 59.907 & 221.750 & 1.425 & 3.925 \\
 &  & 40 & 306.675 & 61.154 & 204.375 & 2 & 4.100 \\
\midrule
H-PoPG & 3 & 5 & 159.575 & 47.926 & 70.025 & 2.150 & 4.625 \\
 &  & 10 & 171.650 & 51.083 & 71.425 & 2.375 & 4.850 \\
 &  & 20 & 178.125 & 52.823 & 70.850 & 2.525 & 5.125 \\
 &  & 40 & 182.100 & 53.999 & 69.550 & 2.525 & 5.275 \\
 & 4 & 5 & 162.500 & 48.484 & 77.700 & 2 & 4.550 \\
 &  & 10 & 174.525 & 51.807 & 79.900 & 2.050 & 4.900 \\
 &  & 20 & 181.850 & 53.850 & 78.550 & 2.225 & 5.200 \\
 &  & 40 & 185.325 & 54.950 & 77.450 & 2.400 & 5.225 \\
 & 6 & 5 & 263.750 & 52.893 & 144.225 & 2 & 4.525 \\
 &  & 10 & 288.350 & 57.552 & 155.250 & 2 & 4.550 \\
 &  & 20 & 300.150 & 59.961 & 149.850 & 2 & 5.875 \\
 &  & 40 & 307.150 & 61.243 & 142.600 & 2.150 & 6.100 \\
\midrule
HapTreeX & 3 & 5 & 159.650 & 48.004 & 62.100 & 2.525 & 4.850 \\
 &  & 10 & 172 & 51.190 & 65.050 & 2.525 & 5.075 \\
 &  & 20 & 178.525 & 52.960 & 66.375 & 2.525 & 5.150 \\
 &  & 40 & 182.450 & 54.106 & 67.475 & 2.525 & 5.350 \\
 & 4 & 5 & 162.525 & 48.509 & 63 & 2.525 & 4.950 \\
 &  & 10 & 174.600 & 51.821 & 66.100 & 2.525 & 5.100 \\
 &  & 20 & 182.025 & 53.887 & 67.900 & 2.500 & 5.325 \\
 &  & 40 & 185.475 & 54.982 & 68.350 & 2.525 & 5.425 \\
 & 6 & 5 & 265.675 & 53.228 & 100.200 & 2.500 & 5.750 \\
 &  & 10 & 290.075 & 57.877 & 105.200 & 2.500 & 6.075 \\
 &  & 20 & 301.950 & 60.296 & 106.925 & 2.525 & 6.450 \\
 &  & 40 & 308.400 & 61.505 & 107.350 & 2.525 & 6.550 \\
\midrule
\bottomrule
\end{tabular}
\end{table*}

\clearpage

\begin{table*}[t]
\centering
\scriptsize
\setlength{\tabcolsep}{4pt}
\renewcommand{\arraystretch}{1.05}
\caption{Detailed haplotype assembly results for allopolyploidy long-read simulated dataset. The values are averaged over subgenomes, mutation rates and samples.
The \textbf{\# Phased} column denotes the number of SNPs assigned to any block and \textbf{\% Phased} is the corresponding fraction of all SNPs ($100\times \#\text{Phased}/\#\text{SNPs}$).
The \textbf{\# Blocks} column gives the number of distinct phased blocks.
Block lengths are computed as the number of consecutive SNPs that are phased together.
\textbf{Block N50} is the block length $L$ such that, the cumulative sum of sorted haplotype blocks (in descending order) is 50\% of the total phased length at $L$. 
\textbf{Max Len.} is the length (in SNP count) of the largest phased block.}
\label{tab:sim_allo_long_collapsed}
\begin{tabular}{l c c r r r r r}
\toprule
\textbf{Method} & \textbf{Ploidy} & \textbf{Cov.} & \textbf{\# Phased} & \textbf{\% Phased} & \textbf{\# Blocks} & \textbf{Block N50} & \textbf{Max Len.} \\
\midrule
\phc{}-long & 3 & 5 & 301.875 & 100 & 1 & 301.875 & 301.875 \\
 &  & 10 & 301.875 & 100 & 1 & 301.875 & 301.875 \\
 &  & 20 & 301.875 & 100 & 1 & 301.875 & 301.875 \\
 &  & 40 & 301.875 & 100 & 1 & 301.875 & 301.875 \\
 & 4 & 5 & 305.675 & 100 & 1 & 305.675 & 305.675 \\
 &  & 10 & 305.675 & 100 & 1 & 305.675 & 305.675 \\
 &  & 20 & 305.675 & 100 & 1 & 305.675 & 305.675 \\
 &  & 40 & 305.675 & 100 & 1 & 305.675 & 305.675 \\
 & 6 & 5 & 307.400 & 100 & 1 & 307.400 & 307.400 \\
 &  & 10 & 307.400 & 100 & 1 & 307.400 & 307.400 \\
 &  & 20 & 307.400 & 100 & 1 & 307.400 & 307.400 \\
 &  & 40 & 205 & 100 & 1 & 205 & 205 \\
\midrule
WhatsHap & 3 & 5 & 289.250 & 94.871 & 197.275 & 1 & 13.525 \\
 &  & 10 & 293.700 & 96.774 & 62.500 & 10.400 & 33.050 \\
 &  & 20 & 298.575 & 98.212 & 22.400 & 27.775 & 53 \\
 &  & 40 & 299.925 & 98.710 & 13.425 & 38.650 & 63.025 \\
 & 4 & 5 & 297.725 & 96.564 & 286.325 & 1 & 4.975 \\
 &  & 10 & 297.175 & 96.682 & 169.400 & 1.825 & 16.025 \\
 &  & 20 & 300 & 97.546 & 44.350 & 12.375 & 30.800 \\
 &  & 40 & 302.750 & 98.350 & 25.900 & 20.900 & 43.375 \\
 & 6 & 5 & 300.040 & 96.931 & 300.040 & 1 & 1 \\
 &  & 10 & 302.640 & 98.121 & 297.960 & 1 & 2.920 \\
 &  & 20 & 300.640 & 97.929 & 154.880 & 2.560 & 13 \\
 &  & 40 & 200.850 & 97.930 & 54 & 6.550 & 17.750 \\
\midrule
H-PoPG & 3 & 5 & 293.350 & 95.998 & 30.575 & 30.250 & 61.950 \\
 &  & 10 & 298.950 & 98.324 & 7.275 & 188.725 & 194.500 \\
 &  & 20 & 299.900 & 98.726 & 2.400 & 284.150 & 285.250 \\
 &  & 40 & 300.225 & 98.851 & 1.775 & 290.975 & 290.975 \\
 & 4 & 5 & 298.750 & 96.833 & 66.300 & 11.800 & 36.050 \\
 &  & 10 & 302.475 & 98.256 & 17.425 & 87.800 & 115.725 \\
 &  & 20 & 303.250 & 98.591 & 4.275 & 268.200 & 269.400 \\
 &  & 40 & 303.700 & 98.740 & 2.375 & 298.850 & 298.850 \\
 & 6 & 5 & 300.840 & 97.183 & 135.320 & 2.920 & 17.480 \\
 &  & 10 & 302.760 & 98.144 & 46.360 & 23.800 & 53.720 \\
 &  & 20 & 303.760 & 98.492 & 7.200 & 232.720 & 237.280 \\
 &  & 40 & 201.650 & 98.331 & 2.350 & 194.550 & 194.550 \\
\midrule
HapTreeX & 3 & 5 & 262.500 & 84.470 & 13.575 & 18.300 & 43.200 \\
 &  & 10 & 293.075 & 95.842 & 6.200 & 118.950 & 143.800 \\
 &  & 20 & 298.500 & 98.043 & 3.125 & 248.575 & 253.900 \\
 &  & 40 & 299.850 & 98.698 & 1.775 & 282.450 & 284.350 \\
 & 4 & 5 & 269.200 & 85.498 & 13.525 & 22.050 & 45.150 \\
 &  & 10 & 296.525 & 95.485 & 6.550 & 92.150 & 125.900 \\
 &  & 20 & 301.600 & 97.936 & 3.150 & 242.800 & 247.300 \\
 &  & 40 & 303.225 & 98.538 & 1.675 & 289.075 & 289.075 \\
 & 6 & 5 & 141.680 & 68.981 & 12.680 & 10.960 & 25.320 \\
 &  & 10 & 157.640 & 76.812 & 6.080 & 48.880 & 66.600 \\
 &  & 20 & 160.320 & 78.191 & 2.080 & 130.200 & 130.200 \\
 &  & 40 & 201.150 & 98.089 & 1.500 & 189.950 & 189.950 \\
\midrule
\bottomrule
\end{tabular}
\end{table*}

\clearpage
\subsection{Standard Metrics}
\label{sec:standardmetrics}
For completeness, we computed standard metrics originally developed for diploid haplotype assembly using the reconstructed haplotypes $H^*$ relative to the true haplotypes $H$ as a function of sequencing coverage.

\subsubsection{Standard MEC}
\label{sec:standardmetrics_mec}
or minimum error correction measures the minimum number of allele flips needed to make the read set consistent with $H$, reported as an average error correction over sequenced SNPs.
In our implementation, MEC was computed in the usual way without penalizing block boundaries: 
a read that spans multiple haplotype blocks is split across the blocks it covers and aligned optimally, then the MEC contributions are averaged over the read's covered columns across all blocks.
\subsubsection{Standard VER}
\label{sec:standardmetrics_mec} 
or vector error rate quantifies phasing errors only on fully phased and genotype-consistent blocks by computing the average number of haplotype switches per SNP position to make each block of $H^*$ match perfectly with the corresponding region of $H$.
We first identified fully phased blocks (regions of $H^*$ with no missing entries) then computed vector error within each such block and aggregated errors across blocks weighted by block length (i.e., by the number of SNPs in the block).
Concretely, standard VER was reported as
$$
\mathrm{VER} \;=\; \frac{\sum_{b \in \mathcal{B}} \mathrm{VE}_b}{\sum_{b \in \mathcal{B}} |b|},
$$
where $\mathcal{B}$ is the set of fully phased blocks, $\mathrm{VE}_b$ is the vector error count for block $b$, and $|b|$ is the number of SNPs in $b$.
Because standard VER is restricted to fully phased blocks, we also reported the \textbf{fully phased fraction}, defined as the fraction of SNPs contained in fully phased blocks:
$$
\mathrm{Coverage}_{\mathrm{VER}} \;=\; \frac{\sum_{b \in \mathcal{B}} |b|}{L},
$$
where $L$ indicates the number of variants.
In the plots, solid lines show the mean over replicates and shaded regions indicate $\pm 1$ standard deviation.

\begin{figure}[!h]
    \centering
    \includegraphics[width=1\linewidth]{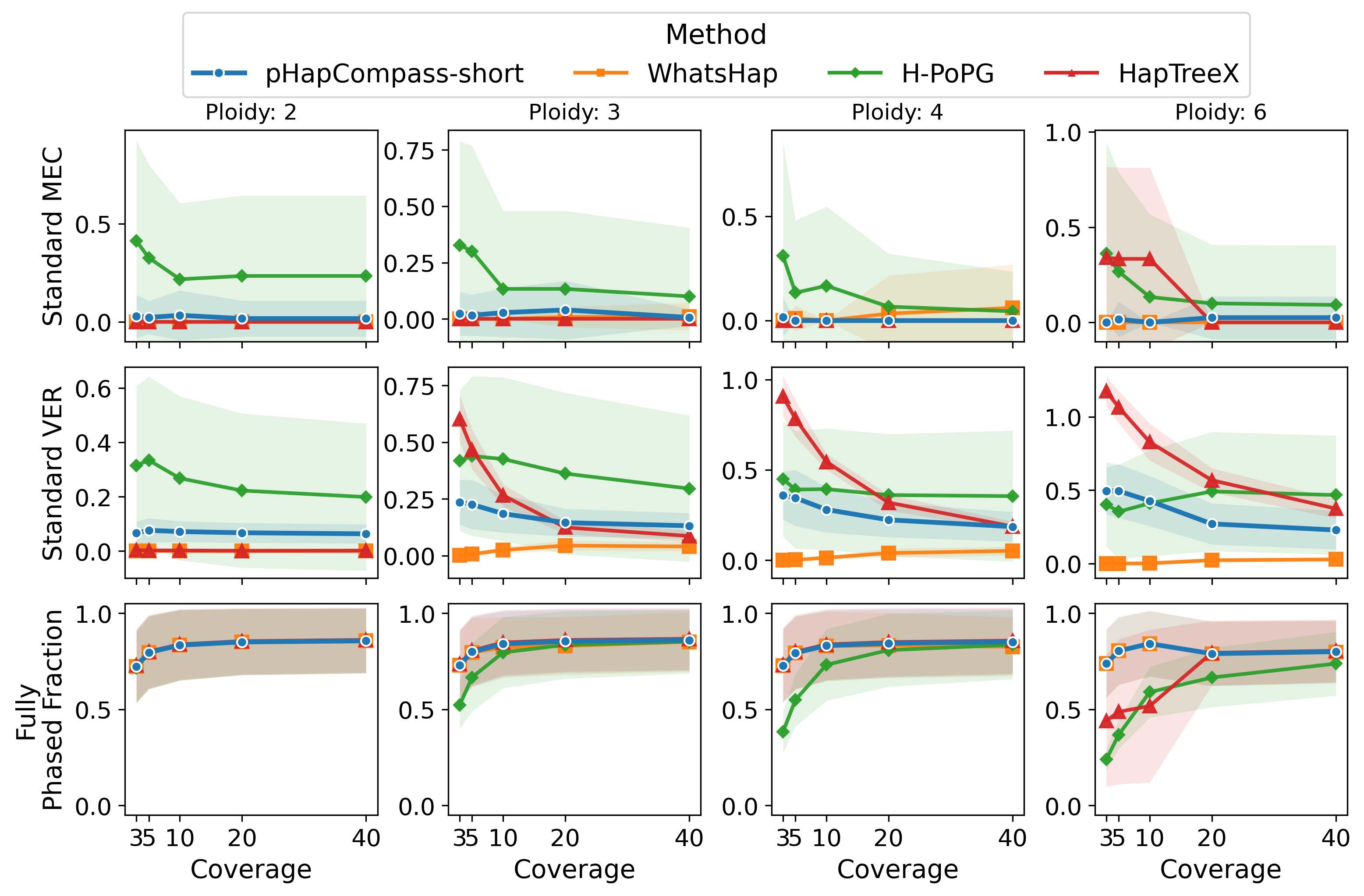}
\caption{\textbf{Autopolyploid short-read standard metrics.} Standard MEC, standard vector error rate (VER), and the fully phased fraction are shown as a function of sequencing coverage for each ploidy (columns). Rows correspond to standard MEC (top), standard VER (middle), and the fully phased fraction (bottom). Curves show the mean across samples and mutation rates; shaded bands denote ±1 standard deviation.}
    \label{fig:auto_short_standard}
\end{figure}
\clearpage

\begin{figure}[!ht]
    \centering
    \includegraphics[width=1\linewidth]{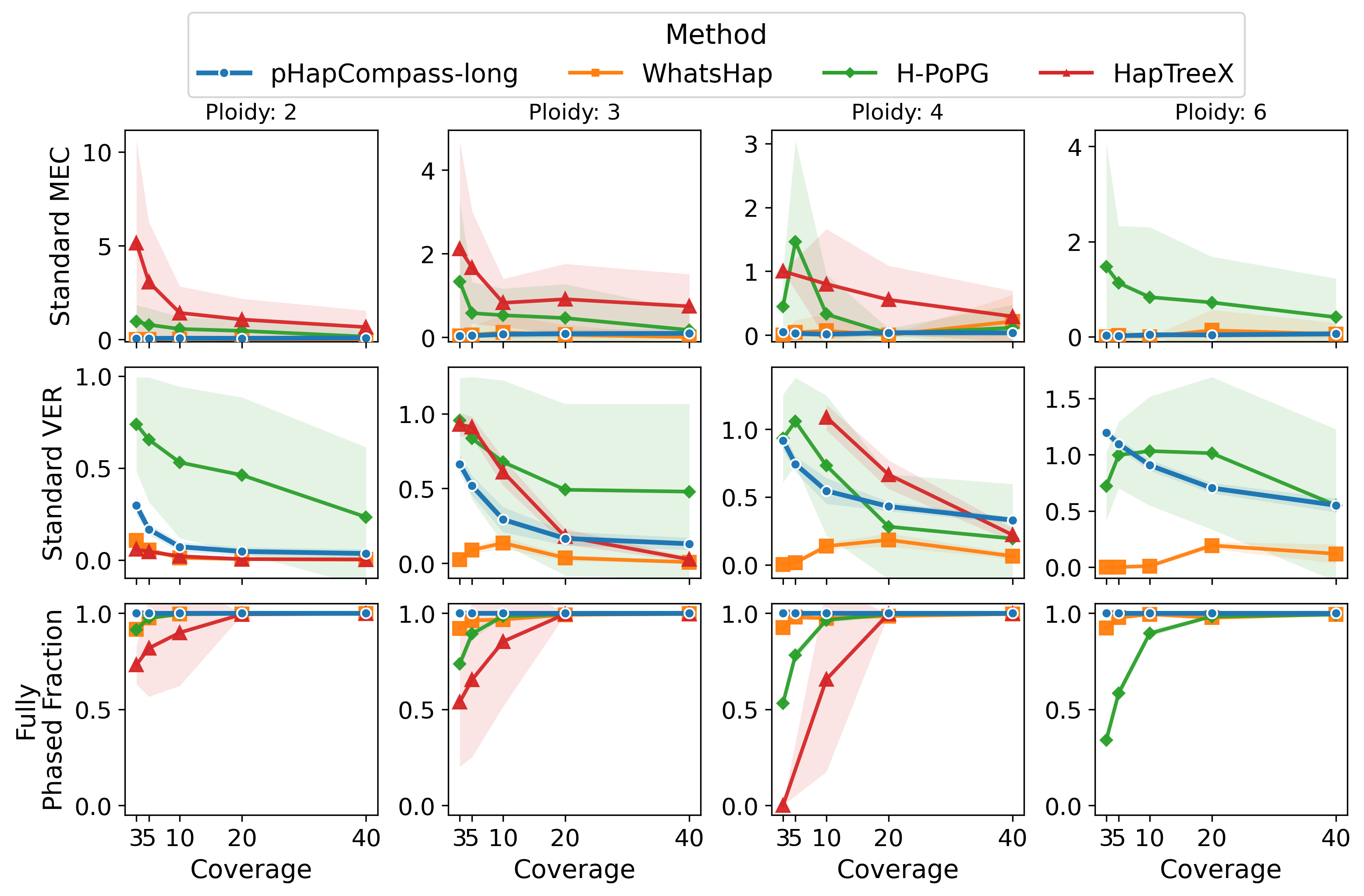}
\caption{\textbf{Autopolyploid long-read standard metrics.} Standard MEC, standard vector error rate (VER), and the fully phased fraction are shown as a function of sequencing coverage for each ploidy (columns). Rows correspond to standard MEC (top), standard VER (middle), and the fully phased fraction (bottom). Curves show the mean across samples and mutation rates; shaded bands denote ±1 standard deviation.}
\label{fig:auto_long_standard}
\end{figure}
\clearpage

\begin{figure}[!ht]
    \centering
    \includegraphics[width=1\linewidth]{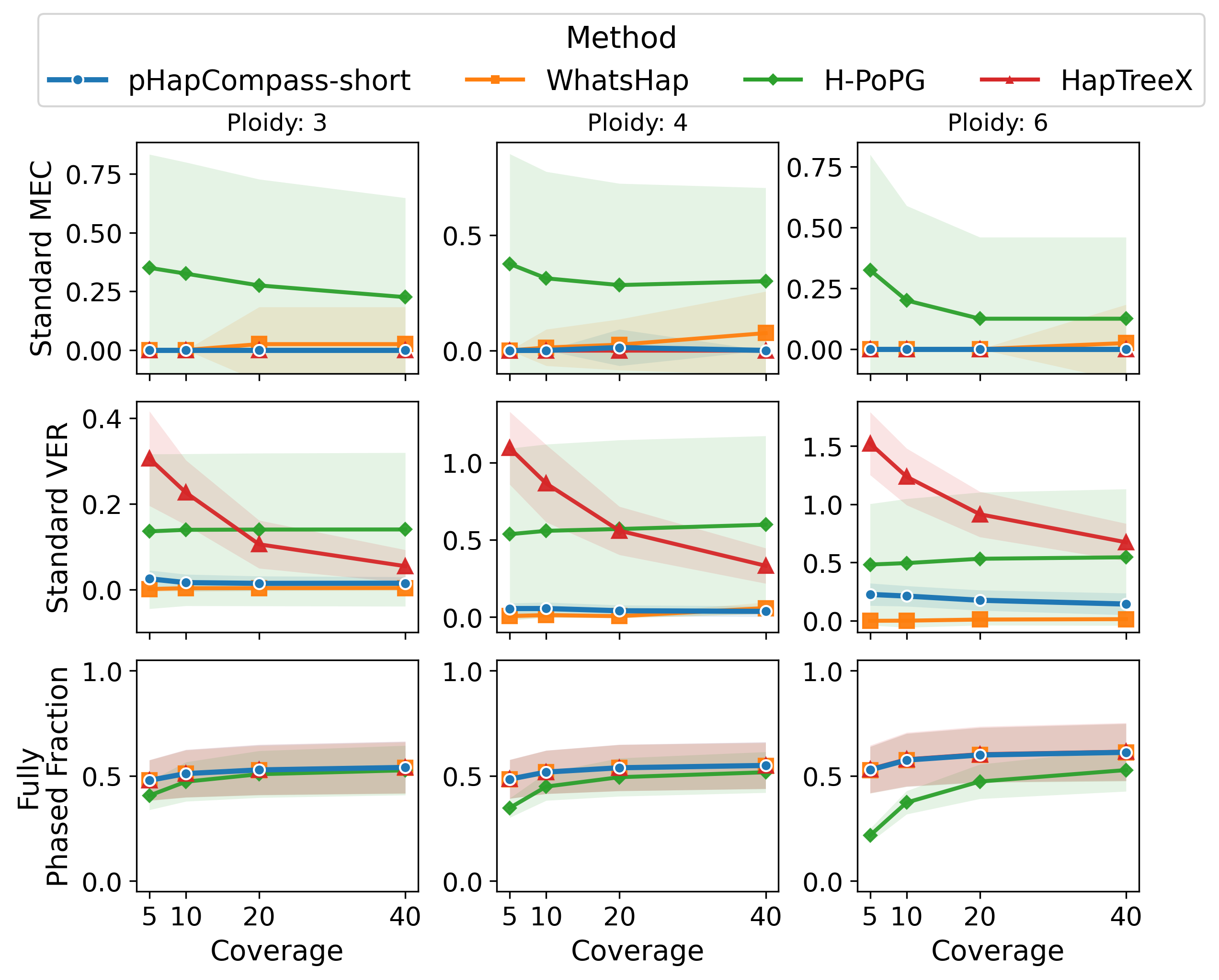}
\caption{\textbf{Allopolyploid short-read standard metrics.} Standard MEC, standard vector error rate (VER), and the fully phased fraction are shown as a function of sequencing coverage for each ploidy (columns). Rows correspond to standard MEC (top), standard VER (middle), and the fully phased fraction (bottom). Curves show the mean across samples and mutation rates; shaded bands denote ±1 standard deviation.}
    \label{fig:allo_short_standard}
\end{figure}
\clearpage

\begin{figure}[!ht]
\centering
\includegraphics[width=1\linewidth]{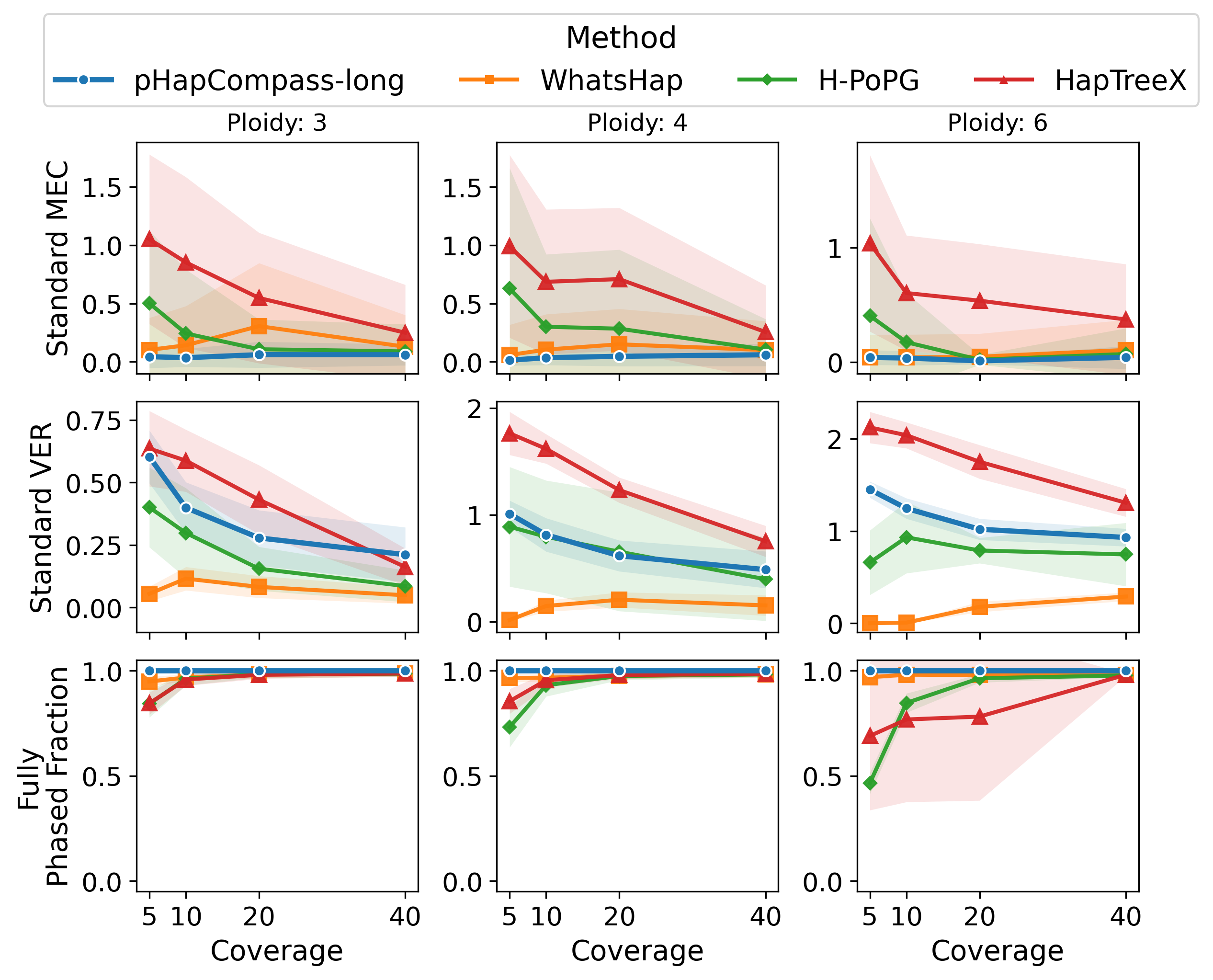}
\caption{\textbf{Allopolyploid long-read standard metrics.} Standard MEC, standard vector error rate (VER), and the fully phased fraction are shown as a function of sequencing coverage for each ploidy (columns). Rows correspond to standard MEC (top), standard VER (middle), and the fully phased fraction (bottom). Curves show the mean across samples and mutation rates; shaded bands denote ±1 standard deviation.}
    \label{fig:allo_long_standard}
\end{figure}
\clearpage

\clearpage
\subsection{Uncertainty Quantification}
\label{sec:supp_Uncertaintyquant}
\begin{figure}[!ht]
    \centering    \includegraphics[width=1\linewidth]{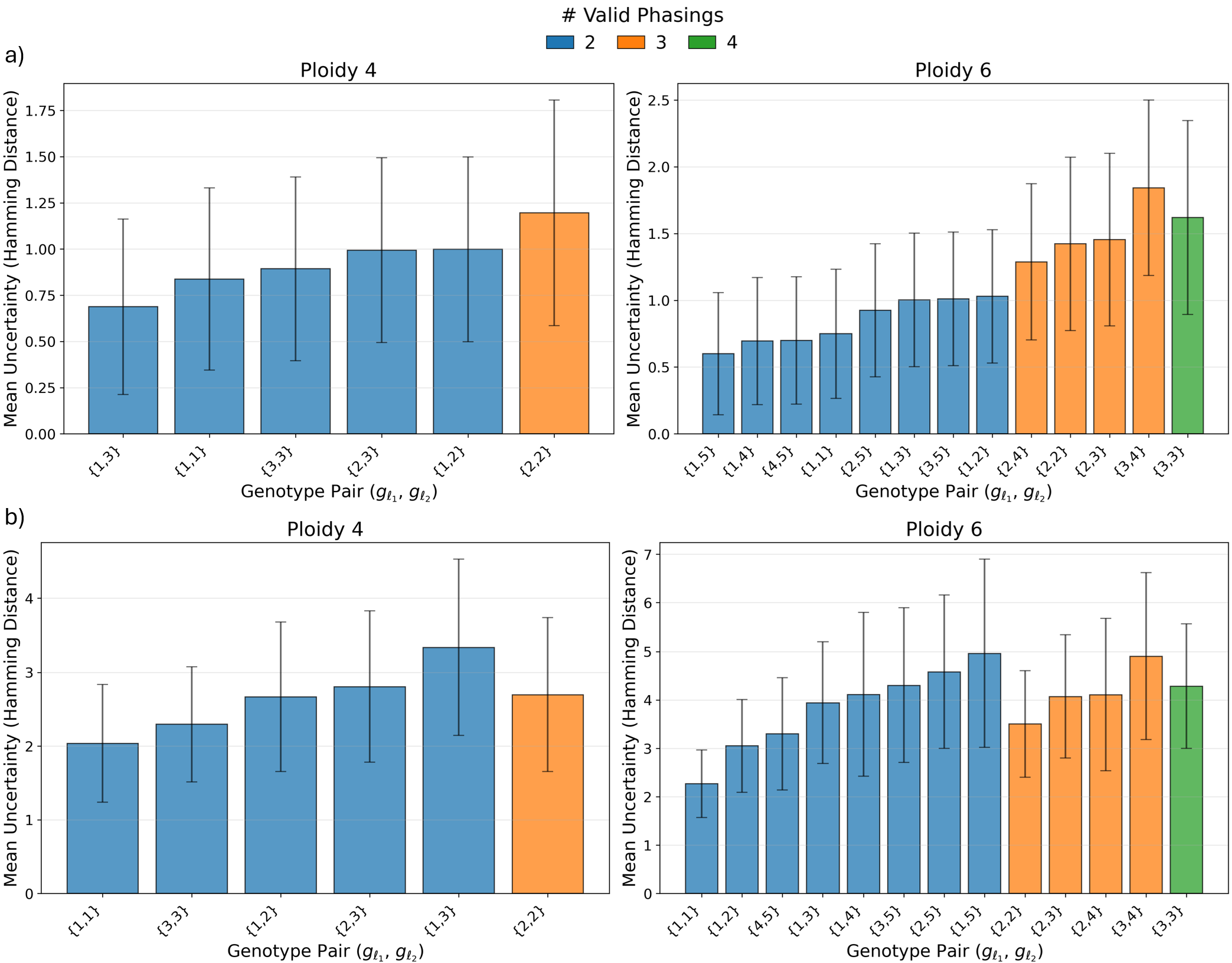}
\caption{\textbf{Uncertainty Versus Genotype Pair.}
Mean phasing uncertainty (Hamming distance) stratified by unordered genotype pairs $\{g_{\ell}, g_{\ell'}\}$, where $g_\ell \in \{1, \ldots, K-1\}$ is the alternate-allele count at SNP $\ell$ for ploidy $K$. Genotype pairs are ordered by the number of valid phasings $|\Phi_{\ell\ell'}| = \min(g_\ell, g_{\ell'}) - \max(0, g_\ell + g_{\ell'} - K) + 1$ (shown by color in legend). Uncertainty is computed as the mean Hamming distance between phasing samples from 10 FFBS runs at each SNP pair, aggregated within each genotype pair and true phasings of that SNP pair; error bars denote $0.5$ standard deviation across SNP pairs in the class. \textbf{a)} short-read data, \textbf{b)} long-read data.}
\label{fig:uncert_2}
\end{figure}

% We examined whether phasing uncertainty correlates with local alternate allele balance by stratifying SNP pairs according to their alternate-allele counts $\{g_{\ell_1},g_{\ell_2}\}$ (order-independent). 
% For each genotype class, we reported the mean pairwise phasing uncertainty measured as the average Hamming distance between posterior phasing samples (FFBS; 10 runs) for SNP pairs at loci $\ell_1,\ell_2$ and true phasing. 
% In the short-read setting (top row), uncertainty was computed only for SNP pairs jointly spanned by at least one read; consequently, the observed genotype pair distribution reflects read-span coverage and can exclude many SNP pairs regardless of their alternate allele count, which can attenuate the apparent relationship between genotype balance and uncertainty.
% In the long-read setting (bottom row), phasing is inferred as a single contiguous block, yielding uncertainty estimates for all SNP pairs; 
% here we observe a clearer increase in uncertainty for more balanced genotype pairs (those closer to $K/2$ alternate alleles), consistent with the hypothesis that locally ambiguous allele balance reduces phasing accuracy.
% Across both settings, some genotype-pair classes are intrinsically rare (and may be absent) in the simulated reference haplotypes used for this experiment (autopolyploidy; $K\in\{4,6\}$; mutation rate $0.01$; first five simulated samples).
We examined whether phasing uncertainty correlates with the number of valid phasings for different genotype pair configurations by stratifying SNP pairs according to their alternate allele counts $\{g_{\ell}, g_{\ell'}\}$ (order-independent). 
For each genotype pair, we reported the mean pairwise phasing uncertainty measured as the average Hamming distance between posterior phasing samples (FFBS; 10 runs) for SNP pairs at loci $\ell, \ell'$ and the true phasing. 
In the short-read setting (Fig.~\ref{fig:uncert_2}a), uncertainty was computed only for SNP pairs jointly spanned by at least one read (which excludes many SNP pairs at longer distances not spanned by a read). 
In the long-read setting (Fig.~\ref{fig:uncert_2}b), phasing is inferred as a single contiguous block, yielding uncertainty estimates for all SNP pairs. 
Across both settings, genotype pairs are ordered by their number of valid phasings (indicated by color in the legend).
Genotype pairs with more valid phasings exhibited higher uncertainty, consistent with the hypothesis that increased combinatorial phasing possibilities reduces phasing accuracy. 
Some genotype pairs are rare (and may be absent) in the simulated reference haplotypes used for this experiment (autopolyploidy; $K \in \{4, 6\}$; mutation rate 0.01; first five simulated samples).

\clearpage

% \section{Appendix \thesection: Experimental Results}
\section{Experimental Results}
\label{sec:expressupp}

\begin{table}[H]
\centering
\small
\caption{Chromosome identifiers and SNP counts for experimental octoploid strawberry dataset. All chromosomes were sparsified to maintain minimum 200~bp distance between consecutive SNPs.}
\begin{tabular}{lc}
\toprule
\textbf{Chromosome} & \textbf{Number of SNPs} \\
\midrule
CM113920.1 & 110,134 \\
CM113924.1 & 133,818 \\
CM113928.1 & 151,179 \\
CM113932.1 & 125,332 \\
CM113936.1 & 127,901 \\
CM113940.1 & 164,036 \\
CM113944.1 & 104,578 \\
\midrule
\textbf{Total} & \textbf{916,978} \\
\bottomrule
\end{tabular}
\label{tab:chromosome_info}
\end{table}

\begin{table}[ht]
\centering
\scriptsize
\caption{Detailed per-chromosome haplotype assembly results for all methods and coverage levels.}
\begin{tabular}{llcccccc}
\toprule
\textbf{Method} & \textbf{Chr.} & \textbf{Coverage} & \textbf{\# Phased variants} & \textbf{MEC} & \textbf{\# Blocks} & \textbf{Block N50 Length} & \textbf{Max Length} \\
\midrule
\phc{}-short & CM113920.1 & 8×  & 2,762    & 0.00 & 402    & 10 & 43 \\
            &           & 16× & 2,943    & 0.00 & 228    & 27 & 91 \\
            & CM113924.1 & 8×  & 112,835  & 0.00 & 20,747 & 7  & 87 \\
            &           & 16× & 123,669  & 0.00 & 14,046 & 17 & 151 \\
            & CM113928.1 & 8×  & 126,413  & 0.00 & 23,945 & 7  & 75 \\
            &           & 16× & 139,576  & 0.00 & 16,222 & 16 & 148 \\
            & CM113932.1 & 8×  & 105,432  & 0.00 & 19,919 & 7  & 55 \\
            &           & 16× & 116,145  & 0.00 & 13,562 & 16 & 142 \\
            & CM113936.1 & 8×  & 109,092  & 0.00 & 19,893 & 7  & 84 \\
            &           & 16× & 119,043  & 0.00 & 12,941 & 18 & 150 \\
            & CM113940.1 & 8×  & 139,367  & 0.50 & 25,454 & 7  & 78 \\
            &           & 16× & 152,564  & 0.00 & 16,898 & 17 & 161 \\
            & CM113944.1 & 8×  & 88,978   & 0.00 & 16,101 & 8  & 54 \\
            &           & 16× & 96,863   & 0.00 & 10,625 & 18 & 190 \\
\midrule
WhatsHap & CM113920.1 & 8×  & 2,784    & 0.88 & 2,784   & 1 & 1 \\
         &           & 16× & 2,951    & 0.88 & 2,951   & 1 & 1 \\
         & CM113924.1 & 8×  & 113,950  & 0.88 & 113,930 & 1 & 2 \\
         &           & 16× & 124,324  & 0.88 & 124,247 & 1 & 2 \\
         & CM113928.1 & 8×  & 127,760  & 0.88 & 127,732 & 1 & 2 \\
         &           & 16× & 140,337  & 0.88 & 140,238 & 1 & 2 \\
         & CM113932.1 & 8×  & 106,464  & 0.88 & 106,415 & 1 & 2 \\
         &           & 16× & 116,764  & 0.88 & 116,672 & 1 & 2 \\
         & CM113936.1 & 8×  & 110,139  & 0.88 & 110,110 & 1 & 2 \\
         &           & 16× & 119,619  & 0.88 & 119,541 & 1 & 2 \\
         & CM113940.1 & 8×  & 140,796  & 0.88 & 140,764 & 1 & 2 \\
         &           & 16× & 153,368  & 0.88 & 153,283 & 1 & 2 \\
         & CM113944.1 & 8×  & 89,796   & 0.88 & 89,773  & 1 & 2 \\
         &           & 16× & 97,374   & 0.88 & 97,320  & 1 & 2 \\
\midrule
H-PoPG & CM113920.1 & 8×  & 2,793    & 0.50 & 1,898   & 2 & 11 \\
       &           & 16× & 2,958    & 0.88 & 2,093   & 1 & 10 \\
       & CM113924.1 & 8×  & 114,153  & 0.00 & 79,366  & 2 & 23 \\
       &           & 16× & 124,283  & 0.00 & 86,687  & 1 & 33 \\
       & CM113928.1 & 8×  & 127,980  & 1.88 & 89,697  & 2 & 28 \\
       &           & 16× & 140,281  & 0.50 & 97,730  & 1 & 29 \\
       & CM113932.1 & 8×  & 106,675  & 0.50 & 73,732  & 2 & 18 \\
       &           & 16× & 116,646  & 0.88 & 80,839  & 1 & 26 \\
       & CM113936.1 & 8×  & 110,279  & 1.88 & 76,238  & 2 & 30 \\
       &           & 16× & 119,582  & 0.88 & 83,846  & 1 & 38 \\
       & CM113940.1 & 8×  & 141,087  & 0.00 & 100,058 & 1 & 18 \\
       &           & 16× & 153,300  & 0.50 & 107,576 & 1 & 40 \\
       & CM113944.1 & 8×  & 89,943   & 0.00 & 64,553  & 1 & 16 \\
       &           & 16× & 97,310   & 1.88 & 68,274  & 1 & 23 \\
\bottomrule
\end{tabular}
\label{tab:experimental_results_detailed}
\end{table}

\clearpage
% \section{Appendix \thesection: Runtime Analysis}
\section{Runtime analysis}
\label{sec:runtime_supp}

Runtime measurements were collected on two computing environments: (1) Intel Xeon Gold 6230 @ 2.10GHz, 8GB RAM, Ubuntu 22.04 for \phc{}-short, WhatsHap, H-PoPG, and most of the \phc{}-long experiments, and (2) Intel Xeon Gold 6242 @ 2.80GHz, 376GB RAM, Ubuntu 24.04 for remaining \phc{}-long experiments. 
HapTree-X was executed on a separate workstation. 
Due to heterogeneous hardware, absolute runtime comparisons across methods are approximate; we focus on relative scaling within each method.
% \phc{}-short demonstrates efficient scaling with runtimes from seconds (diploid) to minutes (K=6). \phc{}-long shows increased runtime due to iterative Gibbs sampling, ranging from minutes to hours depending on ploidy and coverage. WhatsHap exhibits exponential scaling with ploidy for autopolyploid configurations, with runtimes exceeding hours at K=6. HapTree-X failed to complete on K=6 autopolyploid data due to segmentation faults.

\begin{figure}[H]
    \centering
    \includegraphics[width=1\linewidth]{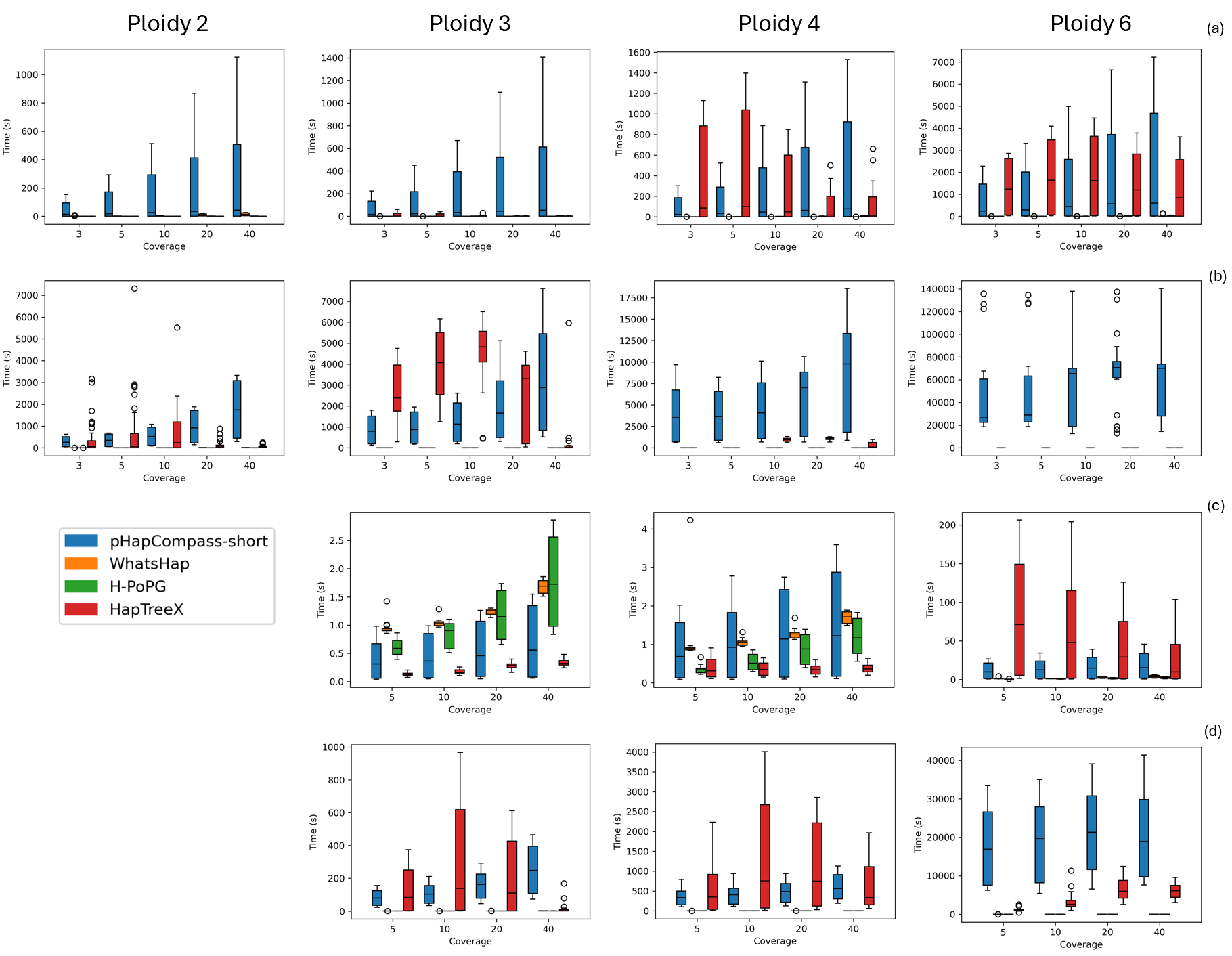}
    \caption{\textbf{Runtime analysis (seconds).} Each method -- \phc{} (blue), WhatsHap (orange), H-PoPG (green), and HapTree-X (red) -- was run across ploidies and coverages for (a) autopolyploid short-read data, (b) autopolyploid long-read data, (c) allopolyploid short-read data, and (d) allopolyploid long-read data. Measurements collected on heterogeneous hardware; values indicate relative scaling patterns.}

    \label{fig:times}
\end{figure}

\clearpage
\subsection{Runtime Analysis (normalized by SNP count)}

To disentangle algorithmic scaling from SNP-density effects, we additionally computed runtimes normalized by the number of SNPs, i.e., seconds per SNP ($t/L$). 
This normalization substantially reduced the apparent gap between auto- and allopolyploid settings observed in Fig.~\ref{fig:times}, indicating that much of the difference is driven by a differing numbers of SNPs (and hence problem size) rather than fundamentally different per-locus computational cost. 
% Across ploidies and coverages, \phc{} typically incurs higher $t/L$ than the competing methods.
% , consistent with the additional probabilistic inference it performs; however, probabilistic reasoning alone does not imply slower runtime in general, as implementation details and the structure of the underlying optimization/sampling procedure can dominate.
% Finally, in \phc{}-long development we observed slow convergence, motivating the inclusion of gradient-based updates within the Gibbs/FFBS framework, which improved practical convergence behavior while preserving the probabilistic formulation.

\begin{figure}[H]
    \centering
    \includegraphics[width=1\linewidth]{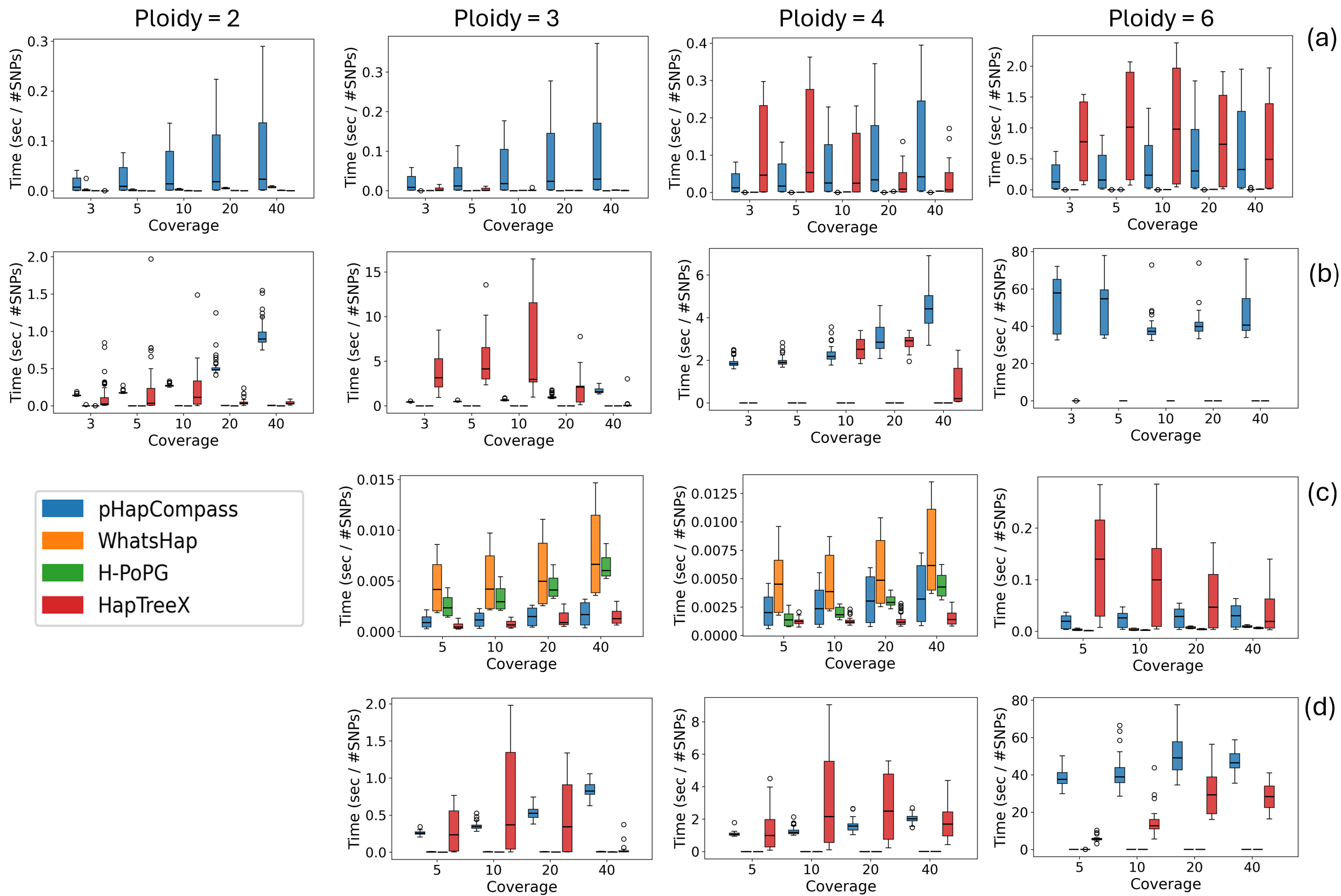}
\caption{\textbf{Runtime analysis (seconds per SNP).} Each method -- \phc{} (blue), WhatsHap (orange), H-PoPG (green), and HapTree-X (red) -- was run across ploidies and coverages for (a) autopolyploid short-read data, (b) autopolyploid long-read data, (c) allopolyploid short-read data, and (d) allopolyploid long-read data. Measurements collected on heterogeneous hardware; values indicate relative scaling patterns.}
\label{fig:times_norm}
\end{figure}

\end{document}